\documentclass[final,5p,times,twocolumn]{elsarticle}

\usepackage[table]{xcolor}
\usepackage{hyperref}
\usepackage{soul}
\usepackage{color,colortbl}
\definecolor{Gray}{gray}{0.9}

\usepackage{tabularray}
\usepackage{amsmath}
\usepackage[noabbrev,capitalise]{cleveref}

\usepackage{enumerate}
\usepackage{paralist}
\usepackage{float}
\usepackage{graphicx}

\usepackage{tcolorbox}

\newcounter{resq}
\newenvironment{resq}[1][]{\refstepcounter{resq}\par\medskip \begin{tcolorbox}[colback=black!10!white]
 \noindent \textbf{RQ~\theresq. #1} \rmfamily}{\end{tcolorbox} \medskip}
 
\newenvironment{ejemplo}[1][]{\par\smallskip \begin{tcolorbox}[boxsep=1pt,left=4pt,right=4pt,top=3pt,bottom=2pt]
 \rmfamily}{\end{tcolorbox} \smallskip}

\newcommand{\cmss}[1]{{\fontfamily{cmss}\fontseries{c}\selectfont{#1}}}

\usepackage{setspace}
\usepackage{tikz}
\usetikzlibrary{trees}
\usepackage{pgfplots}
\pgfplotsset{compat=1.18}

\usepackage{soul}
\usepackage{amsmath}

\definecolor{RoyalBlue}{RGB}{76,110,177}

\definecolor{celesteA}{RGB}{78,172,222}
\definecolor{celesteB}{RGB}{95,181,225}

\definecolor{aguaA}{RGB}{73,161,154}
\definecolor{aguaB}{RGB}{80,172,168}

\definecolor{verdeC}{RGB}{131,180,97}

\definecolor{limaAA}{RGB}{187,202,64}
\definecolor{limaA}{RGB}{152,170,54}

\definecolor{naranjaB}{RGB}{221,125,55}
\definecolor{naranjaC}{RGB}{226,145,77}
\definecolor{naranjaD}{RGB}{234,168,106}

\definecolor{fucsiaC}{RGB}{215,90,140}

\definecolor{purpuraA}{RGB}{115,49,131}
\definecolor{purpuraB}{RGB}{130,74,142}

\definecolor{azulP}{RGB}{100,147,197}

\definecolor{violetaP}{RGB}{172,118,151}
\definecolor{rosaP}{RGB}{237,126,126}
\definecolor{naranjaP}{RGB}{233,206,164}
\definecolor{amarilloP}{RGB}{234,237,191}
\definecolor{verdeP}{RGB}{206,225,193}
\definecolor{turquesaP}{RGB}{144,186,185}

\newcommand{\reducedstrut}{\vrule width 0pt height .9\ht\strutbox depth .9\dp\strutbox\relax}

\newcommand{\ttestingD}[2]{%
  \begingroup
  \setlength{\fboxsep}{.2pt}%
  \colorbox{azulP!30}{\reducedstrut{#1}$_{\text{#2}}$\/}%
  \endgroup
}

\newcommand{\tdebugD}[2]{%
  \begingroup
  \setlength{\fboxsep}{.2pt}%
  \colorbox{turquesaP!30}{\reducedstrut{#1}$_{\text{#2}}$\/}%
  \endgroup
}

\newcommand{\tvmf}[1]{%
  \begingroup
  \setlength{\fboxsep}{.2pt}%
  \colorbox{verdeC!30}{\reducedstrut{#1}\/}%
  \endgroup
}

\newcommand{\tprogver}[1]{%
  \begingroup
  \setlength{\fboxsep}{.2pt}%
  \colorbox{rosaP!30}{\reducedstrut{#1}\/}%
  \endgroup
}

\newcommand{\tstatic}[1]{%
  \begingroup
  \setlength{\fboxsep}{.2pt}%
  \colorbox{violetaP!30}{\reducedstrut{#1}\/}%
  \endgroup
}

\newcommand{\ttesting}[1]{%
  \begingroup
  \setlength{\fboxsep}{.2pt}%
  \colorbox{azulP!30}{\reducedstrut{#1}\/}%
  \endgroup
}

\newcommand{\tdebug}[1]{%
  \begingroup
  \setlength{\fboxsep}{.2pt}%
  \colorbox{turquesaP!30}{\reducedstrut{#1}\/}%
  \endgroup
}

\newcommand{\fig}[1]{%
\textcolor{gray}{(\autoref{#1})}
}

\newcommand{\gen}[1]{%
  $\rightarrow$ 
  {\fontfamily{cmss}\fontseries{c}\selectfont 
  \sethlcolor{celesteB!60}%
  \hl{$\langle$#1$\rangle$\/}}
  $\rightarrow$ 
}

\newcommand{\eval}[1]{%
  $\rightarrow$ 
  {\fontfamily{cmss}\fontseries{c}\selectfont 
  \sethlcolor{limaAA!60}%
  \hl{$\langle$#1$\rangle$\/}}  $\rightarrow$ 
}

\newcommand{\abstr}[1]{%
  $\rightarrow$ 
  {\fontfamily{cmss}\fontseries{c}\selectfont 
  \sethlcolor{fucsiaC!45}%
  \hl{$\langle$#1$\rangle$\/}}  $\rightarrow$ 
}

\newcommand{\exec}[1]{%
  $\rightarrow$ 
  {\fontfamily{cmss}\fontseries{c}\selectfont 
  \sethlcolor{purpuraB!35}%
  \hl{$\langle$#1$\rangle$\/}}  $\rightarrow$ 
}

\newcommand{\ntestingD}[3]{%
  \begingroup
  \setlength{\fboxsep}{.2pt}%
  \fcolorbox{black}{azulP!30}{\reducedstrut\textbf{#2}$_{\text{#3}}$\/}~\cite{#1}%
  \endgroup
}

\newcommand{\ntestingU}[1]{%
  \begingroup
  \setlength{\fboxsep}{.2pt}%
  \fcolorbox{black}{azulP!30}{\reducedstrut\textbf{#1}\/}~\cite{#1}%
  \endgroup
}

\newcommand{\ntesting}[2]{\ntestingD{#1}{#1}{#2}}

\newcommand{\testingD}[3]{%
  \begingroup
  \setlength{\fboxsep}{.2pt}%
  \colorbox{azulP!30}{\reducedstrut\textbf{#2}$_{\text{#3}}$\/}~\cite{#1}%
  \endgroup
}

\newcommand{\testingU}[1]{%
  \begingroup
  \setlength{\fboxsep}{.2pt}%
  \colorbox{azulP!30}{\reducedstrut\textbf{#1}\/}~\cite{#1}%
  \endgroup
}

\newcommand{\testingUD}[2]{%
  \begingroup
  \setlength{\fboxsep}{.2pt}%
  \colorbox{azulP!30}{\reducedstrut\textbf{#2}\/}~\cite{#1}%
  \endgroup
}

\newcommand{\testing}[2]{\testingD{#1}{#1}{#2}}

\newcommand{\ndebugD}[3]{%
  \begingroup
  \setlength{\fboxsep}{.2pt}%
  \fcolorbox{black}{turquesaP!30}{\reducedstrut\textbf{#2}$_{\text{#3}}$\/}~\cite{#1}%
  \endgroup
}

\newcommand{\ndebug}[2]{\ndebugD{#1}{#1}{#2}}

\newcommand{\debugD}[3]{%
  \begingroup
  \setlength{\fboxsep}{.2pt}%
  \colorbox{turquesaP!30}{\reducedstrut\textbf{#2}$_{\text{#3}}$\/}~\cite{#1}%
  \endgroup
}

\newcommand{\debug}[2]{\debugD{#1}{#1}{#2}}

\newcommand{\debugU}[1]{%
  \begingroup
  \setlength{\fboxsep}{.2pt}%
  \colorbox{turquesaP!30}{\reducedstrut\textbf{#1}\/}~\cite{#1}%
  \endgroup
}

\newcommand{\nvmfD}[3]{%
  \begingroup
  \setlength{\fboxsep}{.2pt}%
  \fcolorbox{black}{verdeC!30}{\reducedstrut\textbf{#2}$_{\text{#3}}$\/}~\cite{#1}%
  \endgroup
}

\newcommand{\nvmfU}[1]{%
  \begingroup
  \setlength{\fboxsep}{.2pt}%
  \fcolorbox{black}{verdeC!30}{\reducedstrut\textbf{#1}\/}~\cite{#1}%
  \endgroup
}

\newcommand{\nvmf}[2]{\nvmfD{#1}{#1}{#2}}

\newcommand{\vmfD}[3]{%
  \begingroup
  \setlength{\fboxsep}{.2pt}%
  \colorbox{verdeC!30}{\reducedstrut\textbf{#2}$_{\text{#3}}$\/}~\cite{#1}%
  \endgroup
}

\newcommand{\vmf}[2]{\vmfD{#1}{#1}{#2}}

\newcommand{\vmfU}[1]{%
  \begingroup
  \setlength{\fboxsep}{.2pt}%
  \colorbox{verdeC!30}{\reducedstrut\textbf{#1}\/}~\cite{#1}%
  \endgroup
}

\newcommand{\nstaticD}[3]{%
  \begingroup
  \setlength{\fboxsep}{.2pt}%
  \fcolorbox{black}{violetaP!30}{\reducedstrut\textbf{#2}$_{\text{#3}}$\/}~\cite{#1}%
  \endgroup
}

\newcommand{\nstatic}[2]{\nstaticD{#1}{#1}{#2}}

\newcommand{\staticD}[3]{%
  \begingroup
  \setlength{\fboxsep}{.2pt}%
  \colorbox{violetaP!30}{\reducedstrut\textbf{#2}$_{\text{#3}}$\/}~\cite{#1}%
  \endgroup
}

\newcommand{\static}[2]{\staticD{#1}{#1}{#2}}

\newcommand{\staticU}[1]{%
  \begingroup
  \setlength{\fboxsep}{.2pt}%
  \colorbox{violetaP!30}{\reducedstrut\textbf{#1}\/}~\cite{#1}%
  \endgroup
}

\newcommand{\nprogverU}[1]{%
  \begingroup
  \setlength{\fboxsep}{.2pt}%
  \fcolorbox{black}{rosaP!30}{\reducedstrut\textbf{#1}\/}~\cite{#1}%
  \endgroup
}

\newcommand{\nprogverUD}[2]{%
  \begingroup
  \setlength{\fboxsep}{.2pt}%
  \fcolorbox{black}{rosaP!30}{\reducedstrut\textbf{#2}\/}~\cite{#1}%
  \endgroup
}

\newcommand{\progverD}[3]{%
  \begingroup
  \setlength{\fboxsep}{.2pt}%
  \colorbox{rosaP!30}{\reducedstrut\textbf{#2}$_{\text{#3}}$\/}~\cite{#1}%
  \endgroup
}

\newcommand{\progver}[2]{\progverD{#1}{#1}{#2}}

\newcommand{\progverU}[1]{%
  \begingroup
  \setlength{\fboxsep}{.2pt}%
  \colorbox{rosaP!30}{\reducedstrut\textbf{#1}\/}~\cite{#1}%
  \endgroup
}

\newcommand{\white}[1]{%
  \begingroup
  \setlength{\fboxsep}{.2pt}%
  \colorbox{white!20}{\reducedstrut\textbf{#1}\/}~\cite{#1}%
  \endgroup
}

\newcommand{\whiteD}[2]{%
  \begingroup
  \setlength{\fboxsep}{.2pt}%
  \colorbox{white!20}{\reducedstrut\textbf{#2}\/}~\cite{#1}%
  \endgroup
}

\newcommand{\nwhite}[1]{%
  \begingroup
  \setlength{\fboxsep}{.2pt}%
  \fcolorbox{darkgray}{lightgray!20}{\reducedstrut\textbf{#1}\/}~\cite{#1}%
  \endgroup
}

\newcommand{\nwhiteD}[2]{%
  \begingroup
  \setlength{\fboxsep}{.2pt}%
  \fcolorbox{darkgray}{lightgray!20}{\reducedstrut\textbf{#2}\/}~\cite{#1}%
  \endgroup
}

\makeatletter

\tikzset{
    tasktree/.style={
        level 1/.style = {cyan},
        level 2/.style = {darkgray},
        grow via three points={one child at (0.6,-0.25) and two children at (0.6,-0.25) and (0.6,-0.75)},
        growth parent anchor=south west,
        parent anchor=south west,
        every child node/.style={anchor=west},
        every node/.style={font=\fontfamily{cmss}\fontseries{c}\selectfont},
        edge from parent path={([xshift=2ex] \tikzparentnode\tikzparentanchor) 
            |- (\tikzchildnode\tikzchildanchor)}
    }
}
\makeatother

\newlength{\htree} 
\usepackage[letterspace=210]{microtype}
\UseTblrLibrary{booktabs}

\ExplSyntaxOn
\NewDocumentCommand\FirstOfTwo{m}{
  \exp_last_unbraced:Ne \use_i:nn { #1 }
}
\NewDocumentCommand\SecondOfTwo{m}{
  \exp_last_unbraced:Ne \use_ii:nn { #1 }
}
\ExplSyntaxOff

\NewTblrTheme{fancy}{  
\SetTblrStyle{firstfoot}{fg=gray}  
\SetTblrStyle{middlefoot}{fg=gray}  
\DeclareTblrTemplate{caption-tag}{default}{}
\DeclareTblrTemplate{caption-sep}{default}{}
\DeclareTblrTemplate{caption-text}{default}{Table~\thetable: \enskip\InsertTblrText{caption}}}

\ExplSyntaxOn
\clist_gput_left:Nn \g__tblr_table_known_keys_clist { midlast-caption }
\keys_define:nn { tblr-outer } { midlast-caption .code:n = \__tblr_set_midlast_caption:n { #1 } }
\cs_new_protected:Npn \__tblr_set_midlast_caption:n #1 {
    \DefTblrTemplate{ middlehead, lasthead }{ default }
    {{Table~\thetable: \enskip #1 \enskip (Continued)
    }}
}
\ExplSyntaxOff

\begin{document}

\begin{frontmatter}
\title{Generative transformations and  patterns in LLM-native approaches for software 
verification and falsification} 
\author[1,2]{V\'{\i}ctor A. Braberman\corref{cor1}} \ead{vbraber@dc.uba.ar} 

\author[1,2]{Flavia Bonomo-Braberman} 

\author[3]{Yiannis Charalambous}

\author[4]{Juan G. Colonna}

\author[3,4]{Lucas C. Cordeiro}

\author[4]{Rosiane~de~Freitas}

\cortext[cor1]{Corresponding author. Departamento de Computaci\'on, FCEyN, Universidad de Buenos Aires. Ciudad Universitaria, Pabell\'on 0+$\infty$. Pres. Dr. Ra{\'u}l Alfons\'{\i}n s/n, (1428) Buenos Aires, Argentina.} 

\affiliation[1]{organization={Departamento de Computación, FCEyN, Universidad de Buenos Aires}, city={Buenos Aires}, country={Argentina}} 

\affiliation[2]{organization={Instituto de Investigación en Ciencias de la Computación (ICC), CONICET-UBA}, city={Buenos Aires}, country={Argentina}} 

\affiliation[3]{organization={Department of Computer Science, The University of Manchester}, city={Manchester}, country={UK}} 

\affiliation[4]{organization={Instituto de Computação (ICOMP-UFAM), Universidade Federal do Amazonas}, city={Manaus, Amazonas}, country={Brazil}} 

\begin{abstract}
The emergence of prompting as the dominant paradigm for leveraging Large Language Models (LLMs) has led to a proliferation of LLM-native software, where application behavior arises from complex, stochastic data transformations. However, the engineering of such systems remains largely exploratory and ad-hoc, hampered by the absence of conceptual frameworks, ex-ante methodologies, design guidelines, and specialized benchmarks. We argue that a foundational step towards a more disciplined engineering practice is a systematic understanding of the core functional units--generative transformations--and their compositional patterns within LLM-native applications.

Focusing on the rich domain of software verification and falsification, we conduct a secondary study of over 100 research proposals to address this gap. We first present a fine-grained taxonomy of generative transformations, abstracting prompt-based interactions into conceptual signatures. This taxonomy serves as a scaffolding to identify recurrent transformation relationship patterns--analogous to software design patterns--that characterize solution approaches in the literature. Our analysis not only validates the utility of the taxonomy but also surfaces strategic gaps and cross-dimensional relationships, offering a structured foundation for future research in modular and compositional LLM application design, benchmarking, and the development of reliable LLM-native systems. 
\end{abstract}
\end{frontmatter}

\section{Introduction}

The emergence of prompting as the dominant paradigm for interacting with large language models (LLMs) has unlocked new capabilities across diverse domains~\cite{emergent}. Seminal work demonstrated that strategically structured prompts---whether through detailed instructions or few-shot demonstrations---can elicit sophisticated reasoning and task performance without the need for model fine-tuning~\cite{Few-Shot-Brown20, nong_chain--thought_2024, CoT-Wei22}. This has led to the rapid proliferation of LLM-native software, where core functionality is achieved by orchestrating LLMs through prompts, often within complex chains of generative processing alongside conventional code~\cite{hassan2024rethinking, DSPy, Langchain}.

However, the engineering discipline for constructing such systems remains in its infancy~\cite{brabermannapoli, hassan2024rethinking}. As a widely cited industry maxim captures, ``Experimenting with LLMs is the only way to build LLM-native apps''~\cite{medium-llm-guide}. We argue this trial-and-error paradigm persists due to a critical gap: the absence of conceptual frameworks to reason about LLM-native software and ex-ante engineering methodologies.

\begin{figure}[t]
\centering
\includegraphics[width=.85\columnwidth]{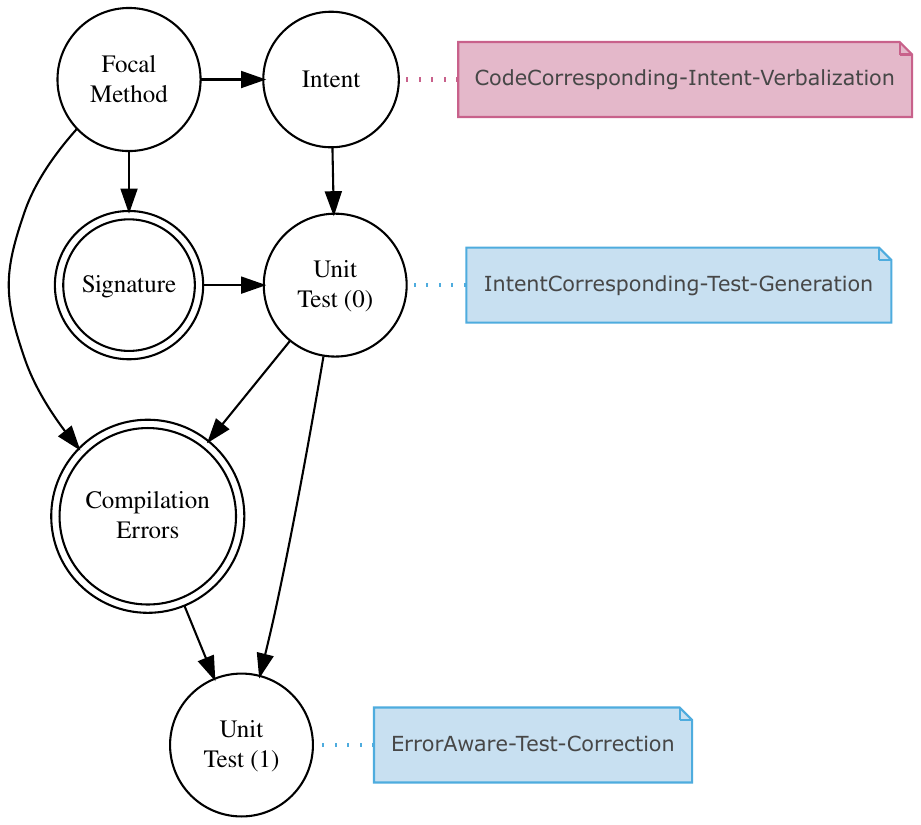}
\caption{
An abstract execution of ChatTESTER~\cite{ChatTester} modeled using our conceptual framework. This Bayesian network depicts data flow through the system, where variables represent either inputs (\texttt{FocalMethod}) or outputs from transformations. Generative transformations (labeled with types like \cmss{IntentCorresponding-Test-Generation}) produce stochastic variables through LLM-mechanized processes, while algorithmic transformations produce deterministic variables (double-circled, e.g., \texttt{CompilationErrors}). Edges denote the conceptual input signature of each transformation; for instance, \texttt{UnitTest(0)} depends on \texttt{Signature} and \texttt{Intent} according to the distribution of an {\cmss{IntentCorresponding-Test-Generation}} transformation. The model reveals two key inter-transformation patterns: \texttt{Use Verbalized SW Entity} (reasoning via intent rather than code) and \texttt{Generate \& Fix} (iterative correction using compilation errors).
}
\label{fig:ChatTester}
\end{figure}

We posit that a more principled engineering approach requires a shift in perspective: from a low-level view of prompts as data, to a view of LLM interactions as generative transformations. These are discrete, prompt-induced, LLM-mechanized operations that convert stochastically input data into output data. As a concrete example,~\autoref{fig:ChatTester} illustrates how the meso-level view and concepts proposed in this paper could be used to precisely and uniformly describe a representative execution --in this case, a solution proposal within scope, ChatTester~\cite{ChatTester}. This transformation-oriented viewpoint resonates with emerging modular and declarative frameworks (e.g.,~\cite{LMQL,cascades,DSPy}) and we argue it can bring the construction of LLM-native software closer to the cognitive way we design traditional systems--reasoning about system behavior through the composition of ``functional units''.

This work serves to demonstrate the viability and utility of the transformation-oriented viewpoint. Through a systematic analysis of the literature, we develop an initial catalog of generative transformations and their composition patterns. We argue this catalog provides immediate value for practitioners while establishing the foundational layer for a more disciplined engineering paradigm--one capable of high-level, abstract reasoning about system behavior. By providing this structured vocabulary, the catalog enables future research directions, such as: (1) specialized benchmarks and theoretical models, (2) transformation-specialized LLMs, and (3) ex-ante methodologies for reliable LLM-native systems design.

To ground our investigation, we focus on the domain of software verification and falsification. This domain is ideal for our study as it represents a critical, rapidly evolving area of software engineering~\cite{huang2024fuzz-survey,wang2023software-survey} with inherent diversity--encompassing fuzzing, formal verification, and test generation. This diversity suggests that LLM-native solutions here are likely to be sophisticated, featuring a rich variety of generative transformations and compositions.

Thus, we first address the following research question:
\begin{resq}\label{rq1}
Can prompt-based interactions with LLMs be abstractly described as the execution of generative transformations with conceptual signatures? What different types of transformations emerge from an analysis of LLM-based software verification proposals?
\end{resq}

To address RQ1, we conducted a systematic analysis of over 100 papers, characterizing prompt-based interactions as data transformations. The diversity of observed transformations motivated us to develop a structured taxonomy and a mapping.

\begin{resq}\label{rq2}
Can the identified transformation types be organized and mapped into a taxonomy that captures their shared characteristics and key differentiators?
\end{resq}

A primary utility of this taxonomy is its ability to scaffold the identification of recurring solution approaches, which we term Inter-Transformation Patterns (by analogy to software design patterns~\cite{DBLP:conf/ecoop/GammaHJV93}).

\begin{resq}\label{rq3}
To what extent can our taxonomy, combined with manual dataflow analysis, systematically surface recurrent solution patterns in existing LLM-native proposals?
\end{resq}

Addressing RQ3 helps validate the utility of our taxonomy while uncovering practical insights for modular LLM application design. To our knowledge, this is the first comprehensive effort to catalog both the transformation space and its associated design patterns.

Finally, we employ our integrated artifacts---the taxonomy, patterns, and their mappings---to conduct a cross-analysis. This analysis aims to identify significant relationships and highlight strategic gaps, thereby revealing current limitations and informing valuable future research directions.

 \subsection{Related Work}

This transformation-oriented viewpoint aligns with and complements emerging frameworks for structuring LLM applications, such as DSPy~\cite{DSPy}. While DSPy provides a programming model and compiler for optimizing the prompts that implement such transformations, our work aims to establish a foundational conceptual taxonomy of the transformations themselves. We seek to answer: What are the fundamental types of generative operations being composed? By systematically cataloging these functional units and their interconnection patterns, we provide a semantic grounding that can inform the design of future programming models, compilers, and evaluation benchmarks. In essence, if DSPy offers a ``how'' for building LLM-native software more robustly, our taxonomy provides a ``what''--a vocabulary for the core stochastic functionalities that such systems are built upon.

Unlike existing taxonomies that focus on broad Natural Language Processing (NLP) tasks, our work uncovers the actual transformations requested in prompt-based interactions.  Existing surveys do not discuss the details of  decomposition of solutions into transformations and the abstraction level of their summarized information is not adequate to answer our questions.
We are not aware of any taxonomy that would serve the preceding purposes. NLP and cognitive tasks appearing in benchmarks like~\cite{DBLP:conf/acl/Honovich0BL23,liu2024agentbench,DBLP:conf/acl/SuzgunSSGTCCLCZ23,valmeekam2023planbench} partially cover some of the transformations encountered, but most identified transformations are novel. 
Traditionally, tasks in the NLP community have been identified and typically classified in benchmarks (e.g.~\cite{DBLP:conf/acl/Honovich0BL23,liu2024agentbench,DBLP:journals/tmlr/SrivastavaRRSAF23,DBLP:conf/acl/SuzgunSSGTCCLCZ23,valmeekam2023planbench}). However, these benchmarks are neither necessarily focused on software engineering problems nor particularly designed to identify transformations that actually appear in the field. This has been partially recognized in some benchmarking papers~\cite{DBLP:conf/iclr/JainHGLYZWSSS25,kayali2024minddatagapbridging,rebmann2024evaluatingabilityllmssolve,DBLP:conf/acl/YanLWL0W0ZZSD24,yang2025structevalbenchmarkingllmscapabilities}.
However, in this paper, we pinpoint some overlaps with our proposed taxonomy. Recently,~\cite{DBLP:conf/iclr/JimenezYWYPPN24} proposes {\it SWE-bench} as an evaluation framework consisting of software engineering problems drawn from real GitHub issues. Some of the difficulties may partially overlap with the problems addressed in the scope of this mapping. However, the focus on GitHub issues implies that problems are closer to the area of software corrective and evolutive maintenance than the area of verification/falsification. More importantly, this is an evaluation framework and thus no transformation elicited by potential solution proposals are presented. 

Several surveys have also been conducted on the use of LLMs in Software Engineering papers (e.g.,~\cite{DBLP:journals/tosem/HeTL25,fan2023large-survey,hou2023large-survey,huang2024fuzz-survey,DBLP:journals/tosem/ZhouCSL25}). Although, to the best of our knowledge, none of the existing surveys performs concrete LLMs transformations identification and mapping, they are relevant related work with some elements in common with this work. 

A systematic review of the literature on the use of LLM in SE (LLM4SE) between 2017 and January 2024 is presented in~\cite{hou2023large-survey}. Key applications of LLM4SE described encompass a diverse range of 55 specific SE problems, grouped into six core SE activities. Problems in the quality assurance category overlap with those in the scope of our study. 
There is no report on how proposals structure solutions into LLM-enabled transformations.

 Within our area of interest, test generation, test adequacy, test minimization, test oracle problem, and test flakiness are problems covered by~\cite{fan2023large-survey}. Unlike this work, the focus is on comparing the effectiveness reported by each approach. Again, there is neither identification nor categorization of concrete LLMs transformations in proposals that use prompts. 

LLM-based fuzzers are analyzed in~\cite{huang2024fuzz-survey}. The focus is on how LLMs are used for fuzzing tests and on comparing LLM-based fuzzers with traditional fuzzers. Although extra detail is given for LLM-based approaches, neither systematic mapping nor taxonomy is proposed.

A survey on testing problems, which also includes debugging and repair proposals, is conducted in~\cite{wang2023software-survey}; 52 papers on testing are covered (33 published since 2022), of which approximately one-third concerned test-based LLM fine-tuning, while the remainder rely upon prompt engineering. The survey analyzes the papers from both the software testing and LLM perspectives. It presents a detailed discussion of the software testing problems for which LLMs are used, including unit test case generation, test oracle generation, system test input generation, bug analysis, and features on how these tasks are solved with the help of LLMs. In particular, it describes the commonly used LLMs, their input, the types of prompt engineering employed, and the techniques that form the concrete cognitive architecture of these proposals (e.g., pure LLMs, LLM + statistical analysis, LLM + program analysis, LLM + mutation testing, LLM + syntactic checking, LLM + differential testing). Yet, there is no focus on which concrete types of transformations are actually elicited.

In~\cite{zhang2023survey}, a systematic survey on LLM-based approaches for SE problems, including empirical evaluations, is presented. It includes 155 studies for 43 specific code-related tasks across four phases within the SE workflow: software 
requirements \& design, software development, software testing and software maintenance. Some SE problems overlap with the scope of our mapping. More precisely, fault localization, vulnerability detection, unit test generation, assertion generation, fuzzing, 
failure-inducing testing, 
penetration testing, 
property-based testing, 
mutation testing, 
GUI testing, bug report detection, bug reproduction, and bug replay.
They report on representative code-related LLMs and, related to our paper, they identify 11 classes of tasks  across four categories defined according to the input/output data type: \emph{Code-Code:} code translation, code refinement, cloze test, 
mutant generation, assertion generation; \emph{Text-Code:} code 
generation, code search; \emph{Code-Text:} code summarization; and \emph{Code-Label:} code classification, clone detection, 
defect detection. 
Almost all the tasks classes are included in our taxonomy. Moreover, their way of classifying tasks/transformations is orthogonal to ours, and we make an explicit sub-classification according to type of input-output of the task for some of the categories. 

Our pattern identification and mapping effort focuses on solutions that can be expressed as the interrelationship of transformations, including the mediation of symbolic computation. This leads to a rather systematic exploration and identification of solution patterns in solution proposals and exemplars. Although our work is inspired by the design patterns of~\cite{DBLP:conf/ecoop/GammaHJV93}, we deliberately deviate from their full documentation style. To maintain a strong, verifiable connection to the recovered transformations and  studies in scope, we focus our documentation on the core elements of name, intent, motivation, and examples. We omit traditional sections like ``Consequences'' as their speculative pros and cons would be premature for the patterns identified in our nascent and rapidly evolving domain.

Our transformation-centric approach provides a distinct analytical lens compared to existing pattern catalogs in academy\footnote{Practitioner guidance for building LLM-enabled applications are typically focused in agents and patterns are either higher-level architectural ones or prompt (or context) guidance (e.g. Open AI's ``A practical guide to building agents'' or Anthropic's ``Effective context engineering for AI agents'').}. For instance,~\cite{white2023promptpatterncatalogenhance} presents a ``Prompt Pattern Catalog'' that structures 16 effective interaction strategies for humans to interact conversationally with LLMs like ChatGPT.
That is, their catalog is not focused on the use of prompts in the context of LLM-native software broadly speaking. Having said that, links with our work can be unveiled when one think on the transformations necessary for implementing an LLM-native application that enables user intervention. For instance, the ``flipped interaction pattern'' can be implemented in LLM-native UI by a combination of our patterns, namely, \texttt{Reactively Consume Input} (in this case, whatever the human has to utter) together with the \texttt{Enrich with Precisions} pattern application for the human task in question.  

Other related work includes~\cite{DBLP:conf/agi/WrayKL25}, which proposes ``Cognitive Design Patterns'' drawn from established cognitive architectures (e.g., SOAR, ACT-R) and maps them to capabilities in agentic LLM systems. Their work demonstrates how high-level cognitive functions, such as ``Observe-Decide-Act'' or ``Episodic Memory'', are instantiated in LLM-native designs. Separately,~\cite{DBLP:journals/jss/LiuLLZZXHW25} presents an ``Agent Design Pattern Catalogue'' comprising 18 architectural patterns and a decision process that address challenges in foundation model-based agent design. Also, it is worth mentioning \cite{DBLP:journals/tmlr/SumersYN024} aimed at providing an architectural description language able to contextualize LLM agent research proposals.  

However, our methodology and results differ fundamentally in both granularity, scope and analytical approach. While~\cite{DBLP:conf/agi/WrayKL25} analyzes systems from a top-down, architecturally-derived perspective, and~\cite{DBLP:journals/jss/LiuLLZZXHW25} catalogs architectural patterns at different levels of abstraction and generality (e.g., agent adapters, reflection, agent goal creation, etc.) that are of interest for a particular application domain, we, instead, adopt a bottom-up, domain-agnostic approach focused on the interrelationship of generative transformations. That is, our transformation-centric analysis captures the emergent behavioral patterns that arise from how operations are sequenced and composed regardless of the application domain.

This transformation composition focus is a rather unique feature that enables a more systematic exploration of the LLM-native design space. It enables the (re)discovery of existing strategies--such as those encapsulated in the prompt pattern catalog--and, crucially, the identification of detailed compositional patterns that were previously uncataloged. This is precisely because it operates at a more granular level than architectural component patterns and is not constrained by pre-existing taxonomies of high-level cognitive functions. By examining how transformations interact behaviorally rather than how components are structured architecturally, our approach captures the emergent solution structures that arise from the unique possibilities of composing LLM-based and symbolic transformations.

\section{Background}

\subsection{Verification and falsification problems}

Testing is a classical and arguably the most popular approach to gaining confidence in software systems before release~\cite{DBLP:books/daglib/0020331,DBLP:books/daglib/0077983}. When emphasizing that testing should be regarded as a way to falsify software and not a way to verify it~\cite{Dijkstra} it becomes clear how research in testing has evolved by focusing on failure detection ability of test suites (automatic) generation proposals~\cite{DBLP:conf/icse/OrsoR14}. Several approaches to test suite generation have been proposed~\cite{DBLP:journals/jss/AnandBCCCGHHMOE13}. Among others, symbolic and random approaches (potentially combined)~\cite{DBLP:journals/cacm/CadarS13,DBLP:conf/pldi/GodefroidKS05,DBLP:conf/icse/PachecoLEB07},
mutation-based ones~\cite{DBLP:journals/tse/JiaH11}, and search-based approaches~\cite{DBLP:conf/sigsoft/FraserA11}. 
A particular technique of testing is fuzzing~\cite{miller_empirical_1990}, which typically involves adaptively generating a series of inputs --usually by mutating some input seeds-- to cause the Program under Test (PuT) to crash or enter a defective state.
It does so by repeatedly stimulating the program and observing its behavior in a sort of feedback loop~\cite{DBLP:journals/tse/BohmePR19,zhu_fuzzing_2022}. 
A typical problem when testing or fuzzing and looking for evidence of (functional) falsification beyond crash or memory corruption is known as the oracle problem. That is, determining, either for a set of inputs or in general terms, the correct behavior that should be exhibited by the program~\cite{barr_oracle_2015}. Metamorphic relation identification has been a way to partially circumvent the oracle problem~\cite{DBLP:journals/csur/ChenKLPTTZ18}. 
Once a failure is detected, fault localization is typically the next problem to be solved. Fault localization aims at listing candidate locations in the source code that could be claimed as likely faulty~\cite{pearson_evaluating_2017,DBLP:journals/tse/WongGLAW16}.
A related defect localization problem is that of Root Cause Analysis (RCA) when system-wide anomalous behavior is found in production (e.g.,~\cite{jayathilaka_performance_2017, soldani_anomaly_2022}). The main challenge in diagnosing the root cause of systems is the interconnected dependence between different services that could be internal or third-party.

Bug reproduction is one of software development's first stages of bug fixing when a bug is reported by software users. It involves reproducing the bug on the developer's system so that it can be further analyzed~\cite{narayanasamy_bugnet_2005}.
An associated challenge is managing multiple reports regarding the same bug, which can slow down the bug-fixing process, as it requires resolving all duplicate entries before attempting to fix the bug itself~\cite{alipour_contextual_2013,deshmukh_towards_2017, alipour_contextual_2016, jalbert_automated_2008}.

There is a wide spectrum of problems derived from the goal of analyzing code without executing it. Static analysis~\cite{DBLP:books/daglib/0098888}, generally speaking, encompasses a variety of problems and approaches ranging from detecting potential vulnerabilities or misuses in source code syntactic structures or ML-representations~\cite{amann_systematic_2019, DBLP:journals/sigplan/HovemeyerP04,   li_large-scale_2021,li_vuldeepecker_2018,s_c_johnson_lint_1978} to sound(y)~\cite{DBLP:journals/cacm/LivshitsSSLACGKMV15} computation of collecting semantics~\cite{chess_static_2004,DBLP:conf/popl/CousotC77, DBLP:conf/oopsla/MartinLL05}. 
Traditional intermediate abstractions like control-flow graphs~\cite{DBLP:conf/comop/Allen70} play central roles in many analyses. Static analysis is not limited to checking for a given property. Well-established research topics include statically built multipurpose program abstractions or extractions. For example, static slicing~\cite{weiser_program_1984} addresses the problem of identifying statically a set of relevant lines for the computation of a value in a given program location.

Program verification~\cite{DBLP:journals/cacm/Hoare69} is an area with a well established tradition and that has covered different classes of programs and properties. Software model checking~\cite{DBLP:books/daglib/0020348,DBLP:books/daglib/0007403-2,DBLP:journals/csur/JhalaM09} and deductive software verification~\cite{dafny,programproofs} are among the main trends for sequential software correctness verification. In many of those approaches, invariant generation~\cite{DBLP:conf/popl/CousotH78} is a key challenge for effective automation.

Recently, some novel application areas and technologies become relevant targets of mentioned verification and testing approaches. Among them: GUI-based/mobile applications (e.g.,~\cite{DBLP:conf/issta/MaoHJ16}), RESTfull services~\cite{DBLP:journals/tosem/GolmohammadiZA24}, smart contracts~\cite{VulSC}, and Machine-Learning systems, software and applications (e.g.,~\cite{DBLP:journals/jss/BraiekK20,DBLP:journals/csr/HuangKRSSTWY20,DBLP:journals/tosem/TangZZZGLGLMXL23,DBLP:journals/tse/ZhangHML22,DBLP:journals/infsof/ZhangL20}). These areas are covered by the studies in scope.

\subsection{Large language models}\label{subsec:subLLMsec}

 Pre-trained Large Language Models (LLMs) (e.g.,~\cite{DBLP:conf/naacl/DevlinCLT19,zhao2023-survey}) are a type of Generative AI based on Deep Neural Networks~\cite{DBLP:books/daglib/0040158}. Initially, these models utilized a Transformer encoder-decoder architecture with cross-attention mechanisms~\cite{vaswani2017attention}. However, many modern LLMs now employ a decoder-only architecture with self-attention layers, which is sufficient for generating output from a given input prompt~\cite{Few-Shot-Brown20}.

Among the most well-established prompting methods used are: Zero-Shot~\cite{Zero-CoT-Kojima22}, Few-Shot~\cite{Few-Shot-Brown20}, Chain-of-Thought~\cite{CoT-Wei22}, Tree-of-Thoughts~\cite{ToT_Yao_2023}, Retrieve Augmented Generation~\cite{RAG2020Patrick} and Reasoning and Acting~\cite{ReAct}. 
Zero-Shot~\cite{Zero-CoT-Kojima22} learning involves prompting an LLM to perform a specific task without providing examples or demonstrations. 
Few-Shot~\cite{Few-Shot-Brown20} learning includes examples of the task within the prompt. These examples serve as demonstrations that condition the model, enabling it to generate sometimes more accurate responses for similar tasks. 
Chain-of-Thought (CoT) is an orthogonal prompt approach that proposes to generate a logical, linear flow of ideas and reasoning~\cite{CoT-Wei22}. Thanks to the autoregressive nature of LLMs, they end up (auto)conditioning themselves by verbalizing utterances that hopefully correspond to steps of an effective reasoning process. 
While CoT is an advanced technique that improves LLM answers, its fundamental concept relies on a linear reasoning process. It does not allow for divergent thoughts. Tree-of-Thought (ToT)~\cite{ToT_Yao_2023} was created to overcome this limitation with the intuition of aggregating different ``points-of-view'' during the reasoning process. In ToT, a query to an LLM can generate several different answers, some better than others. These answers, in turn, can each generate more than one response, forming a tree of thoughts that helps reasoning through different paths. In this tree, the initial query generates the root, and the leaves of each node are the intermediate thoughts. Thus, to form the final response, intermediate prompts corresponding to some path within this tree must be used. The final performance of ToT depends largely on the path-selection strategy. 
While the mentioned techniques improve the generated answers, they depend exclusively on the information in the prompt or the knowledge in the LLM itself. RAG~\cite{RAG2020Patrick} overcomes this limitation by enabling access to the latest information for generating reliable outputs via retrieval-based generation. In RAG, an augmented prompt with contextual information from an additional database helps reduce hallucinations. It makes the LLM adaptive to situations where facts evolve over time by simply updating the knowledge in this database. This approach also avoids the need for LLM fine-tuning, reducing computational demands. RAG can also be combined with other prompting techniques, such as CoT, to condition the reasoning process on the given context. For tasks like code generation, the database can be enhanced with examples of code or test cases, further improving the accuracy and reliability of the newly generated codes.
 
ReAct~\cite{ReAct} combines reasoning and acting in an interleaved manner, where the performed reasoning steps can ask an LLM to perform some tasks, usually involving the operation of external tools, and then reason again about the retrieved output from such tools. This allows dynamic reasoning and planning, like an AI agent, by interacting with external environments and incorporating the additional information from those environments into the context prompt. In the context of code generation, this could allow executing some code snippets or using verification tools, observing the output, and planning the next reasoning step until generating the final answer.
For an extensive survey on prompt methods the reader can refer to~\cite{DBLP:journals/csur/LiuYFJHN23}.

\section{Methodology}

A method that is partially inspired on those of systematic mapping~\cite{petersen2008systematic} was followed as a way to come up with the main contributions of this work, namely: transformation identification, transformations type taxonomy, interconnection patterns, and observed interrelationships. Due to the research questions, in-depth analysis of papers and taxonomy and patterns definitions played a central role that surpassed other typical aspects of systematic mappings such as demographics on areas, topics or venues. Given the application focus (software testing and verification), we went through abstracts of papers published or announced in Software Engineering venues (ICSE, FSE, ASE, ISSTA, ICST, MSR, SANER, ICSME, ICPC, GECCO, TOSEM, TSE, ESE, JSS, STVR, SPE, IST, etc). The same was done for Programming Language and Formal Verification venues (e.g., PLDI, POPL, OOSPLA, ECOOP FM/FV: CAV, TACAS, VMCAI, FM, FMCAD), Operating Systems and Security venues (USENIX, IEEE SSP), and also major AI venues (AAAI, IJCAI, ICML, ICLR, NeurIPS). Arguably, arXiv was a major source of papers. We performed searches using technology keywords such as ``LLM'', ``language'', ``ChatGPT'', ``context'', ``prompt'' and application related keywords such as ``testing'', ``verification'', ``formal'', ``check'', ``static'', ``fuzzing'', ``debugging'', etc. (see SE Problems in \autoref{fig:OverSEprobs}). We mainly found validation research and solution proposal papers~\cite{petersen2008systematic}. From the SE-related surveys on using LLMs, we snowballed one level and added missing papers. The time scope for the first collection was essentially from 2022 until February 2024 (later expanded to June 2024 and tested with late October 2024's conference papers, as mentioned below). Then, we carefully analyzed papers by reading both abstract and content, discarding those that did not use prompts or deal with different SE problems (approx. $140$ papers were discarded, and $75$ remained). 

We first read the papers to identify the tool/approach name given and the actual SE problem solved. Those identified problems were classified into the SE problems taxonomy, and papers were mapped into it as shown in \autoref{fig:OverSEprobs}\footnote{Some papers were mapped into more than one SE problem when adequate.}. This SE problems taxonomy is similar to other taxonomies found in the literature for the overlapped categories (e.g.~\cite{zhang2023survey})\footnote{Due to the exclusion criteria, automated program repair is not shown as a debugging subproblem.}. From there, as the figure shows, by reading each paper, we performed the laborious identification of transformations, their conceptual intent and signatures and integration/orchestration notes (Tables~1--5 in v2 of this preprint~\cite{braberman2024tasks}). \hyperref[abstractview]{Subsection~\ref*{abstractview}} further explains our viewpoint for identification.

\begin{figure}[!ht]
 \centering
 \hypertarget{OverSE}{}
 \resizebox{\columnwidth}{!}{%
 \begin{tikzpicture}[tasktree]
 
\node [draw,RoyalBlue] {\textbf{SE\phantom{j}Problems}}
    child [missing] {}
    child [azulP] {node {\textbf{Testing} }
        child [azulP] {node [darkgray]  {\hyperlink{UTG}{Unit Test Generation}: \white{TestGen-LLM}, \white{FSML}, \white{ChatGPTTests},}}
        child [white] {node [darkgray] { \white{CodeT}, \white{ChatUniTest}, \white{CodeCoT}, \white{AgentCoder}, }}
        child [white] {node [darkgray] {\white{CodaMOSA}, \white{TestChain}, \white{CoverUp}, \white{TestPilot},}}
        child [white] {node [darkgray] {\white{EASEeval}, \nwhite{HITS}, \white{TELPA},  \nwhite{LLM-UTG}, \white{ChatTester} }}
        child [azulP] {node [darkgray]  {\hyperlink{FITG}{Failure-Inducing Test Generation}: \white{DiffPrompt}, \white{AID}, \white{secTests}}}
         child [azulP] {node [darkgray]  {\hyperlink{RT}{Regression Testing}: \white{SymPrompt}}}
        child [azulP] {node [darkgray]  {\hyperlink{IG}{Input Generation}: \nwhite{DirInput}, \white{RESTGPT}, \white{InputBlaster}, }}
        child [white] {node [darkgray] {\white{PBT-GPT}, \white{mrDetector}}}
       child [azulP] {node [darkgray]  {\hyperlink{DMG}{DataSet/Mutant Generation}: \white{FSML}, \white{MuTAP}, \white{BugFarm},}}
       child [white] {node [darkgray] {\whiteD{MuBert}{$\mu$BERT}, \white{CHEMFUZZ}, \white{FormAI}}} 
        child [azulP] {node {\textbf{Fuzzing}}
            child [azulP] {node [darkgray] {\hyperlink{GF}{General Fuzzing}: \white{OSS-Fuzz}, \white{ChatFuzz}, \white{Fuzz4All}, \white{UGCTX}}}
            child [azulP] {node [darkgray] {\hyperlink{KF}{Kernel Fuzzing}: \white{KernelGPT}}}
            child [azulP] {node [darkgray] {\hyperlink{CF}{Compiler/Simulator Fuzzing}: \white{SearchGEM5}, \white{WhiteFox}}}
            child [azulP] {node [darkgray] {\hyperlink{PF}{Protocol/Parser Fuzzing}: \white{FuzzingParsers}, \white{ChatAFL}}}
            child [azulP] {node [darkgray] {\hyperlink{DLF}{DL-Libraries Fuzzing}: \white{TitanFuzz}, \white{FuzzGPT}, \nwhite{Magneto}}}
        edge from parent}
        child [missing] {}
        child [missing] {}
        child [missing] {}
        child [missing] {}
        child [missing] {}   
        child [azulP] {node [darkgray]  {\hyperlink{GUI}{GUI Testing}: \white{QTypist}, \white{GPTDroid}, \white{AXNav}}}
        child [azulP] {node [darkgray]  {\hyperlink{FunT}{Functional Testing}: \white{TARGET}, \white{ScenicNL}, \nwhite{SoVAR},}} 
        child [white] {node [darkgray] {\white{LLMeDiff}, \white{SysKG-UTF},  \nwhite{LeGEND}}}
        child [azulP] {node [darkgray]  {\hyperlink{PenT}{Penetration Testing}: \white{PentestGPT}, \whiteD{pwnd}{Pwn'd}, \nwhite{AdvSCanner}}}
        child [azulP] {node [darkgray]  {\hyperlink{OraP}{Oracle Problem}: \white{FSML}, \white{SELF-DEBUGGING}, }}
        child [white,draw opacity=0] {node [darkgray, draw opacity=1] {\white{nl2postcondition}, \white{TOGLL}, \white{Eywa},  \white{PropertyGPT}, }}
        child [white,draw opacity=0] {node [darkgray, draw opacity=1] {\white{ClarifyGPT},  \white{CEDAR}, \white{PROSPER}, \white{EMR}, }}
        child [white,draw opacity=0] {node [darkgray, draw opacity=1] {\white{GameBugDescriptions}, \white{MetaMorph}, \white{ALGO}}}
    edge from parent}
    child [missing] {}
    child [missing] {}
    child [missing] {}
    child [missing] {}
    child [missing] {}
    child [missing] {}
    child [missing] {}
    child [missing] {}   
    child [missing] {}
    child [missing] {}
    child [missing] {}
    child [missing] {}
    child [missing] {}
    child [missing] {}
    child [missing] {}   
    child [missing] {}
    child [missing] {}
    child [missing] {}   
    child [missing] {}
    child [missing] {}
    child [missing] {}
    child [missing] {}
    child [missing] {}
    child [missing] {}    
    child [turquesaP] {node {\textbf{Debugging}}
        child  [turquesaP] {node [darkgray]  {\hyperlink{BugR}{Bug Reproduction}: \white{AdbGPT}, \white{CrashTranslator}, \white{LIBRO}, }}
        child  [turquesaP,opacity=0] {node [darkgray,opacity=1]  {\nwhite{SemSlicer}}}
        child  [turquesaP] {node [darkgray]  {\hyperlink{DupB}{Bug Report Analysis}: \nwhite{FAIL}, \white{Cupid}}}
        child  [turquesaP] {node [darkgray]  {\hyperlink{FLoc}{Fault Localization}: \white{SELF-DEBUGGING}, \white{AutoFL}, \white{AutoSD}, }}
        child [white,draw opacity=0] {node [darkgray, draw opacity=1] {\white{LLM4CBI}, \white{ChatGPT-4(Log)}}}
        child  [turquesaP] {node [darkgray]  {\hyperlink{RCA}{Root Cause Analysis}: \white{RCACopilot}, \white{x-lifecycle}, \nwhite{LasRCA},}} 
        child [white,draw opacity=0] {node [darkgray, draw opacity=1] {\white{RCAAgents}, \white{LM-PACE}, \white{inContextRCA}}}
    edge from parent} 
    child [missing] {}
    child [missing] {}
    child [missing] {}
    child [missing] {}
    child [missing] {}
    child [missing] {}
    child [missing] {}
    child [verdeC] {node {\textbf{Vulnerability/Misuse/Malware/Fault Detection}}
        child  [verdeC] {node [darkgray] {\hyperlink{VulD}{Vulnerability Detection}: \white{ChatGPT4vul}, \white{VulBench},  \white{NLBSE24},}} 
        child [white,draw opacity=0] {node [darkgray,draw opacity=1] {\white{VulDetect}, \white{GRACE}, \white{VSP}, \white{AIagent}, \nwhite{VFFinder},  }}
        child [white] {node [darkgray] {\white{DLAP}, \white{MultiTask}, \white{PromptEnhanced}, \nwhite{VulAdvisor},  }}
        child [white] {node [darkgray] {\nwhiteD{RustC4}{RustC$^{\text{4}}$}, \white{ChatGPT(Plus)} }}
        child  [verdeC] {node [darkgray] {\hyperlink{Line}{Line-Level/Edit-Time Fault/API Misuse Prediction}: \white{FLAG}, \white{EditTime},}} 
        child [white,draw opacity=0] {node [darkgray, draw opacity=1] {\white{WitheredLeaf}, \white{LLMAPIDet}}}
        child  [verdeC] {node [darkgray] {\hyperlink{Smart}{Vulnerability Detection for Smart Contracts}: \white{ChatGPTSCV}, \white{SmartAudit},}} 
        child [white,draw opacity=0] {node [darkgray, draw opacity=1] { \white{GPTLens}, \white{GPTScan}, \white{LLM4Vuln}, \nwhite{Skyeye}, \nwhite{Sleuth}}}
        child [verdeC] {node [darkgray] {{Vulnerability Detection for Packages}: \nwhite{SpiderScan}}} 
    edge from parent}
    child [missing] {}
    child [missing] {}
    child [missing] {}
    child [missing] {}
    child [missing] {}
    child [missing] {}
    child [missing] {}
    child [missing] {}
    child [missing] {}
    child  [violetaP] {node {\textbf{Static Analysis}}
        child  [violetaP] {node [darkgray] {\hyperlink{CFG}{Call-Graph/CFG Construction}: \white{CFG-Chain}}}
        child  [violetaP] {node [darkgray] {\hyperlink{UBI}{Use Before Initialize}: \white{LLift}}}
        child  [violetaP] {node [darkgray] {\hyperlink{RLeak}{Resource Leak}: \white{SkipAnalyzer}, \white{InferROI}}}
        child  [violetaP] {node [darkgray] {\hyperlink{DFA}{Data-Flow Analysis}: \white{LLMDFA}}}
        child  [violetaP] {node [darkgray] {\hyperlink{Taint}{Taint Analysis}: \whiteD{EV}{E\&V}, \white{LATTE}}}
        child  [violetaP] {node [darkgray] {\hyperlink{Slicing}{Static Slicing}: \white{SimulinkSlicer}}}
        child  [violetaP] {node [darkgray] {\hyperlink{FixAA}{Fix Acceptance Check}: \white{CORE}}}
        child  [violetaP] {node [darkgray] {{Toxicity Analysis}: \nwhite{ToxicDetector}}}
    edge from parent}
    child [missing] {}
    child [missing] {}
    child [missing] {}
    child [missing] {}
    child [missing] {}
    child [missing] {}
    child [missing] {}
    child [missing] {}
    child [rosaP] {node {\textbf{Program Verification}}
        child  [rosaP] {node [darkgray] {\hyperlink{PV}{Program Verification}:  \white{AlloyRepair}, \white{ChatInv}, \white{Loopy},  }}
        child [white,draw opacity=0] {node [darkgray, draw opacity=1] {\white{SpecGen}, \white{Dafny-Synth}, \nwhite{ESBMCibmc}, \white{Clover},  }}
        child [white,draw opacity=0] {node [darkgray, draw opacity=1] {\white{AutoSpec}, \nwhite{Lam4Inv}, \white{Lemur}, \white{RustProof} }}
    edge from parent};
\end{tikzpicture}
 }%
 \caption{Overview of the SE problems considered in this paper, and the tools corresponding to each of them. Framed tools are the ASE'24 ones, that were added at the validation step.}\label{fig:OverSEprobs}
 \end{figure}

The initial list of transformations ($200$~approx. with repetitions) was then clustered into categories explained in \autoref{sec:rq2}. Variation points and hierarchy decisions lead to the trees seen (Figs.~\ref{fig:TG}--\ref{fig:PG} and Figs.~\ref{fig:GCG}--\ref{fig:FR} in Appendix). Another author, in a ``tabula rasa'' modality (i.e., no expertise in topics and without reading previously the papers nor elaborated tables or taxonomy), was able to map the $200$ transformations recovered into the hierarchy to validate that both the table and the taxonomy were ``natural'' and easy to understand resources. Interestingly, during the final steps of this elaboration, we updated the list to beginning of June 2024 with the more recent papers (36 papers) that were not in the original table. We were  able to seamlessly add them (and map their $83$ transformations) albeit we decided to change some names of transformations and internal structure of taxonomy trees to adhere to traditional terminology in those areas in which certain types of transformations already exist (e.g., classification tasks).
This update thus served as another validation step. After that, we went through tables, trees, and the mapping matrix (\autoref{tab:correlation}) and validated that taxonomy at least helped detect and exhibit some known and potentially unknown patterns of LLM usage. Finally, we also check that NLP tasks~\cite{DBLP:conf/acl/SuzgunSSGTCCLCZ23,NLP-periodic-table} (e.g.~relation extraction~\cite{DBLP:conf/acl/SuzgunSSGTCCLCZ23,DBLP:conf/acl/YaoYLHLLLHZS19}) relevant to the surveyed works were either included into the taxonomy or not featured by any of the approaches.
A version of this work was uploaded on September 8th, 2024, as a technical report~\cite{braberman2024tasks}. A couple of weeks later, we performed the identification of transformations of solution proposals appearing in  papers that had just been announced as accepted to ASE~2024 and mapped them into the baselined taxonomy trees to double check whether types were already seen or the versatility of trees to accommodate novel variants. That mapping is highlighted using, in the trees of the paper and the ones of \ref{sec:AppendixTrees}, squared boxes around the names of the ASE~2024 tools that were not previously analyzed as arXiv preprints.  

After this, we updated the publication venues of all the papers that were originally retrieved from  arXiv\footnote{This is why some papers of very recent conference editions  and journal volumes appear as part of the papers in scope while not necessarily all papers of those venues that satisfied the criteria were analyzed.}. That step included changing some of the tool names when they were modified in the published version. 

Finally, by analyzing how different types of transformations were interconnected in each solution proposal (within the subset of peer-reviewed papers), we identified a set of recurrent patterns, which we termed \textbf{Inter-Transformation Patterns}. We further conjectured which limitations of LLMs are addressed by each of these patterns, and then examined the relationships between patterns and transformation categories.

\subsection{Inclusion and exclusion criteria}

As said, our technological focus is on using LLMs that employ prompts. This excludes proposals that are based primarily on either fine-tuning LLMs (see for instance~\cite{DBLP:conf/ijcai/NiuL0022}) or general transformer architectures (e.g.,~\cite{AlqarniA22,CiborowskaD22,FuzzGPT,TOGA,Baldur,LineVul,Muffin,VulBERTa,HuLXLXY23,DBLP:conf/icse/KharkarMJLSCS22,MohsenHWMM23,LEVER,VULGEN,LLMInv,CAT-LM,hashtroudi2023automated,LAST,tufano2021unit,venkatesh2024emergence,ZhangCZP23,TroBo}, etc.), or leverage embeddings  of underlying transformers (e.g.~\cite{hossain2024deep,LLMAO}, etc.). Selected Software Engineering topics are by no means all areas addressed by LLM-enabled in-context/prompting approaches. Yet, they currently constitute a significant amount of works in SE venues\footnote{For instance, they made up approximately 40\% of LLM-enabled prompt-based approaches of FSE~2024 main track papers.}. As a notable excluded topic, we do not cover papers whose main SE goal was program generation/synthesis or program repair, which are allegedly ``natural'' areas of application of generative AI~\cite{fan2023large-survey,hou2023large-survey,wan2023-survey,DBLP:conf/icse/XiaWZ23,zhang2023survey}. The other excluded topic is code reviews. Yet, there are two important caveats regarding exclusions. On the one hand, some approaches studied elicit code generation or repair transformations to achieve testing, fault localization or verification goals and we, consequently, do report and analyze those papers. On the other hand, we also report papers whose primary goal is program generation, debugging with repair, or code review but they prompt testing, verification or fault localization tasks to achieve their end-to-end goals. Thus, the exclusion is limited to works whose primary SE goal is out of scope and do not prompt LLMs to solve Software Engineering problems within the scope.
The other important exclusion criterion is the date: this paper version covers papers that appeared in journals, conferences, or arXiv up to the first week of June~2024, and papers already announced in conferences up to July~2024.

We are aware of other comparative analyses that we do not include here because their prompts are based on literature already reported in our work, for example~\cite{DBLP:journals/tse/TangLZL24}. Last but not least, it can be noticed that some traditional SE falsification and verification areas like run-time verification, symbolic execution and (some) model checking challenges are not covered: under the defined LLM and timeline related criteria, no studies were found that could directly be mapped as addressing those topics as their primary contribution focus.

\label{abstractview}

\section[Transformations]{Conceptual decomposition of studies into transformations}

\subsection{Extraction process}

LLM-enabled application proposals can be abstractly characterized as software that interacts with LLMs to generate intermediate data or final results.
This does not preclude integration with classical software tools (compilers, static analyzers, ad-hoc Deep Learning models, vector databases, etc.) as noted by other surveys~\cite{huang2024fuzz-survey}.

We claim that (1) all those LLM based interactions can be understood as  a set of ``transformations''  from some (intermediate) inputs to some (intermediate) output, (2) those transformation types, 
while very rich, can be categorized, and (3) the same can be said about patterns of how transformations are related to solve problems in contemporary literature.

~\autoref{fig:ChatTester} and ~\autoref{fig:whitefox} exhibit transformations and their interrelationships. In particular, they depict, as  Bayesian Networks~\cite{DBLP:books/daglib/0023091}, the (causal) dependencies of (random/deterministic) conceptual variables in representative abstract runs of a couple of solution proposals. The (conditional) probability distribution of generated data is actually governed by either LLM-mechanized or algorithmic ``transformations''. 
Such a graphical representation serves as a framework guiding the extraction, identification, and categorization of transformations --treated as a phenomenon of interest--, particularly those mechanized by LLMs.
That is, our first step was to conceptually decompose each  solution proposal in the literature in scope by identifying the typology of generative transformations and their conceptual signatures that the solution  leverages.
For such aim, we manually analyzed relevant papers and associated repositories to elucidate such discrete functional units. This was done by reading paper descriptions, prompts and algorithms and then performing: 
\begin{compactenum}
    \item detection of all potential intermediate and final results that would computed by algorithmic manipulation or filled in by LLMs utterances in representative schematic runs of the solution proposal (the data variables);
    \item identification of all the elicitation elements (e.g., instructions, examples, prefixes, etc.) in prompts that indicate what is expected as utterances for such intermediate/final results -our granularity aim;
    \item elucidation of the input signature of such transformations in terms of involved data variables that would condition the generation of utterances;
    \item naming those transformations in a way to convey expected ``functionality''.
\end{compactenum}
This analysis presented several challenges we overcome.
First, the LLMs manifest a great versatility observed in a wide variability of questions, instructions, directions, and example-based elicitation across the literature. Second, solution proposals exhibited significant heterogeneity in their presentation, with diverse prompting strategies and, often, incomplete descriptions in papers of key aspects. That is, while prompts being first-class concepts are typically sketched, their abstract description of functional units we seek is either missing or imprecise 
and terminology varies substantially between studies.
Thirdly, there is a wide variety of  ways to build prompts and elicit generations during inference time. Moreover, prompts and transformations are not necessarily related  1-to-1\footnote{Some role-based agentic presentations or declarative frameworks like~\cite{DSPy} might ease the identification of transformations and signatures, in general, this is not trivial.}. In fact, a statically or programatically built prompt prefix might lead to the generation of chained transformations during inference-time (and, also, we needed to elucidate actual conceptual signatures embedded into the context). 
Last but not least, the autoregressive nature of LLMs--including their potential for step-by-step, inference-time reasoning generation, as demonstrated by Chain-of-Thought approaches~\cite{CoT-Wei22}--may lead to self-generated transformative stages that cannot be statically recovered\footnote{Yet we do capture transformations suggested by tool developers for the elicited CoT--a pretty usual practice in papers in scope (e.g., ``think step by step, first summarize API usage, then ...'').}. Those transformations are auto-elicited during runtime to serve higher-level transformations designed by tool developer-- which are the recovering effort focus. Interestingly,~\cite{bogdan2025thoughtanchorsllmreasoning} proposes a categorization of sentences in CoT trajectories (Problem Setup, Plan Generation, Fact Retrieval, Active Computation, Uncertainty Management, Result Consolidation, Self Checking, and  Final Answer Emission) which could complement our catalog of higher-level developers-guided transformations.  
As said,  there are many means in which those prompts are built and, potentially, some details might even be defined with the help of LLM-based/optimization mechanisms before  applications runtime~\cite{promptbreeder,DBLP:journals/corr/abs-2402-09497,DSPy,li2021prefixtuning,APE}\footnote{Very few were encountered in the literature under review.}. 

As said input and output conceptual types were elucidated and annotated. For reactive transformations (e.g.,~\cite{ReAct}), the nature of interactions with the environment (that involve further input and output) was also recovered. Input signature was identified by reflecting and discussing carefully about conceptual data that would be necessary to perform the intended transformation. This work is relatively easy when  prompts sent to LLMs relate 1-to-1 to transformations. As said, when prompts accumulate conversation history comprising several transformations this signature elucidation is harder: we need to reflect on which data on the prefix the transformation (conceptually) should be function of. That is, in probabilistic  terms, output should be conditionally independent~\cite{DBLP:books/daglib/0023091}--given the identified input signature--with respect to the rest of data conveyed in the prompt. Of course, LLM's mechanisms are opaque, and it could be the case that even conceptually irrelevant data in context, due to the attention mechanisms, generates an internal representation that acts as a latent variable which might affect the transformation's distribution on outputs beyond identified natural signature data. Yet, we believe this identification is crucial since it is worth understanding, in the future, if those latent variables help or hurt or are neutral to the quality of implemented transformations.  

Through this analysis, we systematically identified and, in many cases, defined transformation types as a way to abstractly describe what was instructed in prompts found in the literature. 
Our naming conventions for transformation types were designed to
accurately reflect their fundamental functional nature and capture their distinguishing characteristics.

\begin{table}
 \centering
 \includegraphics[width=\columnwidth]{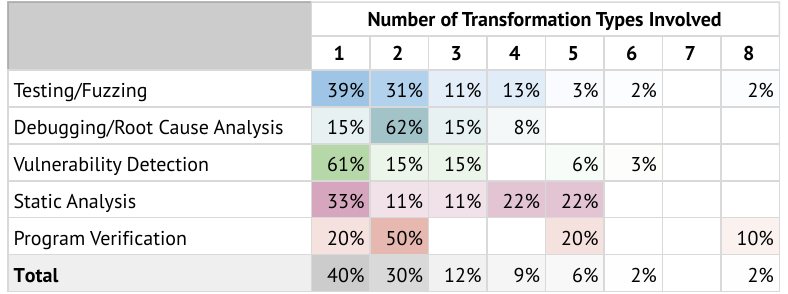}
 \caption{Number of transformation types involved in each solution approach. The analysis is performed for each type of software engineering problem, and the entries of the table express the percentage of solution proposals of that type that feature that number of different types of transformations. When a paper proposes more than one approach for the same problem, these are considered individually, counting the number of types of transformations involved in each of the approaches.}
 \label{tab:complex}
\end{table}

We abstracted away low-level implementation details (e.g., decoding strategies, prompt engineering specifics like wording~\cite{DBLP:conf/acl/LuBM0S22}, etc.) to focus on generalizable transformative intents and transformations elicited by LLM prompts, enabling a clearer identification of commonalities and ``functional'' variation points.

We use a structured naming convention for transformations. Two parts were mandatory: [Property/Criterion]-[Action/transformation type].
The [Action/Transformation] type describes the fundamental abstract operation being performed (e.g., Generation, Mutation, OneClass-Classification).
[Property/Criterion] describes the conditions, constraints, or goals that are the guiding principles of the transformation. This sets the rationale or intended effect (e.g., IntentCorresponding in IntentCorresponding-Code-Generation, CoverageAugmenting in CoverageAugmenting-Test-Generation, VulnerabilityProneness in VulnerabilityProneness-Code-OneClass-Classification, etc.).
In most cases, we also added as an additional descriptor the type of artifact being produced, modified or analyzed. This artifact could refer to the type of object that the transformation yields as a way to subtype generative actions (e.g. Test in CoverageAugmenting-Test-Generation, Invariant-Lemma in Invariant-Lemma-Generation)  or the type of object subject to the transformation (e.g.Code in VulnerabilityProneness-Code-OneClass-Classification) according to what we believe was more relevant to make the transformation description more informative. 

This sort of descriptors helped us to cluster and categorize transformations based on shared functional characteristics. Notation also simplified the elaboration of the taxonomy's dimensions.  It was also key to achieve inter-annotator agreement by resolving disagreements and inconsistencies via discussion and refinement.

\subsection{Results}

The following is an illustrative example of transformations recovered from a couple of solution proposals from different SE problems: unit test generation~\cite{ChatTester}, fuzzing~\cite{Fuzz4All}, penetration testing~\cite{PentestGPT}, and vulnerability detection for smart contracts~\cite{GPTLens}.

\begin{ejemplo} \scriptsize 
{{\fontfamily{cmss}\fontseries{c}\selectfont 
 \textbf{ChatTester}~\cite{ChatTester}:} 
\begin{compactenum}[1.]
    \item focal meth code \abstr{CodeCorresponding-Intent-Verbalization} intent
    \item focal meth sig + intent \gen{IntentCorresponding-Test-Generation} unit tests
    \item unit test + error msgs \gen{ErrorAware-Test-Correction} unit test
\end{compactenum}}

{{\fontfamily{cmss}\fontseries{c}\selectfont
 \textbf{Fuzz4All}~\cite{Fuzz4All}:} 
\begin{compactenum}[1.]
    \item doc + (examp)$^*$ + (specs)$^*$ \abstr{Usage-Summarization} distilled usage/funct
    \item usage/funct \gen{UsageSastifying-Input-Generation} fuzz inputs
    \item usage/funct + fuzz input  \gen{Input-Variation-Generation} fuzz input
\end{compactenum}}
{{\fontfamily{cmss}\fontseries{c}\selectfont 
\textbf{PentestGPT}~\cite{PentestGPT}:} 
\begin{compactenum}[1.]
    \item user-intents \exec{IntentCorresponding-Plan-Generation} penetration task tree (ptt)
    \item testing results + ptt \exec{Elements-Update} ptt
    \item ptt + updated ptt \eval{Validity-Transition-Analysis} result
    \item  ptt \exec{RulesSatisfying-Information-Extraction} (potential next task)$^+$
    \item  ptt + tasks \exec{MostPromising-NextTask-Selection} sugg next task
    \item  sub-task + avail tools \exec{ToolsConstrained-Plan-Generation} seq of steps [CoT]
    \item step \exec{Plan-Refinement} cmd [CoT]
    \item raw user intent | test outp \abstr{Summarization} condensed info 
\end{compactenum}}
{{\fontfamily{cmss}\fontseries{c}\selectfont 
\textbf{GPTLens}~\cite{GPTLens}:} 
\begin{compactenum}[1.]
    \item src-code \eval{Rationalized-VulnerabilityProneness-Code-MultiLabel-Classification} (vul + funct name + reason)$^*$ (auditor)
    \item src-code + vul + funct name + reason  \eval{Audits-Qlty(Correctness, Severity, Profitability)-Assessment} score + explan (critic)
\end{compactenum}}
\end{ejemplo}   

Colors are indicative of the transformation category as it will be explained in the next sections. Numbers are later used as sub-indexes when mapping transformations into dimensions choices when taxonomy is presented.

The outcome of this extraction and naming process for the initial set of literature in scope can be found in Tables~1--5 in the v2 of this preprint~\cite{braberman2024tasks}.

Despite the richness and variety of approaches, we recovered from the initial set of $111$ reported papers their underlying transformations ($284$ in total)\footnote{Some recurring ones.} and presented them homogeneously, no matter sophistication of proposal  or how it was reported.  Thus, this work provides some evidence that transformation types, currently, constitute a natural blueprint of what cognitive capabilities are being leveraged in an  LLM-enabled approach, at least for contemporary solution proposals for software testing and verification problems.

\autoref{tab:complex} summarizes the distribution of problem decomposition complexity of solution proposals measured as the number of different types of transformations elicited per proposal. Solution proposals are grouped according to software engineering problem addressed. A paper might propose more than one solution. This is particularly the case of seminal exploratory works which analyze the proficiency of models when prompted for a studied SE task. That explains the higher concentration of 1-transformation approaches particularly in unit test generation and vulnerability proneness classification, SE categories that encompass several validation research papers. Having said that, we can see that the majority of LLM-enabled solution proposals require two or more types of transformation, typically interrelated. Inter-transformation patterns in \autoref{sect:patterns} many times explain the reasons for the existence of designs featuring more than one transformation type in solution proposals. For instance, a large number of the 2-transformation approaches actually feature a generative transformation followed by a corrective stage (see \texttt{Generate and Fix}).  
Also, approaches for SE problems differ in complexity. For instance, while testing and fuzzing exhibit a similar distribution of the entire population, static analysis proposals seem to require a larger number of transformations.  
That is, even if many approaches leverage in low-level transformations dynamically elicited during thought trajectories, 
non-trivial problem decomposition into two or more transformations and architecture is statically identifiable.
That is, for many SE~problems covered by this report, it seems that ingenious problem decomposition and some degree of tool orchestration is currently necessary, possibly because it has not been observed yet breakthrough  behavior~\cite{DBLP:journals/tmlr/SrivastavaRRSAF23} that would enable solving the actually required end-to-end transformation as a single one. This phenomenon has been also dubbed ``the  compositional gap'' in the NLP literature~\cite{DBLP:conf/emnlp/PressZMSSL23} . This topic is revisited in detail in \autoref{sect:patterns}. In a few words, the number of transformations involved is the result of two antagonist forces: the sophistication of the goals of LLM-native applications and the ability of LLM to proficiently solve higher-level transformations with a single instruction.

Identified transformations are rich in nature and functional features and, to the best of our knowledge, most of them were not previously contained in existing taxonomies or benchmarks. They possibly fall somewhere in between open-ended and directed generation tasks~\cite{DBLP:conf/iclr/HoltzmanBDFC20}. Next section will provide an overview of type categories by means of the taxonomy we elaborated.

\section[Transformation taxonomy]{Transformation taxonomy}\label{sec:rq2}

\begin{figure}[htbp]
 \centering
 \hypertarget{OverLLM}{}
 \resizebox{\columnwidth}{!}{%
  
\begin{tikzpicture}[tasktree]
 
\node [draw,RoyalBlue] {\textbf{Transformation Taxonomy}}
    child [missing] {}
    child [celesteA] {node {\textbf{Generative}  \phantom{Ag}}
        child [celesteA]  {node {Code Generation \phantom{Ag}}
            child [celesteA] {node [darkgray] {\textbf{General-Code Generation} \fig{fig:GCG}}}
            child [white,draw opacity=0] {node [gray] {E.g. {IntentCorresponding-Code-Generation}}}
            child [celesteA]  {node [darkgray] {\textbf{Domain-Specific-Code Generation} \fig{fig:SCG}}}
            child [white,draw opacity=0] {node [gray] {E.g. {InputGenerator-Generation}, {DriverCode-Generation/Correction}}}
            child [celesteA]  {node [darkgray] {\textbf{Test Generation} \fig{fig:TG}}}
            child [white,draw opacity=0] {node [gray] {E.g. {Basic\&Edge-TestCase-Generation}, {CoverageAugmenting-Test-Generation}}}
        edge from parent}
        child [missing] {}
        child [missing] {}
        child [missing] {}
        child [missing] {}
        child [missing] {}
        child [missing] {}
        child [celesteA]  {node [darkgray] {\textbf{Annotation Generation} \fig{fig:AnnG}}} 
        child [white,draw opacity=0] {node [gray] {E.g. {Assertion-Generation}, {Invariant-Lemma-Generation}}}
        child [celesteA]  {node [darkgray] {\textbf{Data Generation} \fig{fig:DataG}}}  child [white,draw opacity=0] {node [gray] {E.g. {ConstraintSatisfying-Input-Generation}, {Basic\&Edge-TestInput-Generation}}}  
    edge from parent}
    child [missing] {}
    child [missing] {}
    child [missing] {}
    child [missing] {}
    child [missing] {}
    child [missing] {}
    child [missing] {}
    child [missing] {}
    child [missing] {}
    child [missing] {}
    child [missing] {}
    child [limaA] {node {\textbf{Evaluative}  \phantom{Ag}}
        child [limaA] {node [limaA] {SW-Entity Analysis}
            child [limaA] {node [darkgray] {\textbf{Behavior Analysis} \fig{fig:BA}}}       
            child [white,draw opacity=0] {node [gray] {E.g. {RootCause-Analysis}, {Behavior-Anomaly-Detection}}}
            child [limaA] {node {Code Analysis}
                child {node {Code Classification  \phantom{Ag}}
                    child {node [darkgray] {\textbf{Direct Code-Classification} \fig{fig:DCC}}}
                    child [white,draw opacity=0] {node [gray] {E.g. {VulnerabilityProneness-Code-OneClass-Classification}}}
                    child {node [darkgray] {\textbf{Rationalized Code-Classification} \fig{fig:RCC}}}
                    child [white,draw opacity=0] {node [gray] {E.g. {Rationalized-VulnerabilityProneness-Code-MultiLabel-Classification}}}
                edge from parent}
                child [missing] {}
                child [missing] {}
                child [missing] {}
                child [missing] {}
                child {node [darkgray] {\textbf{Code Scaling} \fig{fig:CSC}}}
                child [white,draw opacity=0] {node [gray] {E.g. {CVSS-Scoring}, {VulnerabilityProneness-Code-MultiLabel-Soft-Classification}}}
                child {node [darkgray] {\textbf{Line Code-Ranking} \fig{fig:LCR}}}
                child [white,draw opacity=0] {node [gray] {E.g. {FaultProneness-CodeLines-Ranking}}}
            edge from parent} 
            child [missing] {}
            child [missing] {}
            child [missing] {}
            child [missing] {}
            child [missing] {}  
            child [missing] {}       
            child [missing] {}
            child [missing] {}       
            child [missing] {}
            child [limaA] {node [darkgray] {\textbf{Task-Solution Analysis} \fig{fig:TSA}}}     
            child [white,draw opacity=0] {node [gray] {E.g. {Self-Validation}, {Answer-Quality-Assessment}}}
        edge from parent}
        child [missing] {}
        child [missing] {}
        child [missing] {}
        child [missing] {}
        child [missing] {}
        child [missing] {}
        child [missing] {}  
        child [missing] {}       
        child [missing] {}
        child [missing] {}
        child [missing] {}
        child [missing] {}
        child [missing] {}
        child [missing] {}
        child [limaA] {node [darkgray] {\textbf{Text Analysis} \fig{fig:TA}}} 
        child [white,draw opacity=0] {node [gray] {E.g. {Docstrings-Equivalence-Checking}, {Ambiguity-Analysis}, 
{EvidenceSupport-Judgment}}}
    edge from parent}
    child [missing] {}
    child [missing] {}
    child [missing] {}
    child [missing] {}
    child [missing] {}
    child [missing] {}
    child [missing] {}
    child [missing] {}
    child [missing] {}
    child [missing] {}
    child [missing] {}
    child [missing] {}
    child [missing] {}
    child [missing] {}
    child [missing] {}
    child [missing] {}
    child [missing] {}
    child [naranjaB] {node {\textbf{Extractive}  \phantom{Ag}}
        child [naranjaB] {node [naranjaD] {SW-Entity Extraction}
            child [naranjaB] {node [darkgray] {\textbf{Code-Elements Identification \& Extraction} \fig{fig:CEI}}}      
            child [white,draw opacity=0] {node [gray] {E.g. {CodeElements-Identification}, {CodeBlocks-Extraction}, {Model-Slicing}}}
        edge from parent}
        child [missing] {}
        child [missing] {}
        child [naranjaB] {node [darkgray] {\textbf{Text-Elements Identification \& Extraction} \fig{fig:TEE}}}   
        child [white,draw opacity=0] {node [gray] {E.g. {VariableNames-Identification}, {StructureCompliant-Information-Extraction}}}
    edge from parent}
    child [missing] {}
    child [missing] {}
    child [missing] {}
    child [missing] {}
    child [missing] {}
    child [fucsiaC] {node {\textbf{Abstractive}  \phantom{Ag}}
        child [fucsiaC] {node {SW-Entity to NL}
            child [fucsiaC] {node [darkgray] {\textbf{SW-Entity Verbalization} \fig{fig:AV}}}  
            child [white,draw opacity=0] {node [gray] {E.g. {CodeCorresponding-Intent-Verbalization}, {Summarization}}}
        edge from parent}
        child [missing] {}
        child [missing] {}
       child [fucsiaC] {node {NL to SW-Entity}        
            child [fucsiaC] {node [darkgray] {\textbf{Formalization} \fig{fig:For}}}
            child [white,draw opacity=0] {node [gray] {E.g. {PostCondition-Formalization}}}
        edge from parent}
        child [missing] {}
        child [missing] {}
        child [fucsiaC] {node {SW-Entity to NL / NL to NL}    
            child [fucsiaC] {node [darkgray] {\textbf{SW-Entity-Property Identification \& Characterization} \fig{fig:PrC}}}
            child [white,draw opacity=0] {node [gray] {E.g. {VariablesSameValue-Identification}, {TaintFlow-Identification}}}
        edge from parent}
        child [missing] {}
        child [missing] {}
        child [fucsiaC] {node {NL to NL}
            child [fucsiaC] {node [darkgray] {\textbf{Focused-Abstractive Summarization} \fig{fig:FAS}}} 
            child [white,draw opacity=0] {node [gray] {E.g. {RootCauseOriented-Summarization}, {Assessment-Yes/No-Summarization}}}
        edge from parent}
    edge from parent}
    child [missing] {}
    child [missing] {}
    child [missing] {}
    child [missing] {}
    child [missing] {}
    child [missing] {}
    child [missing] {}
    child [missing] {}
    child [missing] {}
    child [missing] {}
    child [missing] {}
    child [missing] {}
    child [purpuraA] {node {\textbf{Executive}  \phantom{Ag}}
        child [purpuraA] {node {Planning}
            child [purpuraA] {node [darkgray] {\textbf{Plan Generation} \fig{fig:PG}}}
             child [white,draw opacity=0] {node [gray] {E.g. {ScenarioCorresponding-Plan-Generation}, {Plan-Refinement}}}
        edge from parent}
        child [missing] {}
        child [missing] {}
        child [purpuraA] {node {Decision Making}
            child [purpuraA] {node [darkgray] {\textbf{What-to-do-Next Generation} \fig{fig:WTD}}}
            child [white,draw opacity=0] {node [gray] {E.g. {ReachabilityOriented-NextAction-Generation}}}
        edge from parent}
        child [missing] {}
        child [missing] {}
        child [purpuraA] {node {High-level Instruction Following}
            child [purpuraA] {node [darkgray] {\textbf{Execution} \fig{fig:PSE}}}
            child [white,draw opacity=0] {node [gray] {E.g. {Step-Computation}, {Trace-Execution}}}
            child [purpuraA] {node [darkgray] {\textbf{Textual Data Manipulation} \fig{fig:TDM}}}
            child [white,draw opacity=0] {node [gray] {E.g. {CFG-Fusion}, {InListCloseMeaning-Elements-Replacement}}}
        edge from parent}
    edge from parent}
    child [missing] {}
    child [missing] {}
    child [missing] {}
    child [missing] {}
    child [missing] {}
    child [missing] {}
    child [missing] {}
    child [missing] {}
    child [missing] {}
    child [missing] {}
    child [missing] {}
    child [aguaA] {node {\textbf{Consultative} \phantom{Ag}}
        child [aguaA] {node [darkgray] {\textbf{Knowledge Distillation} \fig{fig:FR}}}
        child [white,draw opacity=0] {node [gray] {E.g. {Example-Recall}, {Definition-Recall}, {ProtocolGrammar-Recall}}}
    edge from parent};
\end{tikzpicture}
 }%
 \caption{Taxonomy based on the transformation types found in the reported papers.}
 \label{fig:OverLLMtasks}
\end{figure}

The proposed taxonomy results from our initial best effort to cluster, brand, and rationalize transformation types implemented by prompts and conversations reported in the reviewed literature. Our taxonomy is organized hierarchically: a directed tree of transformation classes, being the types closer to the root the most general ones (\emph{Transformation Categories\/}). Specificity is expressed in two different ways: either by ``subtyping'' using directed edges in top levels of the taxonomy (see \autoref{fig:OverLLMtasks}) and, for more concrete levels of the taxonomy (\emph{Transformation Classes\/}), types came up as combinations of choices of the identified dimensions (that is, externally-visible variation points) as a way to capture the specificity of the transformation being proposed by the analyzed studies (Figs.~\ref{fig:TG}--\ref{fig:PG} and~\ref{fig:GCG}--\ref{fig:FR}). 

The initial rationale for the taxonomy is to highlight commonalities between transformations requested by prompts while preserving the richness of some relevant details. Taxonomy is also meant to be ``useful'' in helping identify some patterns of use of LLMs and prompting and also making explicit some less explored (or unexplored) transformation types. Ultimately, taxonomy might also help to understand how to leverage warnings and advances of the natural language generation community (e.g.,~\cite{DBLP:journals/corr/abs-2403-00025}) to build LLM-enabled applications in a principled way (see \autoref{sec:prelim-conclus}).

\begin{figure}[tbh]
\centering
\includegraphics[width=\columnwidth]{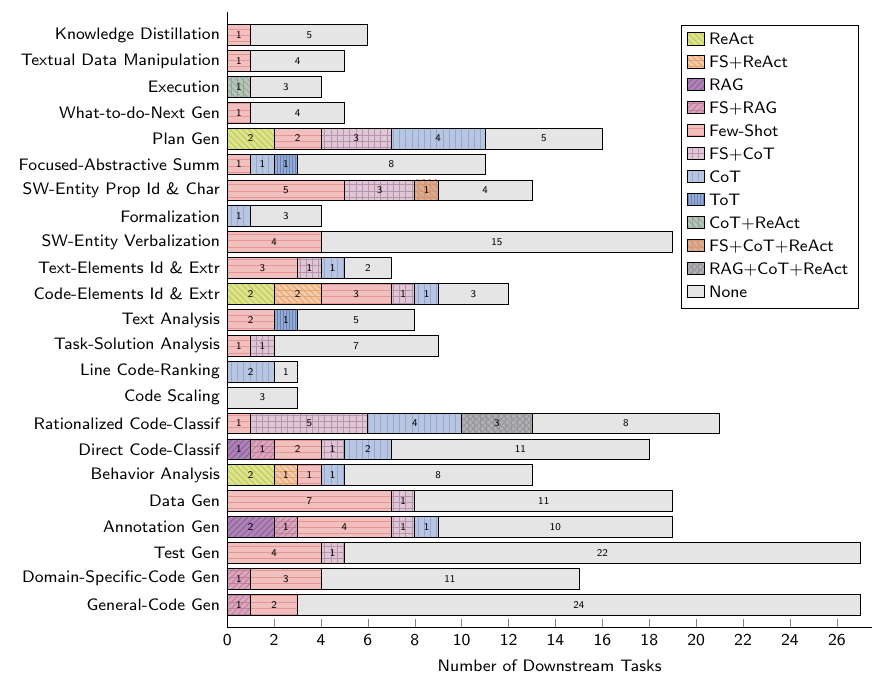} 
\caption{Number of transformations using each prompting technique or combination of prompting techniques, aggregated by transformation category.}
 \label{fig:chart-prompt}
\end{figure}

Our taxonomy was deliberately designed to be extensible along two complementary dimensions. First, intra-category extensibility: within each category of transformations, new properties and options can be incorporated to capture emerging types of generative transformations without disrupting the existing structure. This ensures that the taxonomy can evolve as LLM-native practices diversify. Second, cross-domain extensibility: while this work focuses on software entities in the domain of verification and falsification, the same organizing principles can accommodate new categories of transformations in other domains (e.g., legal document generation, scientific writing). We thus view the taxonomy not as a fixed classification, but as a scaffold for reasoning, intended to adapt and expand as new classes of transformations and patterns are observed.

The top-level branching in the taxonomy responds to identifying different high-level conceptual operations being performed. Those actions typically require, for humans, different cognitive abilities and, for the ``pre-LLMs era'', have involved different research communities and used quite different theories, algorithms, and/or training approaches. For instance, distinguishing generation vs. evaluation of a software entity--although being both solved by generative utterances--is in line with the perception of two different human tasks, problem hardness, different performance evaluation methods,  and different existing SE-tool categories. As we will show later, categories were initially leveraged to detect design patterns, pinpoint  opportunities, and to highlight commonalities and differences between SE approaches.

\paragraph*{Generative}
Generative transformations are defined by their primary function: creating  or  modifying existing software artifacts. The classification criterion is that the main generated output must be a software entity (SW-entity). This category is subdivided by artifact type: \textit{code} (Figs~.\ref{fig:GCG},~\ref{fig:SCG}), \textit{tests} (\autoref{fig:TG}), \textit{code annotations} (\autoref{fig:AnnG}), and \textit{data} (\autoref{fig:DataG}). Code generation is further distinguished by whether the functionality is predetermined (domain-specific, e.g., fuzzing drivers) or emergent from input parameters (general).

\paragraph*{Evaluative} 
Evaluative transformations analyze properties of input software entities, producing judgments, scores, or qualitative assessments. The classification criterion requires the output to be conceptually analytical rather than generative. Subcategories are defined by output characteristics: \textit{direct classification} (categorical labels without justification, akin classical discriminative ML tasks~\cite{DBLP:journals/pieee/LinWHZX20,9448435}, \autoref{fig:DCC}), \textit{rationalized classification} (judgments with natural language explanations, \autoref{fig:RCC}), \textit{scaling/ranking} (numerical or ordinal assessments, Figs.~\ref{fig:CSC},~\ref{fig:LCR}), \textit{behavior analysis} (assessment of satisfaction of dynamic properties from behavioral descriptions, \autoref{fig:BA}), \textit{task-solution analysis} (evaluation of other transformations' outputs, \autoref{fig:TSA}), and \textit{text analysis} (general NL assessment, \autoref{fig:TA}).

\paragraph*{Extractive} 
Extractive transformations identify and retrieve specific elements from larger contexts without significant abstraction or reformulation. The classification criterion is verifiable by checking whether output elements exist verbatim or near-verbatim within the input. This category includes code element identification (\autoref{fig:CEI}) and text extraction operations (\autoref{fig:TEE}), where the transformation performs location and retrieval rather than synthesis.

\paragraph*{Abstractive} 
Abstractive transformations construct higher-level representations through synthesis, summarization, or formalization. The classification criterion requires that outputs cannot be found verbatim in inputs and represent conceptual abstractions. Subcategories include \textit{verbalization} (converting SW-entities to natural language descriptions, \autoref{fig:AV}), \textit{formalization} (producing formal specifications, \autoref{fig:For}), \textit{property characterization} (identifying abstract properties, \autoref{fig:PrC}), and \textit{focused abstractive summarization} (NL summarization tasks~\cite{DBLP:conf/emnlp/RushCW15}, \autoref{fig:FAS}).

\paragraph*{Executive} 
Executive transformations produce operational instructions for goal-directed behavior. The classification criterion is met when outputs specify actions, plans, or state transitions rather than static artifacts. This encompasses \textit{plan generation} (creating operational sequences, \autoref{fig:PG}), \textit{decision-making} (selecting next actions, \autoref{fig:WTD}), and \textit{instruction following}~\cite{DBLP:conf/icml/HuangAPM22} (executing imperative operations, Figs.~\ref{fig:PSE},~\ref{fig:TDM}).

\paragraph*{Consultative} 
Consultative transformations retrieve and present information from the model's pre-trained parameters on a given subject without relying on provided context~\cite{NLP-periodic-table}. The classification criterion requires that the prompt segment eliciting the transformation does not contain the answer at all nor an input to process, necessitating recall of compressed knowledge like definitions, specifications, or guidance (\autoref{fig:FR}).

~\autoref{fig:chart-prompt} indicates both the number of transformations mapped into each category and some proxy indication on how challenging was for authors to implement the expected behavior by indicating the used prompting strategy/strategies.

\subsection{Dimensions of transformation classes}

Figures~\ref{fig:TG}--\ref{fig:PG}  provide examples of how transformation subtypes for a specific category are defined using property dimensions and choices. They also show how the relevant studies are mapped onto these choices, which are represented as the leaves of the decision trees. The featured choices are those actually observed in our analysis; however, for some dimensions, more choices are naturally possible. This potential for unobserved choices is a key feature of this representational method.

\begin{figure}[ht]
    \centering
    \resizebox{!}{27\htree}{%
    \begin{tikzpicture}[tasktree]
 
\node [draw,celesteA] {\textbf{Test\phantom{j}Generation}}
    child [missing] {}
    child [celesteA] {node {Generation criteria\phantom{j}}
        child [celesteA] {node [darkgray] {Differential: \testing{MuTAP}{3}}}
        child [celesteA] {node [darkgray] {Natural: \testing{FSML}{UTG}, \testing{ChatGPTTests}{1}, \testing{ChatUniTest}{1}, \testing{MuTAP}{1}, } 
            child [white] {node [darkgray] {\phantom{urau} \testing{DiffPrompt}{3}, \testingU{TestPilot}, \testingU{EASEeval}, \ntestingU{LLM-UTG} }}
            child [celesteA] {node [darkgray] {Bug revealing: \testing{FuzzGPT}{4} }}   
        edge from parent}
        child [missing] {}
        child [missing] {}
        child [celesteA] {node [darkgray] {Coverage improving: \testingU{TestGen-LLM}, \testing{ChatGPTTests}{2},  \testing{CoverUp}{1}, }}
        child [white] {node [darkgray] {\phantom{Coverage improving:} \testing{CoverUp}{2}, \testingU{TELPA}}}
        child [celesteA] {node [darkgray] {Targeted: \testingU{CodaMOSA}}}
        child [celesteA] {node [darkgray] {Intent corresponding:  \testingU{CodeT}, \testingU{CodeCoT}, \testingU{AgentCoder}, \testing{ChatTester}{2}}}
        child [celesteA] {node [darkgray] {Parameterized: \testing{PBT-GPT}{3}}}
        child [celesteA] {node [darkgray] {Property based: \testing{PBT-GPT}{4}}}
        child [celesteA] {node [darkgray] {Similarity inspired: \testing{FuzzGPT}{3}}}
        child [celesteA] {node [darkgray] {Mimicking test: \testingU{secTests}}}
        child [celesteA] {node [darkgray] {Bug reproducing: \debugU{LIBRO}}}
    edge from parent}
    child [missing] {}
    child [missing] {}
    child [missing] {}
    child [missing] {}
    child [missing] {}
    child [missing] {}
    child [missing] {}
    child [missing] {}
    child [missing] {}
    child [missing] {}
    child [missing] {}
    child [missing] {}
    child [missing] {}
    child [celesteA] {node {In-filling context}
        child [celesteA] {node [darkgray] {Given prefix: \testing{FuzzGPT}{4}}}
    edge from parent}
    child [missing] {}
    child [celesteA] {node {Corrective mode\phantom{j}}
        child [celesteA] {node [darkgray] {Feedback based\phantom{j}}
            child [celesteA] {node [darkgray] {Compilation errors: \testing{ChatUniTest}{2}, \testing{MuTAP}{2}, \testing{CoverUp}{2}, \testingU{TestPilot}, }} 
             child [white] {node [darkgray] {\phantom{Compilation errors:} \testing{ChatTester}{3}}}
            child [celesteA] {node [darkgray] {Assertion failure: \testing{ChatUniTest}{2}, \testingU{TestPilot}, \testing{ChatTester}{3}}}
            child [celesteA] {node [darkgray] {Examples to augment: \testingU{TELPA}}}
            child [celesteA] {node [darkgray] {Coverage info: \testing{CoverUp}{1}, \testing{CoverUp}{2}}}
            edge from parent}
    edge from parent}
    child [missing] {}
    child [missing] {}
    child [missing] {}
    child [missing] {}
    child [missing] {}
    child [missing] {}
    child [celesteA] {node {Rationalization\phantom{j}}
        child [celesteA] {node [darkgray] {Bug description: \testing{FuzzGPT}{3}}}
    edge from parent};
\end{tikzpicture}
    }%
    \caption{Test Generation. .
    }\label{fig:TG}
\end{figure}

     \begin{figure}[ht]
    \centering
    \resizebox{!}{52\htree}{%
    \begin{tikzpicture}[tasktree]
 
\node [draw,limaA] {\textbf{Rationalized\phantom{j}Code-Classification}}
    child [missing] {}
    child [limaA] {node {Analysis criteria}
        child [limaA] {node [darkgray] {Vulnerability proneness:  \vmf{EditTime}{3}, \vmf{EditTime}{4}, \vmf{SmartAudit}{2}, \vmf{SmartAudit}{3}, }
            child [white] {node [darkgray] {\phantom{lerability proneness:} \vmfU{VulBench}, \vmf{GPTLens}{1},  \vmf{VulDetect}{1}, \vmf{VulDetect}{2},  }}
            child [white] {node [darkgray] {\phantom{lerability proneness:} \static{LATTE}{4}, \vmf{VSP}{1}, \vmf{VSP}{2},  \vmf{LLM4Vuln}{3},  }}
            child [white] {node [darkgray] {\phantom{lerability proneness:}  \vmf{LLM4Vuln}{3w/PreCoT}, \vmf{LLM4Vuln}{3w/PostCoT}, \vmf{DLAP}{3} }}
            child [limaA] {node [darkgray] {Null-dereference: \static{SkipAnalyzer}{1}}}
            child [limaA] {node [darkgray] {Leak presence: \static{SkipAnalyzer}{2}}}
            child [limaA] {node [darkgray] {Data-flow CWI: \vmf{VulDetect}{3} \phantom{j}}}
            child [limaA] {node [darkgray] {Entity inconsistency: \vmf{WitheredLeaf}{1}}}
        edge from parent}
        child [missing] {}
        child [missing] {}
        child [missing] {}
        child [missing] {}
        child [missing] {}
        child [missing] {}
        child [missing] {}
        child [limaA] {node [darkgray] {Malware proneness:  \nvmf{Sleuth}{2} }}
        child [limaA] {node [darkgray] {Fixability analysis}
            child [limaA] {node [darkgray] {Rename: \vmf{WitheredLeaf}{3}}} 
        edge from parent}  
    edge from parent}
    child [missing] {}
    child [missing] {}
    child [missing] {}
    child [missing] {}
    child [missing] {}
    child [missing] {}
    child [missing] {}
    child [missing] {}
    child [missing] {}
    child [missing] {}
    child [missing] {}
    child [limaA] {node {Nature of classification \phantom{j}}
        child [limaA] {node {Multiplicity}
            child [limaA] {node [darkgray] {OneClass (Binary): \vmf{EditTime}{3}, \vmf{EditTime}{4}, \vmf{WitheredLeaf}{1}, }}
            child [white] {node [darkgray] {\phantom{OneClass (Binary):} \vmf{WitheredLeaf}{3}, \static{SkipAnalyzer}{1}, \static{SkipAnalyzer}{2}, }}
            child [white] {node [darkgray] {\phantom{OneClass (Binary):} \vmf{VSP}{2}, \nvmf{Sleuth}{2}, \vmf{DLAP}{3} }}
            child [limaA] {node [darkgray] {MultiLabel \phantom{Ag}}   
                    child  [limaA] {node [darkgray] {Open-ended:  \vmf{SmartAudit}{2}, \vmf{SmartAudit}{3},  \vmf{GPTLens}{1},   }}
                    child [limaA] {node [darkgray] {Close-ended: \vmf{ChatGPTSCV}{1},  \vmfU{VulBench}, \vmf{VulDetect}{1}, \vmf{VulDetect}{2}, } 
                        child [white,draw opacity=0] {node [darkgray] {\phantom{le-endedi} \vmf{VulDetect}{3}, \static{LATTE}{4},  \vmf{VSP}{1}, }}
                        child [limaA] {node [darkgray] {In-context defined classes: \vmf{LLM4Vuln}{3}, \vmf{LLM4Vuln}{3w/PreCoT}, }}
                        child [white,draw opacity=0] {node [darkgray] {\phantom{In-context defined classes:} \vmf{LLM4Vuln}{3w/PostCoT} }}
                    edge from parent}
                edge from parent}
        edge from parent}
    edge from parent}   
    child [missing] {}
    child [missing] {}
    child [missing] {}
    child [missing] {}
    child [missing] {}
    child [missing] {}
    child [missing] {}
    child [missing] {}
    child [missing] {}
    child [missing] {}
    child [limaA] {node {Analysis granularity}
        child [limaA] {node [darkgray] {Snippet:
\vmf{EditTime}{3}, \vmf{EditTime}{4}, \vmf{ChatGPTSCV}{1}, \vmf{SmartAudit}{2}, }}
child [white] {node [darkgray] {\phantom{Snippet:} \vmf{SmartAudit}{3},  \vmfU{VulBench}, \vmf{GPTLens}{1}, \vmf{VulDetect}{1}, }}
            child [white] {node [darkgray] {\phantom{Snippet:} \vmf{VulDetect}{2}, \vmf{VulDetect}{3}, \static{LATTE}{4},  \static{SkipAnalyzer}{1}, }} 
            child [white] {node [darkgray] {\phantom{Snippet:} \static{SkipAnalyzer}{2}, \vmf{VSP}{1}, \vmf{VSP}{2}, \vmf{LLM4Vuln}{3},  }}
            child [white] {node [darkgray] {\phantom{Snippet:} 
 \vmf{LLM4Vuln}{3w/PreCoT}, \vmf{LLM4Vuln}{3w/PostCoT}, \vmf{DLAP}{3} }}
        child [limaA] {node [darkgray] {Line: \vmf{WitheredLeaf}{1}, \vmf{WitheredLeaf}{3} }}  
    edge from parent}
    child [missing] {}
    child [missing] {}
    child [missing] {}
    child [missing] {}
    child [missing] {}
    child [missing] {}
    child [limaA] {node {Extra input}
        child [limaA] {node [darkgray] {Code intent: \vmf{SmartAudit}{3}, \vmf{LLM4Vuln}{3w/PreCoT}, \nvmf{Sleuth}{2}}}
        child [limaA] {node [darkgray] {Taint-info: \static{LATTE}{4}}}
        child [limaA] {node [darkgray] {Review guidance: \vmf{DLAP}{3}}}
    edge from parent}
    child [missing] {}
    child [missing] {}
    child [missing] {}
        child [limaA] {node {Corrective mode \phantom{Ag}}
        child [limaA] {node [darkgray] {Self-validation: \vmf{VulDetect}{3} \phantom{Ag}}}
    edge from parent}
    child [missing] {}
    child [limaA] {node {Reactivity}
        child [limaA] {node [darkgray] {Request further information: \vmf{LLM4Vuln}{3}, \vmf{LLM4Vuln}{3w/PreCoT}, }} 
        child [white,draw opacity=0] {node [darkgray] {\phantom{Request further information:} \vmf{LLM4Vuln}{3w/PostCoT}}}
    edge from parent}
    child [missing] {}
    child [missing] {}
    child [limaA] {node {Rationalization \phantom{j}}
        child [limaA] {node [darkgray] {Assessment/description: \vmf{ChatGPTSCV}{1}, \vmf{VulDetect}{1}, \vmf{VulDetect}{2}, }} 
        child [white,draw opacity=0] {node [darkgray] {\phantom{Assessment/description:} \vmf{VulDetect}{3}, \static{LATTE}{4}, \static{SkipAnalyzer}{1}, }}
        child [white,draw opacity=0] {node [darkgray] {\phantom{Assessment/description:} \static{SkipAnalyzer}{2},  \vmf{VSP}{2}}}
        child [limaA] {node [darkgray] {Thought step by step: \vmf{ChatGPTSCV}{1}, \vmfU{VulBench},  \static{SkipAnalyzer}{1}, }}
        child [white,draw opacity=0] {node [darkgray, draw opacity=1] {\phantom{Thought step by step:} \static{SkipAnalyzer}{2}, \vmf{VSP}{1}, \vmf{VSP}{2}, \nvmf{Sleuth}{2}}}
        child [limaA] {node [darkgray] {Reason: \vmf{GPTLens}{1}, \vmf{VulDetect}{1}, \vmf{VulDetect}{2}, \vmf{VulDetect}{3}, }}
        child [white,draw opacity=0] {node [darkgray] {\phantom{Reason:} \vmf{LLM4Vuln}{3}, \vmf{LLM4Vuln}{3w/PreCoT}, \vmf{LLM4Vuln}{3w/PostCoT},}} 
        child [white,draw opacity=0] {node [darkgray] {\phantom{Reason:} \vmf{DLAP}{3} }}
        child [limaA] {node [darkgray] {Fix: \vmf{WitheredLeaf}{3}, (\vmf{SmartAudit}{3}), \vmf{LLM4Vuln}{3w/PostCoT}}}
        child [limaA] {node [darkgray] {Exploit: \vmf{LLM4Vuln}{3w/PostCoT}}}
    edge from parent};
\end{tikzpicture}
 
  
    }%
    \caption{Rationalized Code-Classification. 
    }\label{fig:RCC}
  \end{figure}

 \begin{figure}[htb]
    \centering
    \resizebox{!}{31\htree}{%
    \begin{tikzpicture}[tasktree]
 
\node [draw,fucsiaC] {\textbf{SW-Entity\phantom{j}Verbalization}}
    child [missing] {}
    child [fucsiaC] {node {Abstraction criteria \phantom{Ag}}
            child [fucsiaC] {node [darkgray] {Summary: \testing{PentestGPT}{8}, \vmf{LLM4Vuln}{1},  \ntesting{HITS}{1}\phantom{Ag}}}
            child [fucsiaC] {node [darkgray] {Intent: \vmf{SmartAudit}{3(CoT)}, \testing{DiffPrompt}{1}, \progver{Clover}{4}, \progver{Clover}{7}, }}
            child [white] {node [darkgray] {\phantom{Intent:} \vmf{LLM4Vuln}{1},   \vmf{LLM4Vuln}{3(PreCoT)}, \vmf{LLMAPIDet}{1}, \vmf{LLMAPIDet}{2},  \phantom{Ag}}} 
            child [white] {node [darkgray] {\phantom{Intent:}  \debug{ChatGPT-4(Log)}{1(CoT)}, \progver{RustProof}{1(CoT)},}} 
            child [white] {node [darkgray] {\phantom{Intent:} \testing{ChatTester}{1}, \vmf{PromptEnhanced}{1}  \phantom{Ag}}}
            child [fucsiaC] {node [darkgray] {Explanation: \testing{SELF-DEBUGGING}{OP1}, \testing{SELF-DEBUGGING}{OP3}, \nvmf{Skyeye}{2}}}
            child [white] {node [darkgray] {\phantom{Explanation:} \nvmf{Sleuth}{1}, \progver{RustProof}{1(CoT)}, \progver{RustProof}{2(CoT)} \phantom{Ag}}}
            child [fucsiaC] {node [darkgray] {Differences: \testing{ClarifyGPT}{3}}}
            child [fucsiaC] {node [darkgray] {Problem solving steps:  \ntesting{HITS}{3}\phantom{Ag}}}
        edge from parent}
        child [missing] {}
        child [missing] {}
        child [missing] {}
        child [missing] {}
        child [missing] {}
        child [missing] {}
        child [missing] {}
        child [missing] {}
        child [missing] {}
        child [fucsiaC] {node {Nature of input entity \phantom{Ag}}
            child [fucsiaC] {node [darkgray] {Intent: \testing{PentestGPT}{8} \phantom{Ag}}}
            child [fucsiaC] {node [darkgray] {Code: \testing{SELF-DEBUGGING}{OP1}, \testing{SELF-DEBUGGING}{OP3}, }
            child [white] {node [darkgray] {\phantom{e:} \vmf{SmartAudit}{3(CoT)}, \testing{DiffPrompt}{1}, \testing{ClarifyGPT}{3}, \vmf{LLM4Vuln}{1}, }} 
                child [white] {node [darkgray] {\phantom{e:}    \vmf{LLM4Vuln}{3(PreCoT)}, \ntesting{HITS}{1}, \ntesting{HITS}{3}, \nvmf{Sleuth}{1}, }}
                child [white] {node [darkgray] {\phantom{e:} \debug{ChatGPT-4(Log)}{1(CoT)}, \testing{ChatTester}{1}, \vmf{PromptEnhanced}{1}  \phantom{Ag}}}
                child [fucsiaC] {node [darkgray] {Commit: \nvmfU{VulAdvisor} }}
                child [fucsiaC] {node [darkgray] {Annotated: \progver{Clover}{4}, \progver{RustProof}{1(CoT)}, \progver{RustProof}{2(CoT)} \phantom{Ag}}}       
            edge from parent}
            child [missing] {}
            child [missing] {}
            child [missing] {}
            child [missing] {}
            child [missing] {}
            child [fucsiaC] {node [darkgray] {Call sequences: \nvmf{Skyeye}{2} \phantom{Ag}}}
            child [fucsiaC] {node [darkgray] {Report: \vmf{LLM4Vuln}{1} \phantom{Ag}}}
            child [fucsiaC] {node [darkgray] {Testing outputs: \testing{PentestGPT}{8}}}
            child [fucsiaC] {node [darkgray] {Assertion: \progver{Clover}{7} \phantom{Ag}}
                child [fucsiaC] {node [darkgray] {Precondition: \progver{RustProof}{1(CoT)}, \progver{RustProof}{2(CoT)} \phantom{Ag}}}
                child [fucsiaC] {node [darkgray] {Postcondition: \progver{RustProof}{2(CoT)} \phantom{Ag}}}
            edge from parent}
            child [missing] {}
            child [missing] {}
            child [fucsiaC] {node [darkgray] {Change action: \vmf{LLMAPIDet}{1} \phantom{Ag}}}
        edge from parent}
        child [missing] {}
        child [missing] {}
        child [missing] {}
        child [missing] {}
        child [missing] {}
        child [missing] {}
        child [missing] {}
        child [missing] {}
        child [missing] {}
        child [missing] {}
        child [missing] {}
        child [missing] {}
        child [missing] {}
        child [missing] {}
        child [fucsiaC] {node {Extra input\phantom{Ag}}
            child [fucsiaC] {node [darkgray] {Execution: \testing{SELF-DEBUGGING}{OP3} \phantom{Ag}}}
            child [fucsiaC] {node [darkgray] {Attack type: \nvmf{Skyeye}{2} \phantom{Ag}}}
            child [fucsiaC] {node [darkgray] {Data flow: \nvmf{Sleuth}{1} \phantom{Ag}}}
        edge from parent};
\end{tikzpicture}

  
    }%
    \caption{SW-Entity Verbalization. 
    }\label{fig:AV}
  \end{figure}

     \begin{figure}[htb]
    \centering
    \resizebox{!}{38\htree}{%
    \begin{tikzpicture}[tasktree]
 
\node [draw,purpuraA] {\textbf{Plan\phantom{j}Generation}}
    child [missing] {}
     child [purpuraA] {node {Generation criteria\phantom{j}}
        child [purpuraA] {node [darkgray] {Intent corresponding: \testing{PentestGPT}{1}}}
        child [purpuraA] {node [darkgray] {Scenario corresponding: \testingD{pwnd}{Pwn'd}{1}}}
        child [purpuraA] {node [darkgray] {Achieve goal: \progver{AlloyRepair}{3}, \testing{TARGET}{1}, \testing{TARGET}{2}, \vmf{LLM4Vuln}{3(exploit)}, }
                child [white] {node [darkgray] {\phantom{ieve goal:} \testing{AXNav}{1}}} 
            child [purpuraA] {node {Constrained\phantom{j}}
                child [purpuraA] {node [darkgray] {Available tools: \testing{PentestGPT}{6}}} 
            edge from parent}  
        edge from parent}
        child [missing] {}
        child [missing] {}
        child [missing] {}
        child [purpuraA] {node [darkgray] {Fusing: \testing{SysKG-UTF}{3}}}
        child [purpuraA] {node [darkgray] {Refinement/Concretization: \testing{PentestGPT}{7}, \debug{AdbGPT}{2}, \testing{SysKG-UTF}{2},}}
        child [white,draw opacity=0] {node [darkgray] {\phantom{Refinement/Concretization:} \testing{SysKG-UTF}{4}, \testing{AXNav}{2}, \vmf{DLAP}{2}}}
    edge from parent}
    child [missing] {}
    child [missing] {}
    child [missing] {}
    child [missing] {}
    child [missing] {}
    child [missing] {}
    child [missing] {}
    child [missing] {}
    child [missing] {}
    child [purpuraA] {node {Domain\phantom{j}}
        child [purpuraA] {node [darkgray] {SW review: \vmf{DLAP}{2}}}
        child [purpuraA] {node [darkgray] {SW-system integration: \testing{PentestGPT}{1}, \testing{PentestGPT}{6}, \testing{PentestGPT}{7}, }}
         child [white] {node [darkgray] {\phantom{SW-system integration:} \debug{AdbGPT}{2}, \testingD{pwnd}{Pwn'd}{1}, \testing{SysKG-UTF}{2},  }}
        child [white] {node [darkgray] {\phantom{SW-system integration:} \testing{SysKG-UTF}{3}, \testing{SysKG-UTF}{4}, \vmf{LLM4Vuln}{3(exploit)}, }} 
        child [white] {node [darkgray] {\phantom{SW-system integration:} \testing{AXNav}{1}, \testing{AXNav}{2}, \testing{AXNav}{4}}}
        child [purpuraA] {node [darkgray] {Repair: \progver{AlloyRepair}{3}}}
        child [purpuraA] {node [darkgray] {Physical: \testing{TARGET}{1}, \testing{TARGET}{2}}}
    edge from parent}
    child [missing] {}
    child [missing] {}
    child [missing] {}
    child [missing] {}
    child [missing] {}
    child [missing] {}
    child [missing] {}
    child [purpuraA] {node {Nature of output\phantom{j}}
        child [purpuraA] {node [darkgray] {Step-by-step plan: \progver{AlloyRepair}{3}, \testing{PentestGPT}{6}, \testing{TARGET}{1}, \testing{TARGET}{2},}}
        child [white] {node [darkgray] {\phantom{Step-by-step plan:} \testingD{pwnd}{Pwn'd}{1}, \testing{SysKG-UTF}{2}, \testing{SysKG-UTF}{3},  }}
        child [white] {node [darkgray] {\phantom{Step-by-step plan:} \testing{SysKG-UTF}{4}, \vmf{LLM4Vuln}{3(exploit)}, \testing{AXNav}{1}, }} 
        child [white] {node [darkgray] {\phantom{Step-by-step plan:} \testing{AXNav}{4}, \vmf{DLAP}{2} }}
        child [purpuraA] {node [darkgray] {Hierarchical plan: \testing{PentestGPT}{1}}}
        child [purpuraA] {node [darkgray] {Command/Action: \testing{PentestGPT}{7}, \debug{AdbGPT}{2}, \testing{AXNav}{2}}}
    edge from parent} 
    child [missing] {}
    child [missing] {}
    child [missing] {}
    child [missing] {}
    child [missing] {}
    child [missing] {}
    child [purpuraA] {node {Corrective mode\phantom{j}}
        child [purpuraA] {node [darkgray] {Feedback based\phantom{j}}
            child [purpuraA] {node [darkgray] {Outcomes: \testing{PentestGPT}{1}, \testing{AXNav}{4}}}
            edge from parent}
            child [missing] {}
        child [purpuraA] {node [darkgray] {Rule based: \testing{TARGET}{2}}}
    edge from parent}
    child [missing] {}
    child [missing] {}
    child [missing] {}
    child [purpuraA] {node {Reactivity}
        child [purpuraA] {node [darkgray] {Tool invocation: \testing{PentestGPT}{1}, \testingD{pwnd}{Pwn'd}{1}}}
    edge from parent}
    child [missing] {}
    child [purpuraA] {node {Rationalization\phantom{j}}
        child [purpuraA] {node [darkgray] {Thought: \testing{PentestGPT}{1}, \testing{PentestGPT}{7}, \debug{AdbGPT}{2}, \testing{AXNav}{1}, }}
        child [white] {node [darkgray] {\phantom{Thought:} \testing{AXNav}{2}}}
        child [purpuraA] {node [darkgray] {Assessment: \testing{SysKG-UTF}{4}}}
        child [purpuraA] {node [darkgray] {Pseudo-code guided assessment: \testing{SysKG-UTF}{3}}}
    edge from parent};
\end{tikzpicture}
    }%
    \caption{Plan Generation. }\label{fig:PG}
  \end{figure}

Transformation types (or subclasses) are defined by means of choices in different dimensions to abstractly capture the functional and externally visible nature of transformations underlying prompting strategies in an attempt to separate the ``what'' from the ``how''. 

Each class exhibits dimensions (variation points) in which choices define potential subclasses of transformations.
Also, in each class leaf, we map transformations that appear in solution proposals that feature the given choice denoted by the leaf. Colors indicate the corresponding SE problem in \autoref{fig:OverSEprobs}, and sub-indexes indicate different transformations performed by the same solution proposal or exploratory study, ordered as they appear in Tables~1--5 in v2 of this preprint~\cite{braberman2024tasks}.

 As it can be observed,  choices for correctness or preference criteria are one of the main properties that define the subclass of transformation. This is, naturally, the major variation point in transformation classes. Yet, some other dimensions define crucial details of the nature/properties of transformations. 
As mentioned above, more specific transformation types of a class are described employing those, potentially orthogonal, dimensions. While some are specific to a transformation class (e.g. \textit{Granularity} \textit{Nature of Classification} in evaluative transformations, \textit{Nature of Property} in Property Identification, etc.), some are common to several classes (e.g., \textit{Nature of (generated) Entity}, \textit{(generation/analysis/correctness) Criteria}, \textit{(problem) Domain}), and a few are applicable, in principle, to all categories:
\textit{Rationalization}, \textit{Reactivity}, and \textit{Corrective feedback}. \emph{In-filling context} stands for a transformation's assumed precondition regarding to which extent and how the container context of the expected solution is provided and related to it. 
\emph{Rationalization} stands for the nature of verbalized thoughts/explanations/argumentation the transformation is supposed to generate (many times, in a Chain-of-Thought generation stream). Note that we consider CoT not only a way to elicit improved inference but also a potentially externally visible output of transformations. In fact, rationalizations could be consumed by humans or other models to assess yielded results. 
Defining a transformation as reactive implies that abstractly may not be simply understood as a simple input-output function but as an (adaptive) goal-oriented trajectory of inputs and outputs. In fact, underlying LLM needs to proficiently interact with its environment to achieve the transformation's goal~\cite{ReAct,DBLP:conf/nips/SchickDDRLHZCS23}. Thus, \emph{Reactivity} choices describe how a transformation is supposed to interact with its environment to achieve its goal. For instance, reactive behavior allows a system to incrementally request missing information. Conversely, using external tools (e.g., symbolic systems) can enhance performance by feeding intermediate results back into the step-by-step inference process of a transformer model. 

Finally, \emph{Corrective feedback} dimension stands for choices for transformations that are meant to correct/repair or improve yielded result based on new evidence (see pattern \texttt{Generate and Fix}). When a transformation is requested with corrective feedback it could be argued it becomes a version of the original transformation in which the elicited generation is cognitively closer to fixing a given solution. Nevertheless, at least this transformation is considered a version of the original one in which hints/feedback are also given to achieve goals.

\section[Characteristics of transformation classes]{Mapping studies into the taxonomy}\label{sec:rq3}

The following are conclusions drawn from Tables and Trees, and from the prompt techniques used by each transformation class, which are summarized in \autoref{fig:chart-prompt}.

\subsection[SE Problems and the use of transformations]{SE Problems and the use of Transformations}\label{sec:probs-and-transf}

\begin{table*}[htb]
 \centering
 \caption{Relation between SE problems and transformation taxonomy.} 
\includegraphics[width=\textwidth]{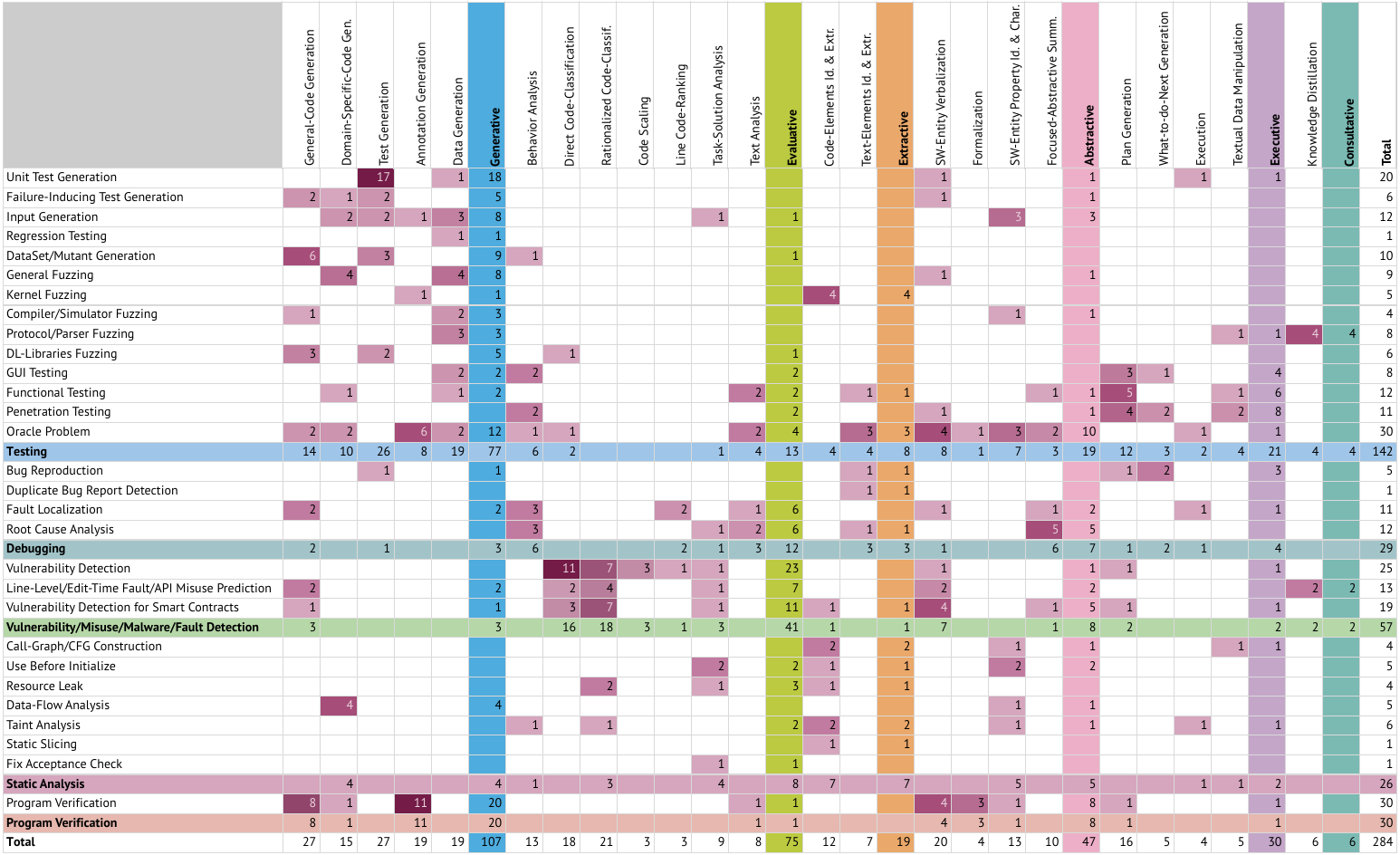}
  \label{tab:correlation}
\end{table*}

\begin{figure}[hbt]
 \centering
 \includegraphics[width=\columnwidth]{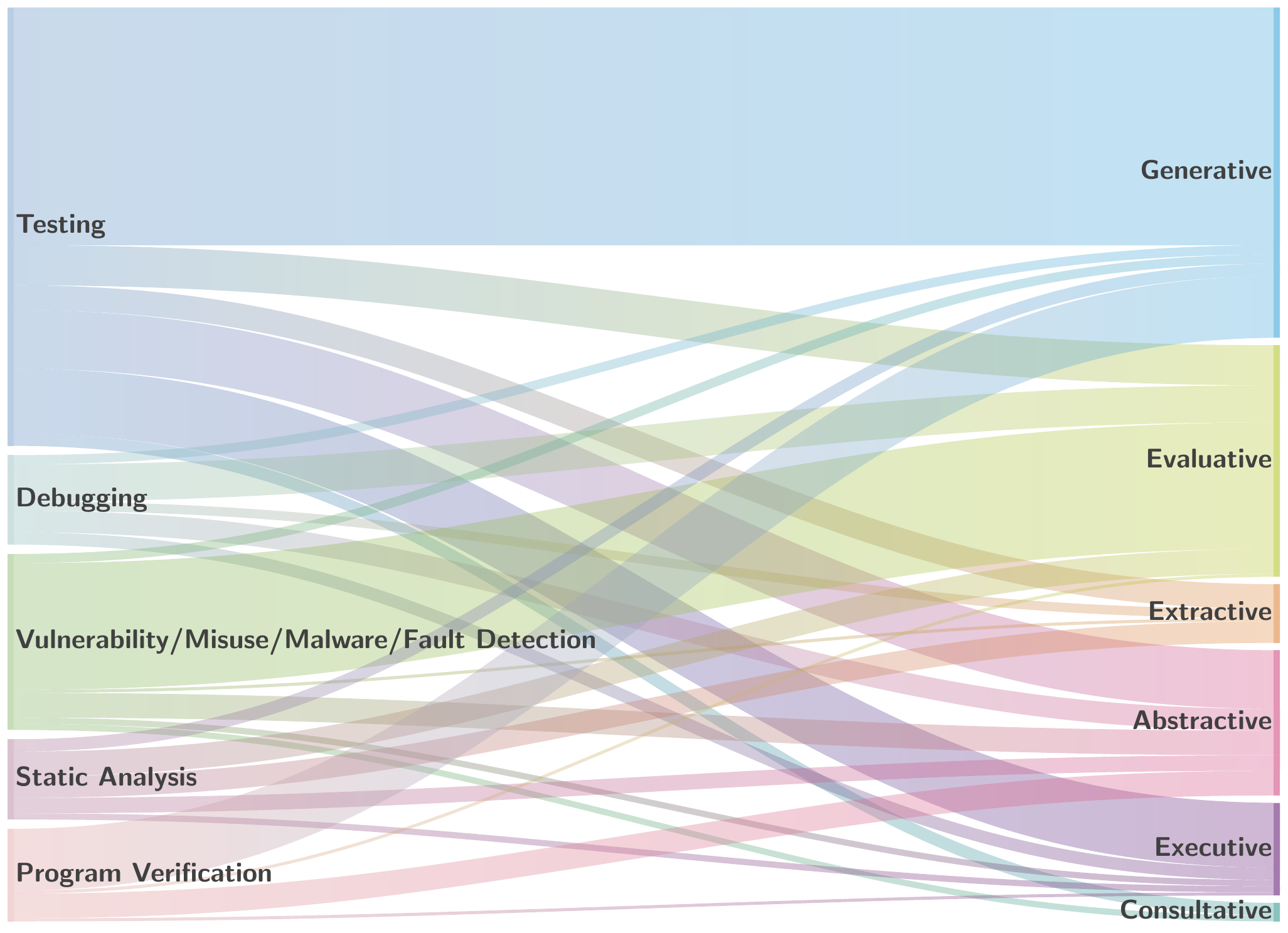}
 \caption{Coarse mapping: SE big areas vs. transf. taxonomy main categories.} 
 \label{fig:correlation-coarse}
\end{figure}

In what follows, we draw preliminary conclusions regarding several aspects of the relationship between SE problem categories and transformation categories (see \autoref{fig:correlation-coarse} and \autoref{tab:correlation}).
Firstly, unit testing approaches are rather homogeneous regarding using generative classes of transformations that are straightforwardly traceable to the final SE problem being addressed. That is, they are typically not broken down into several transformations.

Input generation approaches use LLMs differently.  Typically, they seem to elicit abstractive generation for verbalized characterization of desired input.
Something similar happens to fuzzing approaches that aim to trigger some specific software behavior. They do not rely on a single staged transformation to get data or input. Section on patterns digs into more detail on this.

On the other hand, functional testing, GUI testing, penetration testing and, in general, system-level testing have a strong usage link with executive transformations (e.g., plan generation, decision making, etc.). In fact they would be the main ``clients'' for language models featuring strong planning abilities.

Fuzzing, particularly when inputs are programs (e.g., DL-fuzzing and compiler fuzzing), is the main ``client'' of code generation transformations. Programs that are requested to be generated are typically small, and that, together with the perceived native ability for such transformations, might explain why design patterns and prompting strategies to transformation elicitation are rather straightforward. 

Vulnerability detection constitutes another rather homogeneous SE problems cluster regarding transformation categories involved (mostly evaluative). Smart contract analysis, which often seeks logical vulnerabilities, leverages rationalized classification more frequently than general vulnerability analysis, which is more oriented to detect code patterns. 

Oracle problems, debugging, and static analysis are SE problems for which existing solutions tend to use a wide spectrum of transformation categories. 

Extractive transformations such as code element identification are highly linked to static analysis (LLM-enabled) approaches. 
Indeed, the only testing approach that uses such ability is a kernel-fuzzing approach~\cite{KernelGPT} that uses extractive operations to fill-in a sort of specification later used for generating relevant system calls. The other exception is~\cite{GPTScan}, which tries to locate code playing a certain role for later static analysis.
On the other hand, extractive operations on Natural Language are elicited by approaches addressing debugging, functional testing, and oracle problems.

Program verification approaches may blend SW-entities generative transformations with verbalization and formalization. 

Unsurprisingly, generative transformations are the most elicited, with test generation at the top. Perhaps it is more interesting to pinpoint that general code generation is utilized in almost all categories (except for static analysis approaches). Instead, static analysis approaches like~\cite{LLMDFA} use domain-specific code generation to get scripts and rules to scan code under analysis.
Rationalized vulnerability detection is also a quite populated category due to the number of exploratory studies eliciting such sort of transformations.

It is worth noting that software-entities verbalization (typically but not restricted to intent generation from code) is a class of transformations extensively used trough almost all SE approaches except static analysis. 
Verbalization transformations are either requested by instructing them as the expected result or as intermediate suggested steps in a Chain of Thought. 
We conjecture verbalization being a good performance ``natural'' task~\cite{DBLP:conf/uss/FangMS0ZFATNWH24} and the generated result is also an effective tool for conditioning subsequent transformations (see also Section~\ref{sect:patterns}).

\onecolumn
\begin{footnotesize}
\begin{tblr}[long, theme = fancy, caption = {{Transformation Categories and Characteristics of Mapped Transformations}. }, 
label = {tab:transformation_categories}]{hlines,stretch=1.5,colspec = {p{0.2\linewidth}p{0.75\linewidth}}, width =\textwidth, rowhead = 1, rowfoot = 0, 
rowsep  = 1pt,}

{\textbf{Transformation \\ Category}} & \textbf{Characteristics of Mapped Transformations} \\
\SetRow{celesteB!45}
Code Generation & The most frequent transformation type is the generation of code that corresponds to a given natural language intent. Contextualized transformations (meaning, transformation stated as a ``fill-in'' or completing a partial solution) seem to be preferred whenever they make sense for the addressed problem. Implementation notes: in general terms,  in this category, transformations seem to be elicited by just conditioning the model with their description/instruction. 
 \\
 \addlinespace
 \SetRow{celesteB!30}
{Domain-Specific \\ Code Generation} & Frequent code types are fuzzing drivers (i.e., a program that can execute library functions by feeding them with inputs provided by a fuzzer) or scripts for scanning code. 
Implementation notes: prompting strategies are similar to those of general code generation. Still, there is a proportionally higher use of the Few-Shot demonstrations as an ICL (In-Context Learning) approach~\cite{Few-Shot-Brown20} (possibly, given the potentially fewer data points during pre-training). Function calling~\cite{FUNCTCALLING} also seems an alternative to the lack of training data for the domain specific language (DSL).\\
\addlinespace
\SetRow{celesteB!45}
Test Generation & Types vary greatly on the generation criteria being the ``natural'' (i.e., no specific criteria beyond being as a ``predictable'' test accompanying the code in context according to the model) the more frequent one. 
Implementation notes: test generation is often elicited without using any particular strategy beyond the instruction itself (see~\cite{zhang2023survey} for a similar observation).
In this case, Few-Shot demos are used to clarify some generation criteria that are not standard or in some studies to evaluate ICL.\\
\addlinespace
 \SetRow{celesteB!30}
{Annotation \\ Generation} & Some transformations mapped seem to pursue ambitious correctness criteria (e.g., being  annotations that enable a symbolic proof). Implementation notes: it seems unlikely that LLM has been pre-trained and/or fine-tuned for such tasks, and it seems to require more sophisticated approaches to adequately prompt the model (more than half of transformations require CoT or some guided decomposition).\\
\addlinespace
\SetRow{celesteB!45}
Data Generation & Transformations are rich in terms of criteria/goal spectrum: from consistency with interactions and environment hints to being variants of some other data exemplar.
Implementation notes: Few-Shot seems to be the more common approach after simply instructing to get results.\\
\addlinespace
 \SetRow{limaAA!30}
Behavior Analysis & A typical transformation is root cause analysis which naturally fits as a reactive task that requests/inspects/analyzes information on demand. 
Transformations in this category sometimes require an extra input to perform the evaluation (code, traces or verbalizations) (see also \autoref{sect:patterns}). 
Implementation notes: In-Context Learning and rationalization are also present in the analyzed literature.\\
\addlinespace
 \SetRow{limaAA!45}
{Direct Code \\ Classification} & 
Security vulnerability proneness is the most frequent analysis domain (e.g., CWE vulnerability presence).
In particular, the most frequent classification transformation is identifying whether a snippet of code belongs to the vulnerability-prone class. 
Close-ended classification means the set of labels/classes of interest is predefined (e.g., the target list of CWE vulnerabilities). While some transformations deal with some (pre)fixed concept of vulnerability proneness (e.g., CWE), in some cases, the concept that defines (the) class is actually characterized verbally in-context, that is, transformations are parametric to pieces of information injected and only available at model's inference time (e.g., brand new definitions of vulnerabilities, a behavioral description, 
etc.). Implementation notes: many of these transformations are frequently elicited just by using a description of what is desired (labels for vulnerabilities assumed ``memorized'' and patterns learned either explicitly or implicitly during pre-training). The use of retrieved examples in-context are among the alternatives used by authors when seeking to condition result to what is known about similar cases.\\
\addlinespace
 \SetRow{limaAA!30}
{Rationalized \\ Code-Classification} &
{``Nature of classification'' alternatives are similar to direct code classification (including also denotation relationship between labels and concepts are provided as an actual input parameter of such transformations during inference time). 
Particular to this category is the variety of types of rationalizations that justify classification: typically an assessment or a reason but also fixes and exploits are possible ways to witness the classification of code into a vulnerability proneness class. 
Implementation notes: rationalized transformations compared to direct ones use more frequently ICL prompting strategies. It is notable the use of CoT to get the rationalization of the classification. In some works, it likely acts as a way to improve quality of classification by (auto)guidance$^\dag$.\\

{\scriptsize \begin{spacing}{1} $^\dag$~It is worth noting that from a wider interpretation of what CoT means, CoT usage might be underreported in this category since we typically do not consider as truly ``CoT-compliant'' prompts that suggest the generation of rationalization \textit{after} verbalizing the classification outcome nor prompts that explicitly instruct directed/guided steps of thought (we consider that as  chained transformations). \end{spacing}}
}\\
\addlinespace
 \SetRow{limaAA!45}
Code Scaling & 
In particular, a couple of works feature transformations that were mapped into \emph{Line-code Ranking} category. Implementation notes: they are typically straightforwardly elicited by instructions.\\
\addlinespace
 \SetRow{limaAA!30}
{Task Solution \\ Analysis} & Transformations found are meant to perform  quality evaluation functions~\cite{shankar2024validatesvalidatorsaligningllmassisted} described in prompts and correctness assessment. Implementation notes: they are typically straightforwardly elicited by instructions. \\
\addlinespace
 \SetRow{limaAA!45}
Text Analysis & It includes transformations that resemble those tasks studied and benchmarked by the Natural Language Processing community (e.g., implicatures, ambiguity detection, etc.~\cite{DBLP:conf/acl/SuzgunSSGTCCLCZ23}). Interestingly, while some transformations are mainly domain-independent, others are supposed to be solved by recalling some domain expertise to leverage missing tacit knowledge.
Implementation notes: 
In-Context Learning~\cite{Few-Shot-Brown20} and debate-based inference~\cite{ToT_Yao_2023} are used in a couple of mapped transformations which, apparently, required certain degree of sophistication to elicit non-erring results.\\ 
\addlinespace
 \SetRow{naranjaC!30}
Code Elements Identification and Extraction & Typically transformations mapped into this category are meant to extract elements that fit some given role (some fixed in tool's design-time (e.g.,~\cite{KernelGPT}) and some configurable during tool's use (e.g.,~\cite{GPTScan})).  One of the salient characteristics of transformations mapped in this category is their reactive nature~\cite{DBLP:conf/nips/SchickDDRLHZCS23,ReAct}, in this case, to incrementally explore code bases. Context-size limitations and attention-mechanism degradation~\cite{DBLP:journals/tacl/LiuLHPBPL24} usually motivate the need for such approaches.
Implementation notes: their reported elicitation typically includes In-Context Learning and CoT.\\
\addlinespace
 \SetRow{naranjaC!40}
Text Elements Identification and Extraction & In the studies under analysis, they are typically used to perform extractive summarization of large NL reports~\cite{liu2019finetunebertextractivesummarization}. Implementation notes: Few-Shot and CoT are used and even combined. \\ 
\addlinespace
 \SetRow{fucsiaC!30}
SW-entity Verbalization & Comprises a set of elicited transformations 
to translate code and other software entities into natural language utterances by instructing them to do so. Intent of code is the most common abstraction goal.
Implementation notes: In-Context Learning seems to have been applied only in cases where the verbalization criteria required further guidance (e.g., verbalization of differences, line-by-line explanations, etc.).\\ 
\addlinespace
 \SetRow{fucsiaC!45}
Formalization & Transformations found typically convert intents into code's pre and post conditions. Implementation notes: most cases are elicited by described instruction and in some cases Few-Shot demonstrations are used. \\
\addlinespace
 \SetRow{fucsiaC!30}
SW-entity Property Identification and Characterization & Transformations are quite varied in terms of nature of abstractions uttered (from input validity constraints to static analysis facts) and input is typically code, but descriptions and runtime information can also be input to the transformation.   
Implementation notes: adequate conditioning (e.g., In-Context Learning) is required since those transformations are, intuitively, far from trained ones.\\ 
\addlinespace
 \SetRow{fucsiaC!45}
{Focused Abstractive \\ Summarization} &  Extracting a summary with a conceptual viewpoint is the most common elicited transformation in this category. Implementation notes: although, in most cases, they  seem to be elicited by just instructing the model to perform adequate summarization (or providing some demonstrations in context), sophisticated (auto)guidance approaches (e.g.,~\cite{ToT_Yao_2023}) are in place when it is likely that the model may produce answers not fully aligned with the purpose of the task.\\
\addlinespace
 \SetRow{purpuraB!20}
Plan Generation & For the studies and proposals in scope, transformations deal with either generating or refining operational descriptions meant to reach goals. Typical domain is that of interacting with a software-enabled system.
Implementation notes: the use of CoT guidance is quite frequent, possibly because LLMs struggle to generate criteria-aligned plans~\cite{valmeekam2023planbench,liu2024agentbench}.\\ 
\addlinespace
 \SetRow{purpuraB!35}
What-to-do Next Generation &  Generally speaking, those transformations are related to interacting with a software-enabled system as the agent's environment, given environment and history information. Implementation notes: in most mapped cases, elicitation is done by just instruction description in prompt.\\
\addlinespace
 \SetRow{purpuraB!20}
Execution & It comprises transformations that deal with program-like instruction(s)/intent computation. This could be either at a level of single instruction or potentially a code snippet or intent. Implementation notes: mostly straightforwardly elicited, use of CoT has been observed too. \\
\addlinespace
 \SetRow{purpuraB!35}
Textual Data Manipulation & Textual data structures are typically used as intermediate memorization/summarization/scratchpad elements incrementally manipulated in long inference processes~\cite{nye2021workscratchpadsintermediatecomputation}.
Implementation notes: mostly elicited by instruction description.\\
\addlinespace 
 \SetRow{aguaB!30}
Knowledge Distillation & Unsurprisingly, seem to mainly be used for recalling memorized knowledge related to SW-related domains (e.g., grammar of a message structure). Implementation notes: mostly elicited by instruction description.\\
\addlinespace
\end{tblr}
\end{footnotesize}

\twocolumn

\section{Inter-transformation patterns in solution proposals}
\label{sect:patterns}

In the previous section we explain how elicited inferences associated to the LLM-generated data  were grouped into taxonomy of  transformations. Now, we pursue the goal of finding behavioral patterns based on transformations, according to the studies in scope.  That is, we aim at identifying what seem to be recurrent behavioral design patterns arising from how LLM-based transformations are related, and the likely problems addressed by them. By ``related'' we adopt a broader view, meaning that generated content of one transformation flows to the input of another one or, broadly speaking, affects the execution of another one. That occurs when staging transformations in a pipeline or chain but also when one transformation prepares data that is eventually consumed by another one or a transformation leads to the invocation of another transformation. The details on how the composition should be architected is not the focus of inter-transformation patterns but the nature and role of transformations involved.
Transformations also include types of algorithmically implemented ones that are instrumental to the solution, like compilers or vector data bases.

\begin{figure}[t]
\centering
\includegraphics[width=.9\columnwidth]{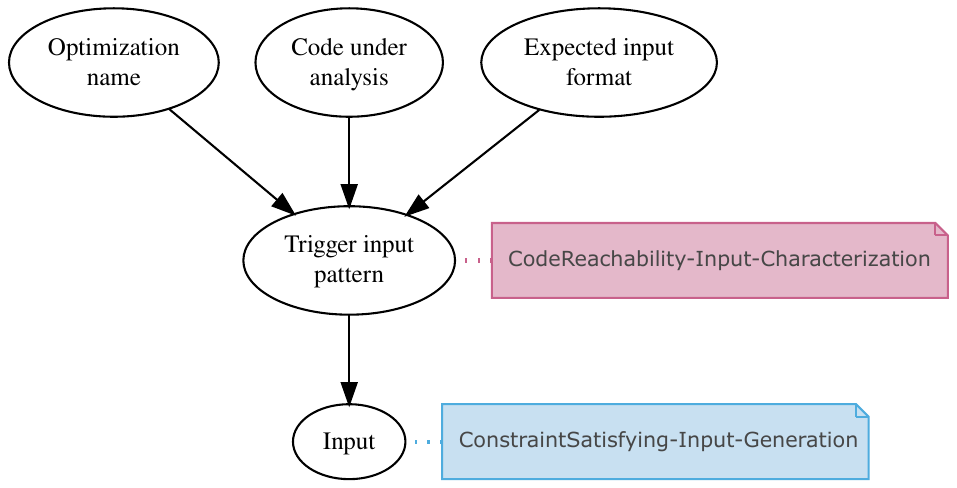}
\caption{An abstract run of WhiteFox~\cite{WhiteFox} exhibiting inter-transformation pattern \texttt{Build Model}.}
\label{fig:whitefox}
\end{figure}

Patterns are, as in the case of Design Patterns~\cite{DBLP:conf/ecoop/GammaHJV93}, potentially reusable solutions to commonly occurring problems, in this case, in many contexts of LLM-native software design. To arrive to such a catalog, we group together and abstract  aspects seen as (potentially) recurrent in the generated data dependencies arising in the transformations of the analyzed approaches. Phenomenology introduced by the notion of transformation and their taxonomy enables to conceptualize generated and computed data dependencies in representative executions of solution proposals as part of a multivariate probabilistic distribution specification as shown in \autoref{fig:ChatTester} and \autoref{fig:whitefox}. 
Such sort of representation sketches helped us to elaborate the inter-transformation pattern catalog. For instance, ChatTester (\autoref{fig:ChatTester}) is an example for the use of two patterns:  \texttt{Use Verbalized SWE} \ and \ \texttt{Generate and Fix}, while WhiteFox (\autoref{fig:whitefox}) features inter-transformation \texttt{Build Model}. Also, taxonomy played a crucial role for discovering them as it is used to label transformations and often permeates into the naming, description and grouping of exemplars found.  
 Given the nature of our work, whether or not they can be considered best practices that may be used to solve stated problems, their pros and cons, or completeness of catalog, is out of scope. Instead, here we map works into identified patterns as a key step forward. That is, we use solution proposals of peer reviewed works\footnote{We believe the focus only on peer reviewed work does not affect the validity of found patterns.} to analyze data dependencies and come up with likely problem solved and solution pattern. 
 Also, low level prompting strategies and linguistic ones are not covered in this catalog either (e.g. ICL strategies) and are considered out-of-scope of this work\footnote{ICL strategies were pinpointed just to provide a sense of eliciting hardness of transformations.}. 

It is worth noting that solution proposals of some peer reviewed papers in scope are not mapped into the pattern table. The reason for this is that they feature single staged transformations for generating SWE. In that group of single staged solution proposals there are four evaluative approaches~\cite{NLBSE24,GPTScan,MultiTask,ChatGPT(Plus)}, 
six extractive transformations~\cite{PROSPER,AdbGPT,SimulinkSlicer,MetaMorph,InferROI,Cupid}, 
five code-generation/code-completion approaches~\cite{FLAG,BugFarm,AID,FormAI,ALGO},
five assertion generation approaches~\cite{nl2postcondition,TOGLL,ChatInv,PropertyGPT,CEDAR},
one data-generation~\cite{QTypist}, 
and one plan-generation~\cite{pwnd}. 
In fact, typically validation studies of evaluative approaches elicit LLMs by a single staged approach as mentioned in previous sections.

 Patterns are grouped into different thematic areas which mainly corresponds to coping with compositional gap~\cite{DBLP:conf/emnlp/PressZMSSL23} and derived topics from broad strategies or other design patterns.  
 \cmss{Dealing with the compositional gap} is arguably the most populated group given that we are particularly interested in the composition of transformations as the common way they appear related. In fact, the variety of strategies found serves as an initial evidence that, as we claim,  compositional gap phenomenon~\cite{DBLP:conf/emnlp/PressZMSSL23} is the main underlying reason for interaction patterns observed when two or more transformations are somehow staged.
 Some other groups refine some of the general patterns presented: \cmss{Reactive Consumption of  Objects}, \cmss{Generating Set of Elements}, \cmss{Generating with Corrective Feedback}, \cmss{RAG-support}. Groups like \cmss{Operating on an Environment}, \cmss{Feedback Processing in Reactive Trajectories}, \cmss{Interaction Memory} are related to subtopics of LLM-enabled transformations typically composed in agentic software.  \cmss{Validation} deals with the general problem of trusting results. Finally, \cmss{Economics} addresses problems derived from the cost of using large foundational models.

Patterns are described in \autoref{tab:patterns} and approaches following them are mapped together with category of involved transformations. 

\subsection{Use of patterns in transformation categories}

\autoref{tab:transformative-goals} highlights how inter-transformation patterns to serve to different categories of end-to-end macro transformations. That is, for each category, we analyze transformations (or staged transformations) in solution proposals which transformative purpose corresponds to that category and then we highlight notable patterns that are either used (or not) in the literature under analysis.

\onecolumn
\begin{footnotesize}
\begin{tblr}[long, theme = fancy, caption = {{Patterns by Topic with References and Observations}. Colors of the tools indicate the corresponding SE-problem in \autoref{fig:OverSEprobs}. }, 
label = {tab:patterns}]{hlines,stretch=1.5,colspec = {p{0.2\linewidth}p{0.75\linewidth}}, width =\textwidth, rowhead = 1, rowfoot = 0, 
rowsep  = 2pt,} 

  \textbf{Pattern Name} & \textbf{Details (Description, Tools, Observations)} \\
  \SetRow{gray!20}
  \SetCell[c=2]{l}{{\small\cmss{\textbf{Dealing with the Compositional-Gap}}}} \\*
   \SetRow{gray!20}
\SetCell[c=2]{l}{LLMs are powerful yet subproblem decomposition might be required for improving performance.} \\*
  \texttt{Enhance} & 
  \begin{minipage}[t]{\linewidth}
    \textbf{Description:} Straightforward use of LLM to solve some code analysis transformations is likely to struggle with inferring the right semantic facts on executions thus leading to hallucinations (e.g., see program state analysis task discussion~\cite{DBLP:journals/tmlr/SrivastavaRRSAF23}). Augmenting signature with structural or verbalized information of existing inputs to improve evaluative transformations. \\
    \textbf{Tools/Observations:}
    \begin{itemize}
      \item (\tvmf{GRACE}, \tvmf{PromptEnhanced}): Code control- and/or data-flow structure for evaluative transformation. 
      \item (\tvmf{PromptEnhanced}, \tvmf{SmartAudit}): SWE verbalization for evaluative transformation.
    \end{itemize}
  \end{minipage} \vspace{0.4pt} 
       \\
    
\texttt{Enrich with Precisions} & 
  \begin{minipage}[t]{\linewidth}
    \textbf{Description:} Performance of generative transformations might benefit from adding more precise verbalized interpretations of NL descriptions (even if those interpretations and their likelihood are inferable from crystallized knowledge).  \\
    \textbf{Tools/Observations:}
    \begin{itemize}
      \item (\ttesting{ScenicNL}): Expert-persona text analysis to derive precise properties of intents before code generation. 
      \item (\ttesting{ClarifyGPT}): Guided user disambiguation to enrich given intent before code generation.
    \end{itemize}
    \end{minipage} \vspace{0.4pt} \\
   
\texttt{Use Verbalized SWE} & 
  \begin{minipage}[t]{\linewidth}
    \textbf{Description:} Leveraging software entities verbalizations (instead of actual SW-entities) as a input of other generative stages might the right way to break up a transformative problem.\\
    \textbf{Tools/Observations:}
    \begin{itemize}
      \item (\ttesting{Fuzz4All}): Usage summarization for data generation stage.
      \item (\ttesting{ClarifyGPT}): Verbalizing differences for text analysis stage.
      \item (\ttesting{ChatTester}): verbalizing intent for test generation stage.
    \end{itemize}
  \end{minipage} \vspace{0.4pt} \\ 
    
  \texttt{Generate and Fix} &
  \begin{minipage}[t]{\linewidth}
    \textbf{Description:} Use correctness feedback when available to fix initial generation. Related topic: {\cmss{Generating with Corrective Feedback}}. \\
    \textbf{Tools/Observations:}
    \begin{itemize}
      \item (\ttesting{PropertyGPT}, \ttesting{KernelGPT}, \tprogver{Clover}, \tprogver{SpecGen}, \tprogver{AutoSpec}, \tprogver{Lemur}, \tprogver{Loopy}): Annotation generation.
      \item (\tprogver{Clover}, \tprogver{SpecGen},  \ttesting{OSS-Fuzz}, \ttesting{UGCTX}, \tstatic{LLMDFA}, \tprogver{AutoSpec}, \tprogver{Lemur}, \tprogver{Loopy}, \ttesting{InputBlaster}): Code generation.
      \item (\ttesting{ChatUniTest}, \ttesting{TestPilot}, \ttesting{ChatTester}): Test generation.
    \item (\tdebug{AutoSD}): Behavior analysis. 
    \item (\ttesting{GPTDroid}, \ttesting{AXNav}, \tdebug{AdbGPT}, \ttesting{Pwn'd}): What-to-do-next, planning.
    \end{itemize}
  \end{minipage}   \vspace{0.4pt}    \\   
   
\texttt{Rationalize then Summarize} & 
  \begin{minipage}[t]{\linewidth}
    \textbf{Description:} Verbose but likely coherent analysis (e.g., rationalized evaluative transformation) is followed by focused summarization to yield the expected concise result.\\
    \textbf{Tools/Observations:}
    \begin{itemize}
      \item (\ttesting{GameBugDescriptions}, \tdebug{AutoFL}): Behavior analysis followed by focused-abstractive summarization. 
      \item (\tvmf{ChatGPTSCV}): Rationalized code-classification postprocessed by focused summarization.
      \item (\tdebug{LM-PACE}): Task-solution analysis (quality assessment) followed by abstractive summarization. Also, text analysis followed by focused-abstractive summarization to yield verdict.
    \end{itemize}
  \end{minipage} \vspace{0.4pt} \\    
   
\texttt{Verbalize Execution} & 
  \begin{minipage}[t]{\linewidth}
    \textbf{Description:} Executive SWE transformation might help to expose subtle semantic aspects to improve later analysis stage.\\
    \textbf{Tools/Observations:}
    \begin{itemize}
      \item (\tdebugD{SELF-DEBUGGING}{FeedbackLoop}): SWE execution uttered to serve a difference analysis stage.
    \end{itemize}
  \end{minipage} \vspace{0.4pt}      \\
   
  \texttt{Tour d'horizon} & 
  \begin{minipage}[t]{\linewidth}
    \textbf{Description:} Elicits overview evaluative transformations. The uttered verbalized result conditions later and more detailed stages. \\
    \textbf{Tools/Observations:}
    \begin{itemize}
      \item (\tvmf{ChatGPT4vul}): Binary classification preceding code ranking.
    \end{itemize}
  \end{minipage}  \vspace{-0.3cm}    \\   
   
\texttt{Elicit Hidden Variable}  & 
  \begin{minipage}[t]{\linewidth}
    \textbf{Description:} Eliciting assumed internal mechanisms for conditioning later verbalized answer. \\
    \textbf{Tools/Observations:}
    \begin{itemize}
      \item (\tvmf{GPTScan}): Hidden thoughts for binary classifications. 
    \end{itemize}
  \end{minipage} \vspace{0.4pt}      \\ 
    
  \texttt{Build Model}  & 
  \begin{minipage}[t]{\linewidth}
    \textbf{Description:} Straightforward use of LLM to solve some code analysis transformations is likely to struggle with inferring the right semantic facts on executions thus leading to hallucinations (e.g., see program state analysis task discussion~\cite{DBLP:journals/tmlr/SrivastavaRRSAF23}). Extracting some useful properties of a SWE-artifact or analyzable models as an initial transformations might be a way to deal with this challenge. \\
    \textbf{Tools/Observations:}
    \begin{itemize}
      \item (\ttesting{EMR}, \ttesting{InputBlaster}): Property identification for domain-specific-code generation.
      \item (\ttesting{WhiteFox}, \ttesting{SymPrompt}, \ttesting{InputBlaster}, \ttesting{RESTGPT}): Constraint-based description of input is uttered for data generation stage.
    \end{itemize}
  \end{minipage}   \vspace{-0.05cm}    \\
    
  \texttt{Transform to Fit}  & 
  \begin{minipage}[t]{\linewidth}
    \textbf{Description:} Transforming input into a more standard problem statement where LLMs might perform better.\\
    \textbf{Tools/Observations:}
    \begin{itemize}
      \item (\ttesting{SearchGEM5}): Parameterized snippet generation and then a data generation stage.
    \end{itemize}
  \end{minipage} \vspace{0.4pt}      \\
   
  \texttt{Divide, Generate and Merge} & 
  \begin{minipage}[t]{\linewidth}
    \textbf{Description:} Modular generation of components of the expected result are later fused and corrected. \\
    \textbf{Tools/Observations:}
    \begin{itemize}
      \item (\ttesting{SysKG-UTF}): Plan generation by refinement of subplans and fusion.
      \item (\ttesting{PBT-GPT}): Deterministic (not LLM-based) fusion of independent generative transformations.
      \item (\ttesting{FuzzingParsers}): LLM-based fusing of generated elements.
    \end{itemize}
   \end{minipage} \vspace{0.4pt}      \\  
   
  \texttt{Extract and Merge} & 
  \begin{minipage}[t]{\linewidth}
    \textbf{Description:} Generation sometimes can be achieved by extracting and building the expected SWE. \\
    \textbf{Tools/Observations:}
    \begin{itemize}
      \item (\ttesting{KernelGPT}): SWE-extraction and then algorithmic construction of result. 
    \end{itemize}
  \end{minipage} \vspace{0.4pt}      \\
      
    \texttt{Extract to Focus} & 
  \begin{minipage}[t]{\linewidth}
    \textbf{Description:} Using LLM-powered extraction  transformations to concentrate the expected transformation on relevant elements of the input SWE. Related topic: {\cmss{Reactive Consumption of Objects}}. \\
    \textbf{Tools/Observations:}
    \begin{itemize}
      \item (\ttesting{SysKG-UTF}): Extractive tasks for planning input.
      \item (\ttesting{EMR}): Text extraction for property identification.
      \item (\tprogver{Dafny-Synth}): Intent to structured NL before formalization.
    \end{itemize}
  \end{minipage} \vspace{0.4pt}      \\
   
 {\texttt{Compute Sequence of \\ Transformations}} & 
  \begin{minipage}[t]{\linewidth}
    \textbf{Description:} Required transformation trajectories can be computed (symbolically or LLM-based) in advance to guide inference steps. \\
    \textbf{Tools/Observations:}
    \begin{itemize}
      \item (\tstatic{LATTE}): Behavior analysis guided by externally fed data-flow steps (symbolically computed).
    \end{itemize}
  \end{minipage} \vspace{0.4pt}      \\
   
    \texttt{Orchestrate Transformations and Tools} & 
  \begin{minipage}[t]{\linewidth}
    \textbf{Description:} It might be not trivial to statically or algorithmically determine when and how to adequately invoke and use transformations and tool results. Let LLM orchestrate and instantiate a subset of available transformations (or tools) to achieve some higher-level transformative goal. That could be done reactively according to results. Related topics: {\cmss{Reactive Consumption of Objects}}, {\cmss{Operating on an Environment}}, {CoT}. \\
    \textbf{Tools/Observations: Only found for the special case \texttt{Reactively Consume Input}.}
  \end{minipage} \vspace{0.4pt} \\
   
  {\texttt{Reactively Consume \\ Input}} & 
  \begin{minipage}[t]{\linewidth}
    \textbf{Description:} Large and complex objects (e.g., repositories, models, etc.)  are inputs of analysis, generation, or retrieval transformations. Bulk treatment might nor work. LLM can guide the exploration and analysis and react to retrieved/queried information. Related topic: {\cmss{Reactive Consumption of Objects}}. \\
    \textbf{Tools/Observations:}
    \begin{itemize}
      \item (\tstatic{LLift}, \ttesting{KernelGPT}): SWE-property identification and SWE-extraction trajectories.
      \item (\tdebug{AutoFL}, \tdebug{RCAAgents}): Behavior analysis from information and code.
      \item (\ttesting{ChatUniTest}): Test generation.
    \end{itemize}
  \end{minipage} \vspace{0.4pt}      \\  
   
  {\texttt{Wrapping Input for \\ Reactive Consumption}} &
  \begin{minipage}[t]{\linewidth}
    \textbf{Description:} An LLM-based transformation might help input consumption by translating requests into lower level operations and/or concrete answers. Related topic:  {\cmss{Reactive Consumption of Objects}}. \\
    \textbf{Tools/Observations:}
    \begin{itemize}
          \item (\tdebug{RCAAgents}, \ttesting{ScenicNL}): Text-analysis transformation to answer open queries on textual feedback or summarized text. 
    \end{itemize}
  \end{minipage} \vspace{-0.2cm}    \\
    \SetRow{gray!20}
  \SetCell[c=2]{l}{{\small\cmss{\textbf{Reactive Consumption of  Objects}}}} \\*
    \SetRow{gray!20}
  \SetCell[c=2]{l}{\begin{minipage}[t]{\linewidth}
  Analysis or retrieval of  relevant information may require dealing with large and complex objects and bulk treatment may not work for several reasons.
  \end{minipage}} \vspace{0.1pt} \\*
 
  \texttt{Extract Incrementally} & 
  \begin{minipage}[t]{\linewidth}
    \textbf{Description:} Reactive information extraction for a next stage of LLM- or algorithmic-based processing. \\
    \textbf{Tools/Observations:}
    \begin{itemize}
      \item (\tstatic{LLift}, \ttesting{KernelGPT}): Code-elements identification \& extraction trajectories.
    \end{itemize}
  \end{minipage} \vspace{0.4pt}      \\
  
 {\texttt{Extract \& Perform} \\ \texttt{Incrementally}} & 
  \begin{minipage}[t]{\linewidth}
    \textbf{Description:} Knowing what to extract might require simultaneous incremental analysis. \\
    \textbf{Tools/Observations:}
    \begin{itemize}
      \item (\tdebug{AutoFL}): Code-fetching for behavior analysis.
      \item (\tdebug{RCAAgents}): Information requests (together with question generation) for behavior analysis.
    \end{itemize}
  \end{minipage} \vspace{0.4pt}      \\  
  \SetRow{gray!20}
  \SetCell[c=2]{l}{{\small\cmss{\textbf{Generating Set of Elements}}}} \\*
    \SetRow{gray!20}
  \SetCell[c=2]{l}{\begin{minipage}[t]{\linewidth} Objects to be generated sometimes are compound of individual SWE (e.g., a test suite). Also, multiple intermediate variations of artifacts may be useful. 
  \end{minipage}} \vspace{0.1pt} \\*
 
  \texttt{Sequence Elements} & 
  \begin{minipage}[t]{\linewidth}
    \textbf{Description:} Generating compound SW entities and set of variants  simply as a trajectory of SWE elements.\\
    \textbf{Tools/Observations:}
    \begin{itemize}
\item (\ttesting{DiffPrompt}, \ttesting{TestGen-LLM},  \ttesting{ChatGPTTests}, \ttesting{CodeT}, \ttesting{ChatUniTest}, \ttesting{CodaMOSA}, \ttesting{TestPilot}, \ttesting{EASEeval}, \ttesting{ChatTester}): Test-generation for test suite construction. 
      \item (\ttesting{Fuzz4All}): Data-generation variations from input.
      \item (\ttesting{$\mu$Bert}): In-filling code-generation.  
      \item (\ttesting{DiffPrompt}, \ttesting{ClarifyGPT}): Intent-corresponding code-generation variations. 
      \item (\ttesting{FuzzingParsers}): Error-triggering variations.
          \end{itemize}
  \end{minipage} \vspace{0.4pt}     \\
   
  {\texttt{Sequence Elements and \\ Guiding Utterances}} & 
  \begin{minipage}[t]{\linewidth}
    \textbf{Description:} Guided variants generation. \\ 
    \textbf{Tools/Observations:}
    \begin{itemize}
      \item (\ttesting{LLMorpheus}, \ttesting{CHEMFUZZ}): Rationalized code variants.
      \item (\ttestingD{FuzzGPT}{FewShot}): Diverse code-generation with knowledge distillation transformation to utter hints intermediate. 
    \end{itemize}
  \end{minipage} \vspace{0.4pt}    \\
   
  {\texttt{Generate Element and \\ Mutate It}} & 
  \begin{minipage}[t]{\linewidth}
    \textbf{Description:} Exemplar-based generation as the first stage of variants generation. \\  
    \textbf{Tools/Observations:}
    \begin{itemize}
      \item (\ttesting{TitanFuzz}, \ttesting{ChatAFL}): Exemplar code-generated and then code-generation of variants by masking and in-filling/completion. 
    \end{itemize}
  \end{minipage} \vspace{0.4pt}    \\
    
  {\texttt{Reactively Sequence \\ Elements}} & 
  \begin{minipage}[t]{\linewidth}
    \textbf{Description:} Feedback-driven solution variation. \\  
    \textbf{Tools/Observations:}
    \begin{itemize}
      \item (\tdebug{LLM4CBI}): Mutation code-generation solutions leveraged by compilation feedback.
    \end{itemize}
  \end{minipage} \vspace{0.4pt}    \\
    \SetRow{gray!20}
  \SetCell[c=2]{l}{{\small\cmss{\textbf{Generating with Corrective Feedback}}}} \\*
   \SetRow{gray!20}
   \SetCell[c=2]{l}{LLMs hallucinate and err; validation signals can feed into corrective generative trajectories.} \\*
    
  \texttt{Fix with Sanity Check Signal} & 
  \begin{minipage}[t]{\linewidth}
    \textbf{Description:} Using sanity check provided by external tool to guide correction trajectory. \\
    \textbf{Tools/Observations:}
    \begin{itemize}
      \item (\ttesting{PropertyGPT}, \ttesting{KernelGPT}, \tprogver{Clover}, \tprogver{SpecGen}): Error-aware correction for annotation generation.  
      \item (\ttesting{ChatUniTest}, \ttesting{ChatTester}, \ttesting{TestPilot}, \ttesting{CoverUp}): Test-generation with compiler feedback.
      \item (\tprogver{Clover}, \tprogver{SpecGen}, \ttesting{OSS-Fuzz}, \ttesting{UGCTX}, \tstatic{LLMDFA}): Compiler-checked code generation.
    \end{itemize}
  \end{minipage} \vspace{0.4pt} \\
   
  \texttt{Fix with Goal Satisfaction Signal} & 
  \begin{minipage}[t]{\linewidth}
    \textbf{Description:} Incorporating verification outcomes into generation/correction trajectory. \\  
    \textbf{Tools/Observations:}
    \begin{itemize}
    \item (\ttesting{CoverUp}): Coverage Information.
    \item (\ttesting{ChatUniTest},  \ttesting{TestPilot}): Assertion failure.
      \item (\tprogver{AutoSpec}, \tprogver{Lemur}, \tprogver{Loopy}): Verified annotation generation.
      \item (\tprogver{Lemur}, \tprogver{AutoSpec}, \tprogver{Loopy}, \ttesting{InputBlaster}): Verified code generation. Effectiveness feedback for the case of InputBlaster (DS-code generation).
      \item (\tdebug{AutoSD}): Reactive trajectory behavior analysis generating hypothesis which are tested.
    \end{itemize}
  \end{minipage} \vspace{0.4pt}   \\
  \SetRow{gray!20}
  \SetCell[c=2]{l}{{\small\cmss{\textbf{RAG-Support}}}} \\*
  \SetRow{gray!20}
  \SetCell[c=2]{l}{Populating DBs for similarity-based information retrieval may leverage LLM-based transformations.} \\*
   
  {\texttt{NL Abstractions for \\ Embeddings}} & 
  \begin{minipage}[t]{\linewidth}
    \textbf{Description:} Improving retrievability through abstraction. \\ 
    \textbf{Tools/Observations:}
    \begin{itemize}
      \item (\tvmf{LLMAPIDet}, \tdebug{x-lifecycle}): Abstractive summarization/verbalization for storing into vectorial DB, later fetched by similarity with verbalized/abstractive summarization of input (\tvmf{LLMAPIDet}, \tdebug{RCACopilot}).
    \end{itemize}
  \end{minipage} \vspace{0.4pt}     \\
     
  {\texttt{Classifications for \\ Embeddings}} & 
  \begin{minipage}[t]{\linewidth}
    \textbf{Description:} Code classification for retrieval enhancement. \\
    \textbf{Tools/Observations:}
    \begin{itemize}
      \item (\ttestingD{FuzzGPT}{FewShot}): Code classification for ulterior demonstration retrieval.
    \end{itemize}
  \end{minipage} \vspace{0.4pt}    \\
    \SetRow{gray!20}
  \SetCell[c=2]{l}{{\small\cmss{\textbf{Validation}}}} \\*
  \SetRow{gray!20}
  \SetCell[c=2]{l}{Ultimately, results are unreliable.} \\*
 
  \texttt{Vote} & 
  \begin{minipage}[t]{\linewidth}
    \textbf{Description:} Using multiple LLM samples to establish consensus through voting mechanisms. \\ 
    \textbf{Tools/Observations:}
    \begin{itemize}
      \item (\tstatic{LLift}): Majority-vote over task solution. 
      \item (\tdebug{LM-PACE}): Multiple answer sampling for distribution analysis.
    \end{itemize}
  \end{minipage} \vspace{0.4pt}   \\
   
  \texttt{Judge} & 
  \begin{minipage}[t]{\linewidth}
    \textbf{Description:} Using LLM-based task evaluation processes to validate outputs. \\
    \textbf{Tools/Observations:}
    \begin{itemize}
      \item (\ttesting{mrDetector}): Data generation with task solution analysis as judgment.
      \item (\tstatic{SkipAnalyzer}, \tvmf{GPTLens}, \tvmf{VulDetect}): Rationalized classifications with task analysis.
       \item (\tstatic{LLift}): SWE-property identification \& characterization  with task-solution analysis.
    \end{itemize}
  \end{minipage} \vspace{0.4pt}   \\
   
  \texttt{Test Symbolically} & 
  \begin{minipage}[t]{\linewidth}
    \textbf{Description:} Employing symbolic methods to verify LLM outputs. \\
    \textbf{Tools/Observations:}
    \begin{itemize}
      \item (\tvmf{GPTScan}): Role-based software identification with symbolic validation.
    \end{itemize}
  \end{minipage} \vspace{0.4pt}   \\
   
  \texttt{Cross Check Consistency} & 
  \begin{minipage}[t]{\linewidth}
    \textbf{Description:} Ensuring consistency across multiple representations or modalities. \\
    \textbf{Tools/Observations:}
    \begin{itemize}
      \item (\tprogver{Clover}): Functional equivalence analysis of verbalized software entities.
    \end{itemize}
  \end{minipage} \vspace{0.4pt}   \\
 
  \texttt{Constrain Generation} & 
  \begin{minipage}[t]{\linewidth}
    \textbf{Description:} Ensuring adherence to formal rules by constraining next-token generation (e.g.~\cite{koo2024automatabasedconstraintslanguagemodel}). \\
    \textbf{Tools/Observations:}
    \begin{itemize}
      \item (\ttesting{ScenicNL}): DS-code generation.
    \end{itemize}
  \end{minipage} \vspace{0.4pt}   \\
  \SetRow{gray!20}
  \SetCell[c=2]{l}{{\small\cmss{\textbf{Operating on an Environment: Approach}}}} \\*
  \SetRow{gray!20}
  \SetCell[c=2]{l}{\begin{minipage}[t]{\linewidth} Problem-domain may require interacting by means of actions with complex environments. Agentic systems are typical exemplars. Related topics: {\cmss{Feedback Processing}}, {\cmss{Interaction Memory}}. \end{minipage}} \vspace{0.1pt} \\*
 
  \texttt{Decide Action as You Go} & 
  \begin{minipage}[t]{\linewidth}
    \textbf{Description:} Reactive action decision-making based on status and trajectory. \\  
    \textbf{Tools/Observations:}
    \begin{itemize}
      \item (\ttesting{Pwn'd}): Reachability-oriented what-to-do-next trajectory.
      \item (\ttesting{GPTDroid}, \tdebug{CrashTranslator}): Goal-oriented what-to-do-next reactive trajectories.
      \item (\tdebug{CrashTranslator}): Chained what-to-do-next generation for refining each trajectory step.
    \end{itemize}
  \end{minipage} \vspace{0.4pt}   \\
   
%
  \texttt{Plan, Refine, Act, Update} & 
  \begin{minipage}[t]{\linewidth}
    \textbf{Description:} Iterative planning with execution feedback. \\
    \textbf{Tools/Observations:}
    \begin{itemize}
      \item (\ttesting{AXNav}): Plan generation, refinement, one step execution, and update based on feedback.  
    \end{itemize}
  \end{minipage} \vspace{0.4pt}   \\
   
  \texttt{Use Given Plan and Refine It} & 
  \begin{minipage}[t]{\linewidth}
    \textbf{Description:} Plan refinement at execution-time.\\
    \textbf{Tools/Observations:}
    \begin{itemize}
      \item (\tdebug{AdbGPT}): Dynamic plan-refinement.
    \end{itemize}
  \end{minipage} \vspace{0.4pt}  \\
    \SetRow{gray!20}
  \SetCell[c=2]{l}{{\small\cmss{\textbf{Feedback Processing in Reactive Trajectories}}}} \\* 
   
  \texttt{Do Bulk Processing} & 
  \begin{minipage}[t]{\linewidth}
    \textbf{Description:} Analyzing feedback with a single staged transformation. \\
    \textbf{Tools/Observations:}
    \begin{itemize}
      \item (\tdebug{AutoSD}): Text-analysis-based judgment on feedback.
      \item (\ttesting{GPTDroid}): Behavior analysis for concluding environment state. 
      \item (\ttesting{AXNav}): Behavior analysis (anomaly detection). 
    \end{itemize}
  \end{minipage} \vspace{0.4pt}   \\
     
  \texttt{Process Feedback Reactively} & 
  \begin{minipage}[t]{\linewidth}
    \textbf{Description:} Feedback is further treated as a large and complex feedback. Related topic: {\cmss{Reactive Consumption of Objects}}. Related patterns: \texttt{Reactively Consume Input}, \texttt{Wrapping Input for Reactive Consumption}. \\
    \textbf{Tools/Observations:}
    \begin{itemize}
      \item (\tdebug{RCAAgents}): Behavior analysis to generate questions and text analysis to generate answers.
    \end{itemize}
  \end{minipage} \vspace{-0.05cm} \\  
  
    \SetRow{gray!20}
  \SetCell[c=2]{l}{{\small\cmss{\textbf{Operating on an Environment: Interaction Memory}}}} \\*
  \SetRow{gray!20}
  \SetCell[c=2]{l}{Exploration may require memory.} \\*
 
  \texttt{Trajectory in Context} & 
  \begin{minipage}[t]{\linewidth}
    \textbf{Description:} Simply maintaining context across interactions acts as history. \\
    \textbf{Tools/Observations:}
    \begin{itemize}
      \item (\tdebug{RCAAgents}, \tdebug{CrashTranslator}, \tdebug{AdbGPT})
    \end{itemize}
  \end{minipage} \vspace{0.4pt}  \\
   
  \texttt{Update ScratchPad} & 
  \begin{minipage}[t]{\linewidth}
    \textbf{Description:} Context required is not necessarily small or attention mechanism degrades. Scratchpads are manipulated by an LLM. \\
    \textbf{Tools/Observations:}
    \begin{itemize}
      \item (\ttesting{AXNav}): Plan updates based on context.
    \end{itemize}
  \end{minipage} \vspace{0.4pt}   \\
   
  {\texttt{Algorithmically \\ Updated ScratchPad}} & 
  \begin{minipage}[t]{\linewidth}
    \textbf{Description:}  Memory/Scratchpad is updated by an algorithmic mechanism. \\ 
    \textbf{Tools/Observations:}
    \begin{itemize}
      \item (\ttesting{GPTDroid}): Symbolic context updates.
    \end{itemize}
  \end{minipage} \vspace{0.4pt}  \\
    \SetRow{gray!20}
  \SetCell[c=2]{l}{{\small\cmss{\textbf{Economics}}}} \\*
  \SetRow{gray!20}
  \SetCell[c=2]{l}{Invoking many times a cutting-edge LLM might be too expensive.} \\*
 
  \texttt{Filter before Expensive Call} & 
  \begin{minipage}[t]{\linewidth}
    \textbf{Description:} Filtering with a cheaper and less capable LLM to get a short list. \\   
    \textbf{Tools/Observations:}
    \begin{itemize}
      \item (\tvmf{GPTScan}): Simpler behavior analysis before a more sophisticated one.
    \end{itemize}
  \end{minipage} \vspace{0.4pt}   \\
\end{tblr}
\end{footnotesize}

\begin{footnotesize}
\begin{tblr}[long, theme = fancy, caption = {{Categories of Final Transformative Goals and  Observation on Patterns Used}. }, 
label = {tab:transformative-goals}]{hlines,stretch=1.5,colspec = {p{0.2\linewidth}p{0.75\linewidth}}, width =\textwidth, rowhead = 1, rowfoot = 0, 
rowsep  = 1pt,} 

\textbf{Category of Final Transformative Goal} & \textbf{Patterns Observations} \\
\SetRow{celesteB!45}
Code Generation &
\texttt{Generate and Fix} is the typical pattern followed to improve the performance of code generation by repairing initial hallucinations. \texttt{Fix with Sanity Check
Signal} depends on the sort of transformation category: for code generation compilation, feedback is the natural option, but other signals are possible (e.g., verifier feedback in~\cite{Clover}).\\
\addlinespace[0.5ex]
 \SetRow{celesteB!30}
DS-Code  Generation&
\texttt{Fix with Goal Satisfaction
Signal} by using execution results is possible as they can be interpreted as a relevant signal for iterative improvement of transformation results.
\\
\addlinespace[0.5ex]
\SetRow{celesteB!45}
Test Generation &
\texttt{Fix with Sanity Check
Signal\/} and \texttt{Fix with Goal Satisfaction
Signal} are both used as specializations of \texttt{Generate and Fix}. A single case of \texttt{Reactively Consume Input} can be found in~\cite{ChatUniTest}.
\texttt{Sequence Elements and
Guiding Utterances} is used by uttering bug description in~\cite{FuzzGPT}. 
\\
\addlinespace[0.5ex]
 \SetRow{celesteB!30}
Annotation  Generation &
Enriching code with annotations uses compilation errors  (i.e., \texttt{Fix with Sanity Check
Signal}) and  verification results (i.e., \texttt{Fix with Goal Satisfaction
Signal})  for patterns that are instances of  \texttt{Generate and Fix}. There is no use of \texttt{Reactively Consume Input} as context is typically feed  entirely into transformations instances found.
 \\
\addlinespace[0.5ex]
\SetRow{celesteB!45}
Data Generation &
\texttt{Use Verbalized SWE\/} and \texttt{Build Model} patterns are followed. 
No use of \texttt{Generate and Fix} was found and there is one case of \texttt{Judge}. No use of \texttt{Reactively Consume Input} as context is bulk processed in the found instances of  data-generation  transformations.
\\
\addlinespace[0.5ex]
 \SetRow{limaAA!30}
Direct Code Classification&
Notably, possibly due to the static-analysis nature of classification and the flexibility of class definition, no instances of \texttt{Generate and Fix} pattern were found. Extra inputs as \texttt{Enhance} (e.g., call graphs, data-flow facts, etc.) are, in some works, provided to improve the quality of classification.\\
\addlinespace[0.5ex]
 \SetRow{limaAA!45}
Rationalized Code Classification & 
Abstractive transformations (e.g., verbalization of software entities) are requested to be uttered before rationalizations (i.e.,  \texttt{Enhance}).
Few use \texttt{Reactively Consume Input} pattern to incrementally explore  code base.  

Generally, evaluative transformations in-scope do not \texttt{Generate and Fix}. For rationalized transformations, self-reflection~\cite{Reflexion} is one of the attempted alternatives to \texttt{Judge}.\\
\addlinespace[0.5ex]
 \SetRow{limaAA!30}
Code Scaling &
There are transformations which use \texttt{Generate and Fix} by means of testing feedback. Some follow \texttt{Use Verbalized SWE} pattern.\\
\addlinespace[0.5ex]
 \SetRow{limaAA!45}
Task Solution Analysis & \texttt{Vote} is used to gain some reliability by consensus. 
\\
\addlinespace[0.5ex]
 \SetRow{limaAA!30}
Text Analysis & No  pattern usage was found to improve performance, validate or fix for this category of transformations.
\\
\addlinespace[0.5ex]
\SetRow{naranjaC!40}
Code Elements Identification and Extraction & 
One of the salient characteristics of its transformations is their reactive nature (i.e., \texttt{Reactively Consume Input}).
\\
\addlinespace[0.5ex]
\SetRow{naranjaC!30}
Text Elements Identification and Extraction & 
No  pattern usage was found to improve performance, validate or fix for this category of transformations. Notably, transformations currently mapped into this category are not designed to be reactive and, thus, they are supposed to work with text that could fit into the LLMs context.\\
\addlinespace[0.5ex]
\SetRow{fucsiaC!45}
SWE Verbalization &
No  pattern usage was found to improve performance, validate or fix for this category of transformations.
\\
\addlinespace[0.5ex]
\SetRow{fucsiaC!30}
Formalization &
Syntactic corrective feedback (i.e., \texttt{Fix with Sanity
Check Signal}) is featured by only one transformation in this category.\\
\addlinespace[0.5ex]
\SetRow{fucsiaC!45}
SWE Property Identification and Characterization &
Only self-validation (e.g.,~\cite{Self-consist,DBLP:journals/corr/abs-2207-05221}) seems to be the way to obtain signal for correction  (i.e., \texttt{Judge\/} and \texttt{Vote}). This might be due to difficulties to get or process  some effective corrective signals. 
\\
\addlinespace[0.5ex]
 \SetRow{purpuraB!20}
Plan Generation &
\texttt{Divide, Generate and Merge} is one pattern observed to get initial plans. 
Plans might be corrected with feedback, either new environmental situation or proposed action results, as different instances of \texttt{Generate and Fix} pattern.\\
\addlinespace[0.5ex]
 \SetRow{purpuraB!35}
What-to-do Next &
Corrective transformations might be feedback with either new environmental situation or proposed action feedback as different instances of \texttt{Generate and Fix} pattern. No orchestration or enhancement of potential assisting tools (e.g., AI Planners, World Models, etc.) have been found.\\
\addlinespace[0.5ex]
 \SetRow{purpuraB!20}
Execution &
Reactivity for using external computation tools is used in this class (\texttt{Orchestrate Transformations and Tools}). No corrective feedback found.\\
\addlinespace[0.5ex]
 \SetRow{purpuraB!35}
Textual Data Manipulation & No  pattern usage was found to improve performance, validate or fix for this category of transformations.
\\
\addlinespace[0.5ex]
\SetRow{aguaB!30}
Knowledge Distillation & No  pattern usage was found to improve performance, validate or fix for this category of transformations.
\\
\end{tblr}
\end{footnotesize}
\twocolumn

\section{Actionable insights}

While some observations on gaps have been pinpointed in preceding sections (e.g., lack of patterns used for some transformative goals in \autoref{tab:transformative-goals}) there are some aspects of the current landscape to which one could add prospective comments and avenues for future work from the identified transformations and patterns. 

\paragraph{Verbalized Abstractions for Enhancement}
Getting useful behavioral abstractions seems an important intermediate step in many LLM-enabled transformations chains (e.g., pattern \texttt{Build Model}). 
We envision that more research would seek for the effective construction (or use) of varied language-oriented abstractions of SW-entities (e.g., GORE~\cite{GORE} artifacts, abstract behavioral models~\cite{EPA}, etc.). Those abstractions should be designed to leverage language analytical inference as a way to deal with the combinatorial nature of behavior (in a more historical vein, compositional deductive analysis frameworks (e.g.,~\cite{DBLP:journals/cacm/Hoare69,10.1145/177492.177726,DBLP:books/daglib/0077033}, etc.) have been helping humans to reason about large state spaces). That would require improving  validation and corrective strategies as well as  modular ways to build them.
As noted, correction in abstractive transformations seems, in general, an area of vacancy when compared with traditional (combinatorial) abstraction construction (e.g. CEGAR~\cite{CEGAR}, where spurious counterexamples are key to interpolation-based refinement), automated error detection (e.g. 
misalignment) and informative corrective feedback in LLM-based abstractions might be challenging. 
We hypothesize that the ability to perform some executive and data generation transformations might be useful to close some feedback loops in static analysis use cases.  

\paragraph {Reactive Consumption of Models}  
In reported transformations, models either built by LLM-transformations or symbolic computation are bulk processed by enhanced transformations. Thus, one can envision an opportunity to reactively consume models when they possess inference capabilities (e.g. probabilistic distribution specifications or programs). Moreover, using LLMs to build world models as an intermediate to reason about relational systems, physical scenes and plans has been advocated by~\cite{wong2023word}  and similar ideas could be empower solutions to sophisticated verification and testing problems: world models (built using abstractive transformations) could be later act as ``external'' components which inference capabilities can be thus leveraged by a reactive transformation to improve their alignment and specificity. 

\paragraph{Improvement and evaluation of transformations and patterns}
A taxonomy can help focus efforts to improve LLM proficiency in classes or instances of transformations that are recurrent and might deserve specific (either existing or novel) strategies to analyze and improve LLM-based transformation performance.
Fine-tuning~\cite{instructionfinetuning}, reasoning bootstrapping~\cite{DBLP:conf/nips/ZelikmanWMG22}, enhanced activation/hallucination detection  (e.g.~\cite{obeso2025realtimedetectionhallucinatedentities,Anthropic}), finer grained prompt tuning (e.g.,~\cite{APE}), prefix tuning~\cite{li2021prefixtuning}, gisting~\cite{DBLP:conf/nips/Mu0G23}, prover-verifier games~\cite{kirchner2024proververifiergamesimprovelegibility}, hypothesis search~\cite{DBLP:conf/iclr/WangZPPHG24}, corrective feedback, external tool integration, specially designed benchmarks (e.g.,~\cite{DBLP:conf/iclr/JimenezYWYPPN24}), evaluation metrics~\cite{DBLP:journals/tist/evalsurvey}, capabilities analysis~~\cite{DBLP:journals/coling/ChangB24}, and learning-problem hardness~\cite{kalai2025languagemodelshallucinate}, etc. are some of the areas that could be explored taking into account characteristics of transformations and patterns\footnote{Natural Language Generation research has been pinpointing, for instance, that the nature of NLG-tasks seem to play an important role when analyzing quality-probability paradox~\cite{10.1162/tacl_a_00502,DBLP:conf/acl/MeisterWPC22} and ultimately deciding an appropriate decoding strategy.}.
For instance, constrained generation and constrained decoding (e.g.,~\cite{MGD,LMQL,DBLP:conf/iclr/Dekoninck0BV24, koo2024automatabasedconstraintslanguagemodel,DBLP:conf/iclr/PoesiaP00SMG22}, etc.) is regarded as a way to build more principled and robust AI-software~\cite{Sigarch-AI,DBLP:conf/iclr/Dekoninck0BV24}. 
Characteristics of transformations might serve to understand in which extent such ideas could be applied and could be pursued. Just as an example, while results of classification transformations, extractive transformations, and the generation of formally defined SW-entities (e.g., assertions) could be potentially constrained by syntactic and even semantic rules (\texttt{Constrain Generation} pattern),
verbalization and transformations yielding NL abstractions, albeit useful for conditioning, are likely harder to be constrained upfront. Hypothesis search~\cite{DBLP:conf/iclr/WangZPPHG24} combined with certified deductive reasoning~\cite{DBLP:journals/tmlr/PoesiaGZG24} might be interesting steps forward for the generation of such sort of entities.

\paragraph{Compositionality} It is hard to forecast how future generation of LLM-enabled tools will look like. However,   
we speculate some  fundamental shortcoming might remain even if models improve their performance in generative transformations. Breaking problems into pieces/transformations (either manually or automatically) before inference time and the use of external symbolic tools seems to be unavoidable due to phenomena like the compositional gap~\cite{DBLP:conf/emnlp/PressZMSSL23} and ``hard to learn'' problems~\cite{kalai2025languagemodelshallucinate}.  
Modularity based in transformations-like concepts can be seen in frameworks like~\cite{DSPy,DBLP:journals/tmlr/SumersYN024,DBLP:conf/chi/WuJD0MTC22}
in which solutions are made up of some sort of declarative modules which instruct ``what'' to do and might let prompting and workflow details be modified as part of an optimization process.
Yet, modularity does not mean truly compositional analysis on solutions is actually enabled. That is, compositional reasoning (e.g.,~\cite{DBLP:journals/cacm/Hoare69}) which is formally and informally ubiquitous in old-plain software, seems to be challenging for LLM-based software~\cite{brabermannapoli}. In a few words, compositional treatment of LLM-native software would mean that local measurable improvements in a constituent transformation would lead to ex-ante guaranteed  end-to-end improvement.  
Catalogs of transformations and pattens of compositions can help to have concrete practical observations to guide approaches address such challenges. 
In a few words, future research on how to elicit,  evaluate, abstractly characterize, and compose transformations in the context of LLM-native software may benefit from identifying recurrent transformations and how they are typically used together in inter-transformation patterns, as done in this work.

\section{Threats to Validity}
 
Naturally, due to the youth of the area and the variety of SE problems and communities, LLM-enabled studies and proposals are presented very differently from one paper to another. 
For instance, LLM interaction sometimes might be described in its low-level raw conversational nature or as just a short hint on how LLM is used. That is, transformations are not always explicitly identified nor functionally described in some presentations. We did our best effort in rationalizing (and filling the unknowns) in the inner workings of studies in scope. However, many details might have been wrongly captured due to our incorrect understanding of those studies and proposals.

 Also, for the sake of simplicity, some details regarding how transformations are organized in a conversation or chain might not, in some cases, be completely or accurately reported. That is, for instance, there might be some imprecision (or even inaccuracy) regarding whether a prompt for a given transformation continues a given conversation (that is a consequence of previous transformations) or, alternatively, a brand new LLM session is launched and only some verbalized results yielded in the previous conversation are copied into the new context.
We are aware that those details might be crucial in practice to achieve good task quality performance~\cite{DBLP:conf/acl/LuBM0S22,Teller,zhao2023-survey} (see \autoref{sec:prelim-conclus}), but we favor highlighting conceptual characteristics of transformations elicited by prompts in studied approaches. 

Mapping from prompts to the identified transformations is not always trivial.
This justifies in part this work but, naturally, it also elevates the risk of being inaccurate or too simplistic in our description. Thus, although we tried to be as faithful as possible to the actual (but still abstract) prompting architecture, by design, this traceability should not be taken for granted. Lack of trivial traceability also occurs when we split transformations instructed by a single prompt to pinpoint the different nature of those transformations. Of course, we do not preclude that best quality performance is achieved when a single prompt elicits all transformations in sequence but that is not the focus of this study, as already mentioned.

In summary, phenomenology based on transformations implies tacitly adhering to a (virtual) modular and functional view of how LLM-enabled approaches interact with LLMs in terms of task elicitation. This is done ``by design'', even though this modularity cannot be taken for granted due to other variables playing a role in the distribution of results. 
However, we believe understanding this abstract decomposition provides a first glimpse of the strategy followed in problem-solving with LLMs.

There is a risk of the taxonomy to be overfit to the reported papers and unable to fit future LLM-enabled works. Beyond being a risk for any taxonomy there are a couple of caveats. Firstly, the property/choices trees are suitable for adding both novel relevant properties and new choices.  It is also worth noting the scope and exhaustiveness of SE topics covered: from falsification to formal verification ones (a scope that is rarely covered by surveys). This means that taxonomy as it is was able to classify a rich variety of approaches for quite different SE problems. All papers satisfying the criteria that were mentioned by existing surveys were reported and classified to avoid cherry-picking. Last but not least, we were able to seamlessly accommodate more than 30\% of papers in this study after having defined the main structure of the taxonomy and there was a final validation step in which we were able to map  ASE'24 easily in categories~found.

\section{Conclusion}\label{sec:prelim-conclus}

Both LLMs foundational research and the engineering of software tools on top of them are areas that are very rapidly evolving. This work addresses an initial question on the current  nature of (human-defined) transformations and their composition in LLM-enabled solutions for some areas related to Software Analysis. It does this by detecting and naming them in the extensively reviewed literature and by building a hierarchical taxonomy of transformation classes that accommodates all instances of transformations seen. We leverage the mapping to explain some existing abstract  patterns  of transformations relationship in the design of such solutions. 

This report does not try to answer another related question: is a transformation viewpoint  the best conceptual tool (adequate level of abstraction) for engineering LLM-enabled solutions? 
We believe the answer to this question greatly depends on the future development of theory to engineer this sort of systems. Theory and tools for engineering good LLM-enabled approaches are at their infancy. It is our vision that more compositional and abstract concepts linked to the idea of transformations are required to actually intellectually control behavior models and probabilistic programs that people are building on top of them~\cite{cascades}. The need of principled engineering might also apply even if, in the future, LLMs themselves address  problems by --autoregressively-- proposing and chaining high-level transformations and tools either as at design time or at inference time.

\subsection*{Acknowledgements}

This work was supported by the Coordenação de Aperfeiçoamento de Pessoal de Nível Superior -- Brazil (CAPES-PROEX), Financing Code 001, and by the Fundação de Amparo à Pesquisa do Estado do Amazonas -- FAPEAM, through the POSGRAD 23-24 project.
This research was carried out within the SWPERFI UFAM-MOTOROLA RD\&I Project ``Técnicas de Inteligência Artificial para Análise e Otimização de Desempenho de Software''. This work was also partially supported by CONICET PIP 11220200100084CO, ANPCyT PICT 2018-3835 and 2021-I-A-00755, UBACyT Grants 20020220300079BA and 20020190100126BA, and A-1-2022-1-173516 IDRC-ANII.

\subsection*{Declaration of generative AI and AI-assisted technologies in the manuscript preparation process}

During the preparation of this work the authors used DeepSeek and ChatGPT in order to improve grammar and style of some sentences. After using this tool/service, the authors reviewed and edited the content as needed and take full responsibility for the content of the published article.


\appendix 
\section{Transformations Taxonomy Trees}\label{sec:AppendixTrees}

We include here the omitted detailed trees of the taxonomy were we have mapped the transformations from the analyzed tools.

\begin{figure}[htb]
    \centering
    \resizebox{!}{55\htree}{%
    \begin{tikzpicture}[tasktree]
 
\node [draw,celesteA] {\textbf{General-Code\phantom{j}Generation}}
    child [missing] {}
    child [celesteA] {node {Generation criteria}
        child [celesteA] {node [darkgray] {Intent corresponding: \testing{TitanFuzz}{1}, \testing{AID}{1}, \testing{DiffPrompt}{2},   \progver{Dafny-Synth}{1}, }
        child [white] {node [darkgray] {\phantom{Int correspondng:} \testing{ClarifyGPT}{1}, \progver{Clover}{3}, \testingU{FormAI}}}
          child [celesteA] {node {Clarifications}
                child [celesteA] {node [darkgray] {Test provided: \progver{Dafny-Synth}{2}}}
               child [celesteA] {node [darkgray] {Signature provided: \progver{Dafny-Synth}{2}, \progver{Dafny-Synth}{5}}}
               child [celesteA] {node [darkgray] {Specification provided: \progver{Dafny-Synth}{5}}}
            edge from parent}
            child [missing] {}
            child [missing] {}                        
            child [missing] {}            
            child [celesteA] {node [darkgray] {Verifiable: \progver{Dafny-Synth}{5}}} 
            child [celesteA] {node [darkgray] {Non-functional requirements satisfying}
                child [celesteA] {node [darkgray] {Complexity: \testingU{ALGO}}}
                child [celesteA] {node [darkgray] {Style: \testing{AID}{1}, \testingU{FormAI}}}
            edge from parent}
        edge from parent}
        child [missing] {}
        child [missing] {}
        child [missing] {}
        child [missing] {}
        child [missing] {}
        child [missing] {}
        child [missing] {}
        child [missing] {}
        child [missing] {}
        child [celesteA] {node [darkgray] {Annotation corresponding: \progver{Clover}{1}, \progver{Clover}{2}}}
        child [celesteA] {node [darkgray] {Repair request corresponding}
            child [celesteA] {node [darkgray] {Defect-description repairing:  \vmf{WitheredLeaf}{3(CoT)}, \debug{AutoSD}{3}, }}
            child [white,draw opacity=0] {node [darkgray] {\phantom{Defect-description repairing:} 
            \vmf{LLM4Vuln}{3(patch)}}}
             child [celesteA] {node [darkgray] {Error-feedback repairing:   \progver{AlloyRepair}{1}}}
              child [celesteA] {node [darkgray] {Suggestion-based repairing:   \progver{AlloyRepair}{2}}}
        edge from parent}
        child [missing] {}
        child [missing] {}
        child [missing] {}
        child [missing] {}
        child [celesteA] {node [darkgray] {Examples based}
            child [celesteA] {node [darkgray] {Translation \phantom{j}}
                child [celesteA] {node [darkgray] {API: \testing{FuzzGPT}{2}}}
            edge from parent}
            child [missing] {}
            child [celesteA] {node [darkgray] {Parametrization: \testing{SearchGEM5}{1}}}
            child [celesteA] {node [darkgray] {Mutation: \testing{FSML}{DS/MG}, \debugU{LLM4CBI}}
                child [celesteA] {node [darkgray] {Bug injecting: \testingU{BugFarm}}}  
                child [celesteA] {node [darkgray] {Different behavior: \testingU{LLMorpheus}}}  
            edge from parent}
            child [missing] {}
            child [missing] {}
            child [celesteA] {node [darkgray] {Repair: \progver{AlloyRepair}{1}, \progver{AlloyRepair}{2}, \vmf{WitheredLeaf}{3(CoT)}, \debug{AutoSD}{3}, }}
             child [white,draw opacity=0] {node [darkgray] {\phantom{Repair:} \testing{AID}{1}, \vmf{LLM4Vuln}{3(patch)} }}
            child [celesteA] {node [darkgray] {Completion}
                child [celesteA] {node [darkgray] {Natural: \vmfU{FLAG}, \testing{TitanFuzz}{2}, \testingUD{MuBert}{$\mu$BERT}, \testing{CHEMFUZZ}{1}}}
            edge from parent}
        edge from parent}
    edge from parent}
    child [missing] {}
    child [missing] {}
    child [missing] {}
    child [missing] {}
    child [missing] {}
    child [missing] {}
    child [missing] {}
    child [missing] {}
    child [missing] {}
    child [missing] {}
    child [missing] {}
    child [missing] {}
    child [missing] {}
    child [missing] {}
    child [missing] {}
    child [missing] {}
    child [missing] {}
    child [missing] {}
    child [missing] {}
    child [missing] {}
    child [missing] {}
    child [missing] {}
    child [missing] {}
    child [missing] {}
    child [missing] {}
    child [missing] {}
    child [missing] {} 
    child [celesteA] {node {Nature of primary generated entity}
        child [celesteA] {node [darkgray] {Code: \vmfU{FLAG}, \testing{FSML}{DS/MG}, \vmf{WitheredLeaf}{3(CoT)}, \testing{SearchGEM5}{1}, }}
        child [white,draw opacity=0] {node [darkgray] {\phantom{Code:} \testing{TitanFuzz}{1}, \testing{TitanFuzz}{2}, \testing{FuzzGPT}{2}, \testingU{BugFarm}, }}
        child [white] {node [darkgray] {\phantom{Code:} \debug{AutoSD}{3},   \testingUD{MuBert}{$\mu$BERT}, \testing{DiffPrompt}{2},  \testing{AID}{1},  }}
        child [white] {node [darkgray] {\phantom{Code:} \progver{Dafny-Synth}{1}, \progver{Dafny-Synth}{2}, \progver{Dafny-Synth}{5},    \testing{ClarifyGPT}{1},   }}
        child [white] {node [darkgray] {\phantom{Code:} \testing{CHEMFUZZ}{1}, \progver{Clover}{1}, \progver{Clover}{2}, \progver{Clover}{3},  }}
         child [white] {node [darkgray] {\phantom{Code:} \vmf{LLM4Vuln}{3(patch)}, \testingU{FormAI},  \testingU{LLMorpheus}, \debugU{LLM4CBI}, }}
         child [white] {node [darkgray] {\phantom{Code:} \testingU{ALGO} \phantom{Ag}}}
        child [celesteA] {node [darkgray] {Model: \progver{AlloyRepair}{1}, \progver{AlloyRepair}{2}}}
    edge from parent}
    child [missing] {}
    child [missing] {}
    child [missing] {}
    child [missing] {}
    child [missing] {}
    child [missing] {}
    child [missing] {}
    child [missing] {}
    child [celesteA] {node {Nature of extra generated entity}
        child [celesteA] {node [darkgray] {Annotations}
            child [celesteA] {node [darkgray] {Pre/Postcondition: \progver{Dafny-Synth}{1}, \progver{Dafny-Synth}{2}}}
            child [celesteA] {node [darkgray] {Loop invariant / Verification annotations: \progver{Dafny-Synth}{5}}}
        edge from parent}
    edge from parent}
    child [missing] {}
    child [missing] {}
    child [missing] {}  
    child [celesteA] {node {In-filling context}
        child [celesteA] {node [darkgray] {Annotated code: \progver{Clover}{1}, \progver{Clover}{2}, \progver{Clover}{3}}}
        child [celesteA] {node [darkgray] {Masked code: \vmfU{FLAG}, \testing{TitanFuzz}{2}, \testingUD{MuBert}{$\mu$BERT}, \testing{CHEMFUZZ}{1},  }}
        child [white,draw opacity=0] {node [darkgray] {\phantom{Masked code:} \testingU{LLMorpheus}}}
    edge from parent}
    child [missing] {}
    child [missing] {}
    child [missing] {}
    child [celesteA] {node {Corrective mode \phantom{j}}
        child [celesteA] {node [darkgray] {Feedback based \phantom{j}}
            child [celesteA] {node [darkgray] {Compiler: \progver{Clover}{2}, \progver{Clover}{3}}}
            child [celesteA] {node [darkgray] {Verifier: \progver{Clover}{1}}}
            child [celesteA] {node [darkgray] {Validity: \progver{AlloyRepair}{1}, \testing{CHEMFUZZ}{1}, \debugU{LLM4CBI}}}
            child [celesteA] {node [darkgray] {Suggested corrections: \progver{AlloyRepair}{2}}}
        edge from parent}
    edge from parent}
    child [missing] {}
    child [missing] {}
    child [missing] {}
    child [missing] {}
    child [missing] {}
    child [celesteA] {node {Rationalization \phantom{j}}
        child [celesteA] {node [darkgray] {Explanation: \testingU{LLMorpheus}}}
    edge from parent};
\end{tikzpicture}
 
    }%
    \caption{General-Code Generation. }\label{fig:GCG}
\end{figure}

\begin{figure}[htb]
    \centering
    \resizebox{!}{40\htree}{%
     \begin{tikzpicture}[tasktree]
 
\node [draw,celesteA] {\textbf{Domain-Specific-Code\phantom{j}Generation}}
    child [missing] {}
        child [celesteA] {node {Generation criteria}
        child [celesteA] {node [darkgray] {Problem-description corresponding}
            child [celesteA] {node [darkgray] {DSL-function description: \testingU{Eywa}}}
            child [celesteA] {node [darkgray] {Function intent: \progver{Dafny-Synth}{3}}}
            child [celesteA] {node [darkgray] {Object description: \static{LLMDFA}{4}, \static{LLMDFA}{5}}}
            child [celesteA] {node [darkgray] {Scenario: \testing{ScenicNL}{4}, \ntesting{LeGEND}{2}}}
            child [celesteA] {node [darkgray] {Constraint: \testing{AID}{2}}}
            child [celesteA] {node [darkgray] {Property: \testing{EMR}{3}}}
            child [celesteA] {node [darkgray] {Rule: \testing{InputBlaster}{4}}}
        edge from parent}
        child [missing] {}
        child [missing] {}
        child [missing] {}
        child [missing] {}
        child [missing] {}
        child [missing] {}
        child [missing] {}
        child [celesteA] {node [darkgray] {Ability to execute code: \testing{OSS-Fuzz}{1}, \testing{OSS-Fuzz}{2}, \testing{PBT-GPT}{2}, }
        child [white] {node [darkgray] {\phantom{ity to execute code:} \testing{UGCTX}{1}, \testing{UGCTX}{2}}}
        child [celesteA] {node [darkgray] {Ability to trigger vulnerability: \ntesting{AdvSCanner}{1}}}}
        child [missing] {}
        child [missing] {}
        child [celesteA] {node [darkgray] {High precision \& recall: \static{LLMDFA}{1}, \static{LLMDFA}{2}}}
    edge from parent}
    child [missing] {}
    child [missing] {}
    child [missing] {}
    child [missing] {}
    child [missing] {}
    child [missing] {}
    child [missing] {}
    child [missing] {}
    child [missing] {}
    child [missing] {}
    child [missing] {}
    child [missing] {}
    child [celesteA] {node {Nature of domain/entity}
        child [celesteA] {node [darkgray] {Scripting for SW analysis}
            child [celesteA] {node [darkgray] {Source/Sink identification: \static{LLMDFA}{1}, \static{LLMDFA}{2}}}
        edge from parent}
        child [missing] {}
        child [celesteA] {node [darkgray] {Metamorphic relation: \testing{EMR}{3}}}
        child [celesteA] {node [darkgray] {Encoders \phantom{j}}
            child [celesteA] {node [darkgray] {Path condition: \static{LLMDFA}{4}, \static{LLMDFA}{5}}}
        edge from parent}
        child [missing] {}
        child [celesteA] {node [darkgray] {Test drivers \phantom{j}}
            child [celesteA] {node [darkgray] {Fuzzing:  \testing{OSS-Fuzz}{1}, \testing{OSS-Fuzz}{2}, \testing{UGCTX}{1}, \testing{UGCTX}{2}}}
            child [celesteA] {node [darkgray] {Input generators: \testing{AID}{2}, \testing{InputBlaster}{4}, \testing{PBT-GPT}{2}}}
        edge from parent}
        child [missing] {}
        child [missing] {}
        child [celesteA] {node [darkgray] {Protocol-related code: \testingU{Eywa}}}
        child [celesteA] {node [darkgray] {Word model: \testing{ScenicNL}{4}, \ntesting{LeGEND}{2}}}
        child [celesteA] {node [darkgray] {Signatures: \progver{Dafny-Synth}{3}}}
        child [celesteA] {node [darkgray] {Adversarial SmartContract: \ntesting{AdvSCanner}{1}}}
    edge from parent}
    child [missing] {}
    child [missing] {}
    child [missing] {}
    child [missing] {}
    child [missing] {}
    child [missing] {}
    child [missing] {}
    child [missing] {}
    child [missing] {}
    child [missing] {}
    child [missing] {}
    child [missing] {}
    child [celesteA] {node {In-filling context}
        child [celesteA] {node [darkgray] {Hint code: \static{LLMDFA}{1}, \static{LLMDFA}{2}}}
        child [celesteA] {node [darkgray] {Constrained decoding: \testing{ScenicNL}{4}}}
    edge from parent}
    child [missing] {}
    child [missing] {}
    child [celesteA] {node {Corrective mode \phantom{j}}
        child [celesteA] {node [darkgray] {Feedback based \phantom{j}}
            child [celesteA] {node [darkgray] {Compiler: \testing{OSS-Fuzz}{2}, \static{LLMDFA}{2}, \static{LLMDFA}{5}, \testing{UGCTX}{2}}}
            child [celesteA] {node [darkgray] {Execution \phantom{j}} 
                child [celesteA] {node [darkgray] {Number of positive/negative: \static{LLMDFA}{2}}}
                child [celesteA] {node [darkgray] {Results: \testing{OSS-Fuzz}{2}, \testing{InputBlaster}{4}, \testing{UGCTX}{2} }}
            edge from parent}
            child [missing] {}
            child [missing] {} 
            child [celesteA] {node [darkgray] {Rationalized failure: \ntesting{AdvSCanner}{1}}}
        edge from parent}
    edge from parent}
    child [missing] {}
    child [missing] {}
    child [missing] {}
    child [missing] {}
    child [missing] {}
    child [missing] {}
    child [celesteA] {node {Reactivity}
        child [celesteA] {node [darkgray] {Function calling: \testing{ScenicNL}{4}}}
    edge from parent};
\end{tikzpicture} 
    }%
    \caption{Domain-Specific-Code Generation. }\label{fig:SCG}
\end{figure}

\begin{figure}[htb]
    \centering
    \resizebox{!}{33\htree}{%
    \begin{tikzpicture}[tasktree]
 
\node [draw,celesteA] {\textbf{Annotation\phantom{j}Generation}}
    child [missing] {}
        child [celesteA] {node {Generation criteria \phantom{j}}
            child [celesteA] {node [darkgray] {Syntactic: \testing{KernelGPT}{5}}}
            child [celesteA] {node [darkgray] {Natural: \testingU{TOGLL}, \progverU{ChatInv}, \testing{PBT-GPT}{1}, \progverU{AutoSpec}}
                child [celesteA] {node [darkgray] {Demonstration similarity: \testing{PropertyGPT}{1}, \testing{PropertyGPT}{2}, }}
                child [white,draw opacity=0] {node [darkgray] {\phantom{Demonstration similarity:} \testing{PropertyGPT}{4}, \testingU{CEDAR}}}
            edge from parent}
            child [missing] {}
            child [missing] {}
            child [celesteA] {node [darkgray] {Natural \& Verifiable: \progver{SpecGen}{1}}}
            child [celesteA] {node [darkgray] {Proof-enabling: \progver{Loopy}{1}, \progver{Clover}{5}, \nprogverU{Lam4Inv}, \progver{Lemur}{1}, }}
            child [white,draw opacity=0] {node [darkgray, draw opacity=1] {\phantom{Proof-enabling:} \progver{RustProof}{2}}}
        edge from parent}
        child [missing] {}
        child [missing] {}
        child [missing] {}
        child [missing] {}
        child [missing] {}
        child [missing] {}
        child [missing] {}
        child [celesteA] {node {Nature of entity \phantom{j}}
            child [celesteA] {node [darkgray] {Precondition: \testing{PropertyGPT}{2}, \progver{SpecGen}{1}, \progver{Clover}{5}, \progverU{AutoSpec}}}         
            child [celesteA] {node [darkgray] {Postcondition: \testing{PropertyGPT}{2}, \progver{SpecGen}{1}, \progver{Clover}{5}, \progverU{AutoSpec}}}
            child [celesteA] {node [darkgray] {Loop invariant: \progver{Loopy}{1}, \nprogverU{ESBMCibmc}, \progverU{AutoSpec}, \nprogverU{Lam4Inv}, }}
            child [white,draw opacity=0] {node [darkgray, draw opacity=1] {\phantom{Loop invariant:} \progver{RustProof}{2}}}
            child [celesteA] {node [darkgray] {Assumption: \progver{Lemur}{1}}}
            child [celesteA] {node [darkgray] {Type state: \testing{KernelGPT}{5}}}
            child [celesteA] {node [darkgray] {Assertion: \testingU{TOGLL}, \testing{PropertyGPT}{1}, \testingU{CEDAR}}}
        edge from parent}
    child [missing] {}
    child [missing] {}
    child [missing] {}
    child [missing] {}
    child [missing] {}
    child [missing] {}
    child [missing] {}
    child [celesteA] {node {In-filling context}
        child [celesteA] {node [darkgray] {Masked assertion: \progverU{ChatInv}}}
        child [celesteA] {node [darkgray] {Annotated code: \progver{Loopy}{1}, \progver{SpecGen}{1}, \testingU{CEDAR}, \progver{Clover}{5}, }}
         child [white] {node [darkgray] {\phantom{Annotated code:} \progverU{AutoSpec}, \progver{Lemur}{1}, \progver{RustProof}{2}}}
        child [celesteA] {node [darkgray] {Unit test:  \testingU{TOGLL}, \testingU{CEDAR}}}
    edge from parent}
    child [missing] {}
    child [missing] {}
    child [missing] {}
    child [missing] {}
    child [celesteA] {node {Corrective mode \phantom{j}}
        child [celesteA] {node [darkgray] {Feedback based \phantom{j}}
             child [celesteA] {node [darkgray] {Compiler failure: \testing{PropertyGPT}{3}, \nprogverU{ESBMCibmc}}}
           child [celesteA] {node [darkgray] {Verification failure: \progver{SpecGen}{2}, \nprogverU{ESBMCibmc}, \nprogverU{Lam4Inv}, }
               child [white,draw opacity=0] {node [darkgray, draw opacity=1] {\phantom{Inductivenesssi:}   \progver{Lemur}{2}, \progver{RustProof}{3}}}
                child [celesteA] {node [darkgray] {Inductiveness: \progver{Loopy}{2}}}
            edge from parent}
            child [missing] {}
            child [missing] {}
            child [celesteA] {node [darkgray] {Validity checker: \testing{PropertyGPT}{4}, \testing{KernelGPT}{5}}}
        edge from parent}
    edge from parent}
    child [missing] {}
    child [missing] {}
    child [missing] {}
    child [missing] {}
    child [missing] {}
    child [missing] {}
    child [celesteA] {node {Rationalization \phantom{j}}
        child [celesteA] {node [darkgray] {Invariant explanation: \progver{RustProof}{2}}}
        child [celesteA] {node [darkgray] {Proof explanation: \progver{RustProof}{2}}}
    edge from parent};

\end{tikzpicture}
 
    }%
    \caption{Code-Given Annotation Generation. }\label{fig:AnnG}
  \end{figure}

  \begin{figure}[htb]
    \centering
    \resizebox{!}{29\htree}{%
      
\begin{tikzpicture}[tasktree]
 
\node [draw,celesteA] {\textbf{Data\phantom{j}Generation}}
    child [missing] {}
    child [celesteA] {node {Generation criteria\phantom{j}}
        child [celesteA] {node [darkgray] {Context compliant}
            child [celesteA] {node [darkgray] {Description: \testing{RESTGPT}{2}}}
            child [celesteA] {node [darkgray] {Usage: \testing{Fuzz4All}{2}}}
            child [celesteA] {node [darkgray] {GUI context: \testing{QTypist}{1}, \testing{QTypist}{2}, \testing{InputBlaster}{2}}}
            child [celesteA] {node [darkgray] {Code + arguments: \testing{SearchGEM5}{2}}}
        edge from parent}
        child [missing] {}
        child [missing] {}
        child [missing] {}
        child [missing] {}
        child [celesteA] {node [darkgray] {Constraints based\phantom{j}}
            child [celesteA] {node [darkgray] {$x~|~P(x)$: \testing{RESTGPT}{2}, \testing{InputBlaster}{2}, \testingU{SymPrompt}, \testing{WhiteFox}{2}}
                child [celesteA] {node [darkgray] {$P \equiv$ code reachability: \ntestingU{DirInput}, \ntesting{HITS}{6}, \ntestingU{Magneto} }}
            edge from parent} 
            child [missing] {}   
            child [celesteA] {node [darkgray] {$x~|~\exists i : x=F(i)$: \testing{SELF-DEBUGGING}{OP2} }}     
            child [celesteA] {node [darkgray] {$x~|~i=F(x)$: \testing{mrDetector}{1}}
                child [celesteA] {node [darkgray] {$x~|~i=F(x,j)$: \ndebug{SemSlicer}{4}}}
                child [celesteA] {node [darkgray] {$x~|~\forall y \  g(i,y)\cong f(x,y)$: \ndebug{SemSlicer}{1}, \ndebug{SemSlicer}{2}}}
            edge from parent}
            child [missing] {}
            child [missing] {}
            child [celesteA] {node [darkgray] {$x~|~$ exercises $(A(x)\Rightarrow C(x))$: \testingU{LLMeDiff}}}     
            child [celesteA] {node [darkgray] {$x~|~$ exercises $(Pre(x) \&$ exercises $Post(x,R))$: \testing{TestChain}{1}, \testing{ClarifyGPT}{2}}}     
        edge from parent}
        child [missing] {}
        child [missing] {}
        child [missing] {}
        child [missing] {}
        child [missing] {}
        child [missing] {}
        child [missing] {}
        child [missing] {}
        child [celesteA] {node [darkgray] {Exemplars compliant}
            child [celesteA] {node [darkgray] {Spec-aware edition:  \testing{ChatAFL}{2}, \testing{Fuzz4All}{3}}}         
            child [celesteA] {node [darkgray] {Variation: \testingU{ChatFuzz}, \testing{Fuzz4All}{4}}
                child [celesteA] {node [darkgray] {Error triggering: \testing{FuzzingParsers}{4}}}
            edge from parent}
            child [missing] {}
            child [celesteA] {node [darkgray] {Natural completion: \testing{ChatAFL}{3}}}
        edge from parent}
    edge from parent}
    child [missing] {}
    child [missing] {}
    child [missing] {}
    child [missing] {}
    child [missing] {}
    child [missing] {}
    child [missing] {}
    child [missing] {}
    child [missing] {}
    child [missing] {}
    child [missing] {}
    child [missing] {}
    child [missing] {}
    child [missing] {}
    child [missing] {}
    child [missing] {}
    child [missing] {}
    child [missing] {}
    child [missing] {}
    child [celesteA] {node {In-filling context}
        child [celesteA] {node [darkgray] {Masked data: \testing{QTypist}{2}}}
        child [celesteA] {node [darkgray] {Data completion: \testing{QTypist}{1}, \testing{ChatAFL}{3}}}
    edge from parent}
    child [missing] {}
    child [missing] {}
    child [celesteA] {node {Corrective mode \phantom{j}}
        child [celesteA] {node [darkgray] {Feedback based \phantom{j}}
            child [celesteA] {node [darkgray] {Execution output: \ntestingU{DirInput}}}
            child [celesteA] {node [darkgray] {Self-refine alternative: \ndebug{SemSlicer}{2}}}
        edge from parent}
    edge from parent};

\end{tikzpicture} 
    }%
    \caption{Data Generation. }\label{fig:DataG}
\end{figure}

\begin{figure}[htb]
    \centering
    \resizebox{!}{26\htree}{%
     
\begin{tikzpicture}[tasktree]
 
\node [draw,limaA] {\textbf{Behavior Analysis}}
    child [missing] {}
    child [limaA] {node {Analysis criteria}
        child [limaA] {node [darkgray] {Anomaly detection}
            child [limaA] {node [darkgray] {Open-ended: \testing{CHEMFUZZ}{2}}
                child [limaA] {node [darkgray] {Attack: \nvmf{Skyeye}{1}, \nvmf{Skyeye}{3} }}
                child [limaA] {node [darkgray] {Vulnerability: \testingD{pwnd}{Pwn'd}{3}}}
                child [limaA] {node [darkgray] {View-point conformance: \testing{GameBugDescriptions}{1}}}
                child [limaA] {node [darkgray] {Self-justified: \testing{AXNav}{3}}}
            edge from parent}
            child [missing] {}     
            child [missing] {}     
            child [missing] {}     
            child [missing] {}     
            child [limaA] {node [darkgray] {Sanity check: \testing{PentestGPT}{3}}
               child [limaA] {node [darkgray] {Execution specification conformance: \staticD{EV}{E\&V}{2}}}
            edge from parent}
        edge from parent}  
        child [missing] {}     
        child [missing] {}     
        child [missing] {}     
        child [missing] {}     
        child [missing] {}     
        child [missing] {}     
        child [missing] {}     
        child [limaA] {node [darkgray] {Root-cause: \debug{x-lifecycle}{2}, \debug{AutoFL}{1}, \debug{AutoSD}{1}, \debug{RCAAgents}{1}, }
            child [white] {node [darkgray] {\phantom{t-cause:} \debug{inContextRCA}{2}}}
            child [limaA] {node [darkgray] {Differences: \debug{SELF-DEBUGGING}{FL}}}
        edge from parent}
        child [missing] {}     
        child [missing] {}
        child [limaA] {node [darkgray] {State: \testing{GPTDroid}{2}\phantom{j} }}
    edge from parent}
    child [missing] {}
    child [missing] {}
    child [missing] {}     
    child [missing] {}
    child [missing] {}
    child [missing] {}
    child [missing] {}
    child [missing] {}
    child [missing] {}
    child [missing] {}
    child [missing] {}
    child [missing] {}
    child [limaA] {node {Extra input}
        child [limaA] {node [darkgray] {Architecture: \debug{x-lifecycle}{2}}}
        child [limaA] {node [darkgray] {Code: \debug{SELF-DEBUGGING}{FL}, \debug{AutoFL}{1}, \debug{AutoSD}{1}}}
        child [limaA] {node [darkgray] {Historic incidents: \debug{RCAAgents}{1}, \debug{inContextRCA}{2}}}
        child [limaA] {node [darkgray] {(hidden) Verbalized call sequence: \nvmf{Skyeye}{3}}}
    edge from parent}
    child [missing] {}
    child [missing] {}
    child [missing] {}
    child [missing] {}
    child [limaA] {node {Reactivity}
        child [limaA] {node [darkgray] {Function calls: \debug{AutoFL}{1}, \debug{RCAAgents}{1}}}
        child [limaA] {node [darkgray] {Questions: \debug{RCAAgents}{1}}}
    edge from parent}
    child [missing] {}
    child [missing] {}
    child [limaA] {node {Rationalization \phantom{j}}
        child [limaA] {node [darkgray] {Applied evaluation criteria: \testing{AXNav}{3}}}
        child [limaA] {node [darkgray] {Execution: \debug{SELF-DEBUGGING}{FL}}}
    edge from parent}
    child [missing] {};
\end{tikzpicture}
    }%
    \caption{Behavior Analysis. }\label{fig:BA}
\end{figure}

    \begin{figure}[htb]
        \centering
    {$\quad$\resizebox{!}{11\htree}{%
     
\begin{tikzpicture}[tasktree]

\node [draw,limaA] {\textbf{Line Code-Ranking}}
    child [missing] {}
    child [limaA] {node {Analysis criteria}
            child [limaA] {node [darkgray] {Fault proneness: \vmf{ChatGPT4vul}{2}, \debug{ChatGPT-4(Log)}{1}, \debug{ChatGPT-4(Log)}{2}}}
    edge from parent}
    child [missing] {}
    child [limaA] {node {Extra input}
            child [limaA] {node [darkgray] {Intent: \debug{ChatGPT-4(Log)}{1}, \debug{ChatGPT-4(Log)}{2}}}
    edge from parent}
    child [missing] {}
    child [limaA] {node {Corrective mode}
            child [limaA] {node [darkgray] {Test results: \debug{ChatGPT-4(Log)}{2}}}
    edge from parent}
    child [missing] {}
    child [limaA] {node {Rationalization}
            child [limaA] {node [darkgray] {Intent: \debug{ChatGPT-4(Log)}{1}, \debug{ChatGPT-4(Log)}{2}}}
            child [limaA] {node [darkgray] {Reason: \debug{ChatGPT-4(Log)}{1}, \debug{ChatGPT-4(Log)}{2}}}
    edge from parent};
\end{tikzpicture}
    }}%
   \caption{Line Code-Ranking. }\label{fig:LCR}
  \end{figure}

      \begin{figure}[htb]
        \centering
    {\resizebox{!}{5\htree}{%
    \begin{tikzpicture}[tasktree]

\node [draw,limaA] {\textbf{Code-Scaling}}
    child [missing] {}
    child [limaA] {node {Scale}
            child [limaA] {node [darkgray] {CVSS: \vmf{ChatGPT4vul}{4}, \vmf{MultiTask}{2}}}
            child [limaA] {node [darkgray] {Vulnerability probability: \vmf{DLAP}{1}}
            edge from parent}
    edge from parent};
\end{tikzpicture}
    }}%
   \caption{Code Scaling. }\label{fig:CSC}
  \end{figure}

    \begin{figure}[htb]
    \centering
    \resizebox{!}{43\htree}{%
     \begin{tikzpicture}[tasktree]

\node [draw,limaA] {\textbf{Direct\phantom{j}Code-Classification}}
    child [missing] {}
    child [limaA] {node {Analysis criteria}
        child [limaA] {node [darkgray] {Vulnerability proneness: \vmf{SmartAudit}{1}, \vmf{ChatGPT4vul}{1}, \vmf{ChatGPT4vul}{3},  }
        child [white] {node [darkgray] {\phantom{nerability proneness:} \vmf{NLBSE24}{1}, \vmfU{GRACE}, \vmfU{AIagent},  \vmf{MultiTask}{1}, }}
        child [white] {node [darkgray] {\phantom{nerability proneness:}  \vmf{MultiTask}{3}, \vmf{MultiTask}{4}, \vmf{PromptEnhanced}{2}, }}
        child [white] {node [darkgray] {\phantom{nerability proneness:}  \vmfU{ChatGPT(Plus)}}}
child [limaA] {node [darkgray] {Similar to demonstrations: \vmfU{GRACE}}}
        edge from parent}
        child [missing] {}
        child [missing] {}
        child [missing] {}
        child [missing] {}
        child [limaA] {node [darkgray] {Malware proneness: \nvmf{SpiderScan}{1}}}
        child [limaA] {node [darkgray] {Used API: \testing{FuzzGPT}{1}, \nvmf{SpiderScan}{2}}}   
        child [limaA] {node [darkgray] {API Misuse: \vmf{LLMAPIDet}{3}}}
        child [limaA] {node [darkgray] {Category of bug: \vmf{WitheredLeaf}{4}}}
        child [limaA] {node [darkgray] {Adherence to description: \vmf{NLBSE24}{2}, \vmf{GPTScan}{1}}}
        child [limaA] {node [darkgray] {Adherence to behavioral description: \testing{SELF-DEBUGGING}{OP7}, \vmf{GPTScan}{2}}}  
    edge from parent}
    child [missing] {}
    child [missing] {}
    child [missing] {}
    child [missing] {}
    child [missing] {}
    child [missing] {}
    child [missing] {}
    child [missing] {}
    child [missing] {}
    child [missing] {}
    child [missing] {}
    child [limaA] {node {Nature of classification \phantom{j}}
        child [limaA] {node {Multiplicity}
            child [limaA] {node [darkgray] {OneClass (Binary): \vmf{SmartAudit}{1},  \vmf{ChatGPT4vul}{1}, \nvmf{SpiderScan}{1},  }
                child [white] {node [darkgray] {\phantom{Training-tiimee}  \vmf{NLBSE24}{1}, \vmfU{GRACE}, \vmf{MultiTask}{1},  }}
                child [white] {node [darkgray] {\phantom{Training-tiimee} \vmf{MultiTask}{3}, \vmf{PromptEnhanced}{2} }}
                child [limaA] {node [darkgray] {In-context defined class: \testing{SELF-DEBUGGING}{OP7}, \vmf{NLBSE24}{2}, }}
                child [white,draw opacity=0] {node [darkgray] {\phantom{In-context defined class:} \vmf{GPTScan}{1}, \vmf{GPTScan}{2}, }}
                child [white,draw opacity=0] {node [darkgray] {\phantom{In-context defined class:} \vmf{LLMAPIDet}{3}, \vmfU{ChatGPT(Plus)}}}
            edge from parent}
            child [missing] {}
            child [missing] {}
            child [missing] {}
            child [missing] {}
            child [missing] {}
            child [limaA] {node [darkgray] {MultiClass/MultiLabel}
                child [limaA] {node [darkgray] {Open-ended:
                \testing{FuzzGPT}{1}}}
                child [limaA] {node [darkgray] {Close-ended: \vmf{ChatGPT4vul}{3}, \nvmf{SpiderScan}{2}, \vmfU{AIagent}, }}
                child [white] {node [darkgray] {\phantom{Close-ended:}  \vmf{MultiTask}{4}}}
                child [limaA] {node [darkgray] {Semi-close-ended: \vmf{WitheredLeaf}{4}}}
            edge from parent}
        edge from parent}
    edge from parent}
    child [missing] {}
    child [missing] {}
    child [missing] {}
    child [missing] {}
    child [missing] {}
    child [missing] {}
    child [missing] {}
    child [missing] {}
    child [missing] {}
    child [missing] {}
    child [missing] {}
    child [missing] {}
    child [limaA] {node {Analysis granularity}
        child [limaA] {node [darkgray] {Snippet:
            \testing{SELF-DEBUGGING}{OP7}, \vmf{SmartAudit}{1}, \testing{FuzzGPT}{1},}}
        child [white] {node [darkgray] {\phantom{Snippet:}    \vmf{ChatGPT4vul}{1}, \vmf{ChatGPT4vul}{3}, \nvmf{SpiderScan}{1}, \vmf{NLBSE24}{1},  }}
            child [white] {node [darkgray] {\phantom{Snippet:} \vmf{NLBSE24}{2}, \vmfU{GRACE}, \vmf{GPTScan}{1}, \vmf{GPTScan}{2},   }} 
            child [white] {node [darkgray] {\phantom{Snippet:} \vmf{LLMAPIDet}{3}, \vmf{MultiTask}{1}, \vmf{PromptEnhanced}{2}, }}
            child [white] {node [darkgray] {\phantom{Snippet:}  \vmfU{ChatGPT(Plus)}}}
        child [limaA] {node [darkgray] {Line: \vmf{WitheredLeaf}{4}, \vmfU{AIagent}, \vmf{MultiTask}{3}}}  
    edge from parent}
    child [missing] {}
    child [missing] {}
    child [missing] {}
    child [missing] {}
    child [missing] {}
    child [missing] {}
    child [limaA] {node {Extra input}
        child [limaA] {node [darkgray] {Code property graph: \vmfU{GRACE}}}
        child [limaA] {node [darkgray] {Intent: \vmf{PromptEnhanced}{2}}}
        child [limaA] {node [darkgray] {Ctrl/Data-flow: \vmf{PromptEnhanced}{2}}}
        child [limaA] {node [darkgray] {Call sequence: \vmf{PromptEnhanced}{2}}}
        child [limaA] {node [darkgray] {Bug reason: \vmf{WitheredLeaf}{4}}}
        child [limaA] {node [darkgray] {Assertion: \testing{SELF-DEBUGGING}{OP7}}}
    edge from parent}
    child [missing] {}
    child [missing] {}
    child [missing] {}
    child [missing] {}
    child [missing] {}
    child [missing] {}
    child [limaA] {node {Rationalization \phantom{j}}
        child [limaA] {node [darkgray] {Background: \vmf{GPTScan}{1}, \vmf{GPTScan}{2}}}
        child [limaA] {node [darkgray] {Execution: \testing{SELF-DEBUGGING}{OP7}}}
    edge from parent};
\end{tikzpicture}

 
    }%
    \caption{Direct Code-Classification. }\label{fig:DCC}
  \end{figure}

\begin{figure}[htb]
    \centering
   \resizebox{!}{18\htree}{%
    \begin{tikzpicture}[tasktree]

\node [draw,limaA] {\textbf{Task-Solution Analysis}}
    child [missing] {}
    child [limaA] {node {Analysis criteria}
        child [limaA] {node [darkgray] {Correctness of analysis: \vmf{VulDetect}{4}, \static{LLift}{2(SelfValid)}, \static{LLift}{3(SelfValid)}}
            child [limaA] {node [darkgray] {False-positive warning filtering: \vmf{WitheredLeaf}{2}, \static{SkipAnalyzer}{3} }}
        edge from parent}
        child [missing] {}
        child [limaA] {node [darkgray] {Quality assessment: \testing{mrDetector}{2}, \debug{LM-PACE}{3}}
            child [limaA] {node [darkgray] {Scored: \vmf{GPTLens}{2}, \staticU{CORE}}}
        edge from parent}
        child [missing] {}
        child [limaA] {node [darkgray] {Reason of failure: \ntesting{AdvSCanner}{2}}}
    edge from parent}
    child [missing] {}
    child [missing] {}
    child [missing] {}
    child [missing] {}
    child [missing] {}
    child [limaA] {node {Domain\phantom{j}}
        child [limaA] {node [darkgray] {Vulnerability analysis: \vmf{WitheredLeaf}{2}, \vmf{GPTLens}{2}}}
        child [limaA] {node [darkgray] {Root cause analysis: \debug{LM-PACE}{3}}}
        child [limaA] {node [darkgray] {Program analysis: \vmf{VulDetect}{4}, \static{LLift}{2(SelfValid)}, \static{LLift}{3(SelfValid)}, }}
        child [limaA, draw opacity=0] {node [darkgray, draw opacity=1] {\phantom{Program analysis:} \static{SkipAnalyzer}{3}}}
        child [limaA] {node [darkgray] {Program repair: \staticU{CORE}}}
        child [limaA] {node [darkgray] {Information retrieval: \testing{mrDetector}{2}}}
        child [limaA] {node [darkgray] {Exploit generation: \ntesting{AdvSCanner}{2}}}
    edge from parent}
    child [missing] {}
    child [missing] {}
    child [missing] {}
    child [missing] {}
    child [missing] {}
    child [missing] {}
    child [missing] {}
    child [limaA] {node {Rationalization\phantom{j}}
        child [limaA] {node [darkgray] {Justification: \vmf{VulDetect}{4}}}
    edge from parent}
    child [missing] {};
\end{tikzpicture}
  
    }%
    \caption{Task-Solution Analysis (some used in the self-reflection framework). }\label{fig:TSA}
\end{figure}

\begin{figure}[htb]
    \centering
    \resizebox{!}{27\htree}{%
     
\begin{tikzpicture}[tasktree]

\node [draw,limaA] {\textbf{Text Analysis}}
        child [missing] {}
        child [limaA] {node {Analysis type}
            child [limaA] {node [darkgray] {Comparative analysis}
                child [limaA] {node [darkgray] {Functional equivalence: \testing{SELF-DEBUGGING}{OP6}, \progver{Clover}{8}}}
                child [limaA] {node [darkgray] {Similarity analysis: \ndebug{FAIL}{4} }}
                child [limaA] {node [darkgray] {Generalized consensus: \nvmf{Skyeye}{4} }}
            edge from parent}
            child [missing] {}
            child [missing] {}
            child [missing] {}
            child [limaA] {node [darkgray] {Ambiguity analysis: \testing{ScenicNL}{2}, \testing{ClarifyGPT}{4}}}
            child [limaA] {node [darkgray] {Support analysis}
                child [limaA] {node [darkgray] {To conclude: \debug{AutoSD}{2}, \ndebug{FAIL}{9}}}
                child [limaA] {node [darkgray] {To perform: \debug{LM-PACE}{1}}}
                child [limaA] {node [darkgray] {To analyze: \ndebug{FAIL}{2}}}
            edge from parent}
            child [missing] {}
            child [missing] {}
            child [missing] {}
            child [limaA] {node [darkgray] {Closed analysis}
                child [limaA] {node [darkgray] {Q\&A: \ndebug{LasRCA}{1}}}
            edge from parent}
            child [missing] {}
            child [limaA] {node [darkgray] {Open analysis}
                child [limaA] {node [darkgray] {Q\&A: \debug{RCAAgents}{2}}
                    child [limaA] {node [darkgray] {Expert completed: \testing{ScenicNL}{3} \phantom{j}}}
                edge from parent}        
            edge from parent}
            child [missing] {}
            child [missing] {}
            child [limaA] {node [darkgray] {Classification \phantom{j}}
                child [limaA] {node [darkgray] {Binary: \ndebug{FAIL}{4}}
                    child [limaA] {node [darkgray] {In-context defined class: \ndebug{SemSlicer}{3} \phantom{j}}}
                edge from parent} 
                child [missing] {}
                child [limaA] {node [darkgray] {Multilabel \phantom{j}}
                    child [limaA] {node [darkgray] {Close ended: \ndebug{LasRCA}{2}, \ndebug{LasRCA}{3} \phantom{j}}}
                    child [limaA] {node [darkgray] {Open ended: \nvmf{Skyeye}{4} \phantom{j}}}
                edge from parent}        
            edge from parent}
        edge from parent}
        child [missing] {}
        child [missing] {}
        child [missing] {}
        child [missing] {}
        child [missing] {}
        child [missing] {}
        child [missing] {}
        child [missing] {}
        child [missing] {}
        child [missing] {}
        child [missing] {}
        child [missing] {}
        child [missing] {}
        child [missing] {}
        child [missing] {}
        child [missing] {}
        child [missing] {}
        child [missing] {}
        child [missing] {}
        child [missing] {}
        child [limaA] {node {Rationalization \phantom{j}}
            child [limaA] {node [darkgray] {Debate: \testing{ScenicNL}{3}}}
            child [limaA] {node [darkgray] {Thought step by step: \ndebug{LasRCA}{2}}}
            child [limaA] {node [darkgray] {Reasons: \ndebug{LasRCA}{3}}}
        edge from parent};
\end{tikzpicture}
  
    }%
    \caption{Text Analysis. }\label{fig:TA}
\end{figure}

\begin{figure}[htb]
    \centering
    \resizebox{!}{29\htree}{%
    \begin{tikzpicture}[tasktree]
 
\node [draw,naranjaB] {\textbf{Code-Elements\phantom{j}Identification \& Extraction}}
    child [missing] {}
    child [naranjaB] {node {Identification criteria \phantom{j}}
        child [naranjaB] {node [darkgray] {Role based \phantom{j}}
            child [naranjaB] {node [darkgray] {Specialized \phantom{j}}
                child [naranjaB] {node [darkgray] {Command values: \testing{KernelGPT}{2}}}
                child [naranjaB] {node [darkgray] {Definition}
                    child [naranjaB] {node [darkgray] {Type: \testing{KernelGPT}{4}}}
                edge from parent}
                child [missing] {}
                child [naranjaB] {node [darkgray] {Type: \testing{KernelGPT}{3}}}
                child [naranjaB] {node [darkgray] {Initializator: \static{LLift}{1}, \testing{KernelGPT}{1}}}
                child [naranjaB] {node [darkgray] {Leakable parameter: \static{LATTE}{1}}}
                child [naranjaB] {node [darkgray] {Resource\_leak-related: \staticU{InferROI}}}
                child [naranjaB] {node [darkgray] {External input: \static{LATTE}{2}}}
                child [naranjaB] {node [darkgray] {Device name: \testing{KernelGPT}{1}}}
                child [naranjaB] {node [darkgray] {Settings to run: \ntesting{HITS}{2}}}
                child [naranjaB] {node [darkgray] {Problem\_solution\_steps-related: \ntesting{HITS}{4}}}
                child [naranjaB] {node [darkgray] {Used variables \& methods: \ntesting{HITS}{5}}}
                child [naranjaB] {node [darkgray] {Used conditions: \ntesting{HITS}{5}}}
            edge from parent}
            child [missing] {}
            child [missing] {}
            child [missing] {}
            child [missing] {}
            child [missing] {}
            child [missing] {}
            child [missing] {}
            child [missing] {}
            child [missing] {}
            child [missing] {}
            child [missing] {}
            child [missing] {}
            child [missing] {}
            child [naranjaB] {node [darkgray] {Scenario defined: \vmf{GPTScan}{3}}}
        edge from parent}
        child [missing] {}
        child [missing] {}
        child [missing] {}
        child [missing] {}
        child [missing] {}
        child [missing] {}
        child [missing] {}
        child [missing] {}
        child [missing] {}
        child [missing] {}
        child [missing] {}
        child [missing] {}
        child [missing] {}
        child [missing] {}
        child [missing] {}
        child [naranjaB] {node [darkgray] {Relevance based: \staticU{SimulinkSlicer}}}
        child [naranjaB] {node [darkgray] {Structural \phantom{j}}
            child [naranjaB] {node [darkgray] {Blocks of code: \static{CFG-Chain}{1}}}
            child [naranjaB] {node [darkgray] {Code of blocks: \static{CFG-Chain}{2}}}
        edge from parent}
    edge from parent}
    child [missing] {}
    child [missing] {}
    child [missing] {}
    child [missing] {}
    child [missing] {}
    child [missing] {}
    child [missing] {}
    child [missing] {}
    child [missing] {}
    child [missing] {}
    child [missing] {}
    child [missing] {}
    child [missing] {}
    child [missing] {}
    child [missing] {}
    child [missing] {}
    child [missing] {}
    child [missing] {}
    child [missing] {}
    child [missing] {}
    child [naranjaB] {node {Reactivity}
        child [naranjaB] {node [darkgray] {Request further information: \static{LLift}{1}, \testing{KernelGPT}{2}, }}
        child [white,draw opacity=0] {node [darkgray] {\phantom{Request further information:}  \testing{KernelGPT}{3}, \testing{KernelGPT}{4}}}
    edge from parent}
    child [missing] {}
    child [missing] {}
    child [naranjaB] {node {Rationalization \phantom{j}}
        child [naranjaB] {node [darkgray] {Background: \vmf{GPTScan}{3}}}
        child [naranjaB] {node [darkgray] {Dependence: \staticU{SimulinkSlicer}}}
    edge from parent};
\end{tikzpicture}
 
    }%
    \caption{Code-Elements Identification \& Extraction. }\label{fig:CEI}
\end{figure}

\begin{figure}[htb]
    \centering
    \resizebox{!}{21\htree}{%
    \begin{tikzpicture}[tasktree]

\node [draw,naranjaC] {\textbf{Text-Elements\phantom{j}Identification \& 
 Extraction}}
    child [missing] {}
    child [naranjaC] {node {Extraction criteria\phantom{j}}
        child [naranjaC] {node [darkgray] {Defined role\phantom{j}}
            child [naranjaC] {node [darkgray] {Keywords: \debugU{Cupid}}}
            child [naranjaC] {node [darkgray] {Variable names: \testingU{MetaMorph}}}
            child [naranjaC] {node [darkgray] {FSM-elements: \testingU{PROSPER}}}
            child [naranjaC] {node [darkgray] {Topic related sentences}
                child [naranjaC] {node [darkgray] {I/O: \testing{EMR}{1}}}
            edge from parent}
            child [missing] {}
        edge from parent}
        child [missing] {}
        child [missing] {}
        child [missing] {}
        child [missing] {}
        child [missing] {}
        child [naranjaC] {node [darkgray] {Best fit \phantom{j}}
            child [naranjaC] {node [darkgray] {Defined structure: \ndebug{FAIL}{5}, \debug{AdbGPT}{1}, \testing{SysKG-UTF}{1}}}
            child [naranjaC] {node [darkgray] {Defined option: \ndebug{FAIL}{8}}}
            child [naranjaC] {node [darkgray] {Input property}
                child [naranjaC] {node [darkgray] {Common root cause: \debug{RCACopilot}{2}}}
            edge from parent}
        edge from parent}
    edge from parent}
        child [missing] {}
        child [missing] {}
        child [missing] {}
        child [missing] {}
        child [missing] {}
        child [missing] {}
        child [missing] {}
        child [missing] {}
        child [missing] {}
        child [missing] {}
        child [missing] {}
        child [naranjaC] {node {Nature of text \phantom{Ag}}
            child [naranjaC] {node [darkgray] {RFC: \testingU{PROSPER} \phantom{j}}}
            child [naranjaC] {node [darkgray] {Options list: \debug{RCACopilot}{2}}}
        edge from parent}
        child [missing] {}
        child [missing] {}
        child [naranjaC] {node {Literality}
            child [naranjaC] {node [darkgray] {Literal: \ndebug{FAIL}{5}, \testing{EMR}{1}, \testing{SysKG-UTF}{1}, \testingU{MetaMorph}, \debugU{Cupid}}}
            child [naranjaC] {node [darkgray] {Transformative\phantom{j}}
                child [naranjaC] {node [darkgray] {Closest meaning: \debug{AdbGPT}{1}}}
            edge from parent}
        edge from parent}
    child [missing] {};
\end{tikzpicture}
 
  
    }%
    \caption{Text-Elements Identification \& Extraction. }\label{fig:TEE}
\end{figure}

\begin{figure}[htb]
    \centering
    \resizebox{!}{34\htree}{%
    \begin{tikzpicture}[tasktree]
 
\node [draw,fucsiaC] {\textbf{SW-Entity Property Identification \& Characterization}}
    child [missing] {}
    child [fucsiaC] {node {Criteria of characterization \phantom{Ag}}
        child [fucsiaC] {node [darkgray] {Descriptions \& domain corresponding: \testing{FSML}{OP}, \testing{SELF-DEBUGGING}{OP5}, \phantom{Ag}}}
        child [white] {node [darkgray] {\phantom{Descriptions \& domain corresponding:} \testing{RESTGPT}{1}, \progver{Dafny-Synth}{4a}, \testing{EMR}{2} \phantom{Ag}} }
        child [fucsiaC] {node [darkgray] {Code corresponding: \static{CFG-Chain}{3}, \static{LLift}{2}, \static{LLift}{3}, \static{LATTE}{3}, \phantom{Ag}}}
        child [white] {node [darkgray] {\phantom{Code corresponding:} \static{LLMDFA}{3},  \testing{WhiteFox}{1} \phantom{Ag}}}
        child [fucsiaC] {node [darkgray] {Run-time context corresponding: \testing{InputBlaster}{1}, \testing{InputBlaster}{3} \phantom{Ag}}}
    edge from parent}
    child [missing] {}
    child [missing] {}
    child [missing] {}
    child [missing] {}
    child [missing] {}                    
    child [fucsiaC] {node {Nature of property \phantom{Ag}}
        child [fucsiaC] {node [darkgray] {Input-validity constraint: \testing{RESTGPT}{1}, \testing{InputBlaster}{1} \phantom{Ag}}} 
        child [fucsiaC] {node [darkgray] {NL-precondition: \progver{Dafny-Synth}{4a} \phantom{Ag}}}
         child [fucsiaC] {node [darkgray] {Precondition constraint to reach code: \testing{WhiteFox}{1} \phantom{Ag}}}
        child [fucsiaC] {node [darkgray] {NL-postcondition: \progver{Dafny-Synth}{4a} \phantom{Ag}}
            child [fucsiaC] {node [darkgray] {Columns of query: \testing{SELF-DEBUGGING}{OP5} \phantom{Ag}}}
        edge from parent}
        child [missing] {}                    
        child [fucsiaC] {node [darkgray] {Qualified post-constraint: \static{LLift}{2} \phantom{Ag}}}          
        child [fucsiaC] {node [darkgray] {Code facts \phantom{Ag}}
            child [fucsiaC] {node [darkgray] {Taint-flow: \static{LATTE}{3} \phantom{Ag}}}
            child [fucsiaC] {node [darkgray] {Same-value: \static{LLMDFA}{3} \phantom{Ag}}}
            child [fucsiaC] {node [darkgray] {Control-flow: \static{CFG-Chain}{3} \phantom{Ag}}}
        edge from parent}
        child [missing] {}
        child [missing] {}
        child [missing] {}  
        child [fucsiaC] {node [darkgray] {Summary}
            child [fucsiaC] {node [darkgray] {May-must: \static{LLift}{3} \phantom{Ag}}} 
        edge from parent}
        child [missing] {}
        child [fucsiaC] {node [darkgray] {Metamorphic \phantom{Ag}}
            child [fucsiaC] {node [darkgray] {Term equivalence: \testing{FSML}{OP} \phantom{Ag}}}
            child [fucsiaC] {node [darkgray] {NL-relation: \testing{EMR}{2} \phantom{Ag}}}
            child [fucsiaC] {node [darkgray] {Bug-revealing mutation: \testing{InputBlaster}{3} \phantom{Ag}}}
        edge from parent}
    edge from parent}
    child [missing] {}
    child [missing] {}                   
    child [missing] {}
    child [missing] {}
    child [missing] {}
    child [missing] {}
    child [missing] {}
    child [missing] {}
    child [missing] {}
    child [missing] {}
    child [missing] {}
    child [missing] {}
    child [missing] {}
    child [missing] {}
    child [missing] {}
    child [missing] {}
    child [fucsiaC] {node {Corrective mode \phantom{Ag}}
        child [fucsiaC] {node [darkgray] {Self-validation: \static{LLift}{2}, \static{LLift}{3} \phantom{Ag}}}
        child [fucsiaC] {node [darkgray] {Feedback based \phantom{Ag}}
            child [fucsiaC] {node [darkgray] {Test results: \testing{InputBlaster}{3}}}
        edge from parent}
    edge from parent}
    child [missing] {}
    child [missing] {}
    child [missing] {}
    child [fucsiaC] {node {Reactivity \phantom{Ag}}
        child [fucsiaC] {node [darkgray] {Requests further info: \static{LLift}{3} \phantom{Ag}}}        
    edge from parent}
    child [missing] {}
    child [fucsiaC] {node {Rationalization \phantom{Ag}}
        child [fucsiaC] {node [darkgray] {Analysis: \testing{FSML}{OP}, \testing{SELF-DEBUGGING}{OP5}}}
        child [fucsiaC] {node [darkgray] {Thought \& assessment: \static{LLMDFA}{3} \phantom{Ag}}}
    edge from parent};
\end{tikzpicture}
 
    }%
    \caption{SW-Entity-Property Identification \& Characterization. }\label{fig:PrC}
  \end{figure}

\begin{figure}[htb]
    \centering
    \resizebox{!}{17\htree}{%
    \begin{tikzpicture}[tasktree]
 
\node [draw,fucsiaC] {\textbf{Formalization}\phantom{j}}
    child [missing] {}
    child [fucsiaC] {node {Abstraction criteria\phantom{j}}
        child [fucsiaC] {node [darkgray] {Intent/Comment corresponding: \testingU{nl2postcondition}, \progver{Clover}{6}, }}  child [white] {node [darkgray] {\phantom{Intent/Comment corresponding:}  \progver{RustProof}{1}, \nprogverUD{RustC4}{RustC$^{\text{4}}$}}}
        child [fucsiaC] {node [darkgray] {Precondition corresponding: \progver{RustProof}{1}}}
        child [fucsiaC] {node [darkgray] {Code corresponding: \progver{RustProof}{1}}}
    edge from parent}
    child [missing] {} 
    child [missing] {} 
    child [missing] {} 
    child [missing] {}        
    child [fucsiaC] {node {Nature of formalization\phantom{j}}
            child [fucsiaC] {node [darkgray] {State-predicate}
                child [fucsiaC] {node [darkgray] {Postcondition: \testingU{nl2postcondition},  \progver{Clover}{6}, \progver{RustProof}{1}}} 
                child [fucsiaC] {node [darkgray] {Precondition:  \progver{Clover}{6}}}
                child [fucsiaC] {node [darkgray] {Assertion: \nprogverUD{RustC4}{RustC$^{\text{4}}$}}}
            edge from parent}
        edge from parent}
        child [missing] {} 
        child [missing] {}
        child [missing] {}
        child [missing] {}
    child [fucsiaC] {node {In-filling context}
        child [fucsiaC] {node [darkgray] {Annotated code: \progver{Clover}{6}, \progver{RustProof}{1}}}
    edge from parent}
    child [missing] {}
    child [fucsiaC] {node {Corrective mode\phantom{j}}
        child [fucsiaC] {node [darkgray] {Feedback based\phantom{j}}
            child [fucsiaC] {node [darkgray] {Compiler: \progver{Clover}{6}}}
        edge from parent}
    edge from parent};
\end{tikzpicture}
    }%
    \caption{Formalization. }\label{fig:For}
  \end{figure}

   \begin{figure}[htb]
    \centering
          {\resizebox{!}{20\htree}{%
    \begin{tikzpicture}[tasktree]
 
\node [draw,fucsiaC] {\textbf{Focused-Abstractive\phantom{j}Summarization}}
    child [missing] {}
    child [fucsiaC] {node {Abstraction criteria\phantom{j}}
        child [fucsiaC] {node [darkgray] {Sign of answer: \debug{LM-PACE}{2}}}
        child [fucsiaC] {node [darkgray] {Scaled assessment: \debug{LM-PACE}{4}}}
        child [fucsiaC] {node [darkgray] {Concept extraction}
            child [fucsiaC] {node [darkgray] {Usage: \testing{Fuzz4All}{1}, \ndebug{FAIL}{3}}}
            child [fucsiaC] {node [darkgray] {Failure described: \testing{GameBugDescriptions}{2}}}
            child [fucsiaC] {node [darkgray] {Faulty culprit: \debug{AutoFL}{2}}}
            child [fucsiaC] {node [darkgray] {Vulnerability described: \vmf{ChatGPTSCV}{2}}}
            child [fucsiaC] {node [darkgray] {Vulnerability mechanism/vector described: \vmf{LLM4Vuln}{2}, \nvmf{VFFinder}{1} \phantom{Ag}}}
            child [fucsiaC] {node [darkgray] {Incident/Root-cause described: \debug{RCACopilot}{1}, \debug{x-lifecycle}{1}, \nvmf{VFFinder}{1}, }}
            child [fucsiaC,opacity=0] {node [darkgray,opacity=1] {\phantom{Incident/Root-cause described:} \debug{inContextRCA}{1}}}
            child [fucsiaC] {node [darkgray] {Relevant report elements: \testing{ScenicNL}{1}, \ntestingU{SoVAR}, \ntesting{LeGEND}{1}}}      
            child [fucsiaC] {node [darkgray] {Output characterizations: \testing{SELF-DEBUGGING}{OP4}}}  
            child [fucsiaC] {node [darkgray] {Reason for toxicity: \nstatic{ToxicDetector}{1}}}  
        edge from parent}
        child [missing] {}
        child [missing] {}
        child [missing] {}
        child [missing] {}
        child [missing] {}
        child [missing] {}
        child [missing] {}
        child [missing] {}
        child [missing] {}
        child [missing] {}
        child [fucsiaC] {node [darkgray] {Timeline: \ndebug{FAIL}{7}}}
    edge from parent}
    child [missing] {}
    child [missing] {}
    child [missing] {}
    child [missing] {}
    child [missing] {}
    child [missing] {}
    child [missing] {}
    child [missing] {}
    child [missing] {}
    child [missing] {}
    child [missing] {}
    child [missing] {}
    child [missing] {}
    child [missing] {}
    child [fucsiaC] {node {Rationalization\phantom{j}}
        child [fucsiaC] {node [darkgray] {Thoughts: \vmf{ChatGPTSCV}{2}}}
        child [fucsiaC] {node [darkgray] {Debate: \testing{ScenicNL}{1}}} 
    edge from parent};
\end{tikzpicture}
 
 
    }}%
    
    \caption{Focused-Abstractive Summarization (previous thoughts or rationalizations). }\label{fig:FAS}
  \end{figure}

    \begin{figure}[htb]
    \centering
    \resizebox{!}{16\htree}{%
     \begin{tikzpicture}[tasktree]

\node [draw,purpuraA] {\textbf{What-to-do-Next\phantom{j}Generation}}
        child [missing] {}
        child [purpuraA] { node {Generation criteria\phantom{j}}
            child [purpuraA] {node [darkgray] {Most promising: \testing{PentestGPT}{5}}}
            child [purpuraA] {node [darkgray] {Coverage oriented: \testing{GPTDroid}{1}}}
            child [purpuraA] {node [darkgray] {Reachability oriented: \testingD{pwnd}{Pwn'd}{2}, \debug{CrashTranslator}{1}, \debug{CrashTranslator}{2}}}
        edge from parent}
        child [missing] {}
        child [missing] {}
        child [missing] {}
        child [purpuraA] { node {Domain\phantom{j}}
            child [purpuraA] {node [darkgray] {SW-system integration: \testing{PentestGPT}{5}, \testingD{pwnd}{Pwn'd}{2}, \debug{CrashTranslator}{1}, }}
            child [white,draw opacity=0] {node [darkgray] {\phantom{SW-system integration:} 
            \debug{CrashTranslator}{2}, \testing{GPTDroid}{1} }}
        edge from parent}
        child [missing] {}     
        child [missing] {}     
        child [purpuraA] {node {Nature of options}
            child [purpuraA] {node [darkgray] {Open-ended: \testingD{pwnd}{Pwn'd}{2}, \debug{CrashTranslator}{1}, \testing{GPTDroid}{1} }}
            child [purpuraA] {node [darkgray] {Close-ended: \testing{PentestGPT}{5}, \debug{CrashTranslator}{2}}}
        edge from parent}        
        child [missing] {}
        child [missing] {}
        child [purpuraA] {node {Given context\phantom{j}}
            child [purpuraA] {node [darkgray] {Enclosing plan: \testing{PentestGPT}{5}}}
            child [purpuraA] {node [darkgray] {Dynamic annotation: \testing{GPTDroid}{1}}}
            child [purpuraA] {node [darkgray] {GUI context: \debug{CrashTranslator}{1}, \debug{CrashTranslator}{2}, \testing{GPTDroid}{1}}}
        edge from parent};
\end{tikzpicture}
    }%
    \caption{What-to-do-Next Generation. }\label{fig:WTD}
  \end{figure}

    \begin{figure}[htb]
        \centering
    {\resizebox{!}{16\htree}{%
     \begin{tikzpicture}[tasktree]
 
\node [draw,purpuraA] {\textbf{Execution}\phantom{j}}
    child [missing] {}
    child [purpuraA]  {node {Domain\phantom{Ag}}    
         child [purpuraA] {node [darkgray] {General code: \debug{SELF-DEBUGGING}{FL(CoT)}}}        
         child [purpuraA] {node [darkgray] {Pseudocode: \staticD{EV}{E\&V}{1}}}
         child [purpuraA] {node [darkgray] {Intent: \testing{TestChain}{2}}}  
         child [purpuraA] {node [darkgray] {Assertions: \testing{SELF-DEBUGGING}{OP7(CoT)}}}  
    edge from parent}
    child [missing] {}
    child [missing] {}
    child [missing] {}
    child [missing] {}
    child [purpuraA]  {node {Granularity}
        child [purpuraA]  {node [darkgray] {Single step: \testing{SELF-DEBUGGING}{OP7(CoT)}, \staticD{EV}{E\&V}{1}}}
        child [purpuraA]  {node [darkgray] {Multiple steps: \debug{SELF-DEBUGGING}{FL(CoT)}, \testing{TestChain}{2}}}
    edge from parent} 
    child [missing] {}
    child [missing] {}
    child [purpuraA]  {node {Corrective mode \phantom{Ag}}
        child [purpuraA]  {node [darkgray] {Feedback based \phantom{Ag}}
            child [purpuraA]  {node [darkgray]  {Verification results: \staticD{EV}{E\&V}{1} \phantom{Ag}}}
            edge from parent}
    edge from parent}
    child [missing] {}
    child [missing] {}
    child [purpuraA]  {node {Reactivity}
        child [purpuraA]  {node [darkgray]  {Request further information: \staticD{EV}{E\&V}{1}}}
        child [purpuraA]  {node [darkgray]  {Request external tool: \staticD{EV}{E\&V}{1}, \testing{TestChain}{2}}
        edge from parent}
    edge from parent};
\end{tikzpicture}
    }}%
    \caption{Execution. }\label{fig:PSE}
  \end{figure}

      \begin{figure}[htb]
        \centering
    {$\quad$\resizebox{!}{15\htree}{%
    \begin{tikzpicture}[tasktree]
 
\node [draw,purpuraA] {\textbf{Textual\phantom{j}Data Manipulation}}
    child [missing] {}
    child [purpuraA] {node {Operation criteria}
        child [purpuraA] {node [darkgray] {Information projection: \testing{PentestGPT}{4}}}
        child [purpuraA] {node [darkgray] {Update: \testing{PentestGPT}{2}}}
        child [purpuraA] {node [darkgray] {Replacement}
                child [purpuraA] {node [darkgray] {Closest meaning: \testing{TARGET}{3}}}
        edge from parent}
        child [missing] {}
        child [purpuraA] {node [darkgray] {Fusion\phantom{j}}
            child [purpuraA] {node [darkgray] {Control flow graph: \static{CFG-Chain}{4}}}
            child [purpuraA] {node [darkgray] {Structure: \testing{FuzzingParsers}{5}}}
        edge from parent}
    edge from parent}
    child [missing] {}
    child [missing] {}
    child [missing] {}
    child [missing] {}
    child [missing] {}
    child [missing] {}
    child [missing] {}
    child [purpuraA] {node {Nature of structure \phantom{Ag}}
        child [purpuraA] {node [darkgray] {Plan: \testing{PentestGPT}{2}, \testing{PentestGPT}{4} \phantom{Ag}}}
        child [purpuraA] {node [darkgray] {Scenario: \testing{TARGET}{3} \phantom{Ag}}}
        child [purpuraA] {node [darkgray] {Graph: \static{CFG-Chain}{4} \phantom{Ag}}}
        child [purpuraA] {node [darkgray] {String: \testing{FuzzingParsers}{5} \phantom{Ag}}}
    edge from parent};
\end{tikzpicture}
    }}%
    \caption{Textual Data Manipulation. }\label{fig:TDM}
  \end{figure}

\begin{figure}[htb]
    \centering
    \resizebox{!}{13\htree}{%
    \begin{tikzpicture}[tasktree]
 
\node [draw,aguaA] {\textbf{Knowledge Distillation}}
    child [missing] {}
    child [aguaA] {node {Domain\phantom{j}}
        child [aguaA] {node [darkgray] {Vulnerability detection: \vmf{EditTime}{1}, \vmf{EditTime}{2}}}
        child [aguaA] {node [darkgray] {Protocol parsing: \testing{ChatAFL}{1}}}
        child [aguaA] {node [darkgray] {Language parsing: \testing{FuzzingParsers}{1}, \testing{FuzzingParsers}{2}, \testing{FuzzingParsers}{3}}}
        child [aguaA] {node [darkgray] {Toxicity: \nstatic{ToxicDetector}{2}}}
    edge from parent}
    child [missing] {}
    child [missing] {}
    child [missing] {}
    child [missing] {}
    child [aguaA] {node {Nature of knowledge}
        child [aguaA] {node [darkgray] {Definitions: \vmf{EditTime}{1}}
            child [aguaA] {node [darkgray] {Grammar/Structure: \testing{FuzzingParsers}{1}, \testing{ChatAFL}{1}}}
        edge from parent}
        child [missing] {}
        child [aguaA] {node [darkgray] {Examples: \testing{FuzzingParsers}{2}, \vmf{EditTime}{2}}
            child [aguaA] {node [darkgray] {Conceptual: \nstatic{ToxicDetector}{2}}}
        edge from parent}
        child [missing] {}
        child [aguaA] {node [darkgray] {Set of items:  \testing{FuzzingParsers}{3}}}
    edge from parent};

\end{tikzpicture}
    }%
    \caption{Knowledge Distillation.}\label{fig:FR}
     \end{figure}


\end{document}